\documentclass[useAMS,usegraphicx]{mn2e}
\usepackage{natbib}  
\title[Archiving Multi-epoch Data]{Archiving multi-epoch data and the
discovery of variables in the near infrared}
\author[Cross et al.]{\parbox[t]{\textwidth}
{N.J.G. Cross$^{1,*}$, R.S. Collins$^{1}$, N.C. Hambly$^{1}$, R. P.
Blake$^{1}$, M.A. Read$^{1}$, E.T.W. Sutorius$^{1}$, R.G. Mann$^{1}$,  P.M.
Williams$^{1}$} \vspace*{6pt} \\ $^{1}$ Scottish Universities' Physics Alliance 
(SUPA),Institute for Astronomy, School of Physics, University of Edinburgh,
Royal Observatory, \\ 
Blackford Hill, Edinburgh EH9 3HJ, UK \\ 
$^{*}$ njc-AT-roe.ac.uk}
    
\begin{document} 

\date{Accepted XXXX. Received XXXX ; in original form XXX }

\pagerange{\pageref{firstpage}--\pageref{lastpage}} \pubyear{2005}

\maketitle 

\begin{abstract}
We present a description of the design and usage of a new synoptic pipeline and
database model for time series photometry in the VISTA Data Flow System (VDFS).
All UKIRT-WFCAM data and most of the VISTA main survey data will be processed 
and archived by the VDFS. Much of these data are multi-epoch, useful for finding 
moving and variable objects. Our new database design allows the users to easily
find rare objects of these types amongst the huge volume of data being produced by
modern survey telescopes. Its effectiveness is demonstrated through examples 
using Data Release 5 of the UKIDSS Deep Extragalactic Survey (DXS) and the
WFCAM standard star data. The synoptic pipeline provides additional quality
control and calibration to these data in the process of generating accurate
light-curves. We find that $0.6\pm0.1\%$ of stars and
$2.3\pm0.6\%$ of galaxies in the UKIDSS-DXS with $K<15$ mag are variable with
amplitudes $\Delta\,K>0.015$ mag.

\end{abstract}


\section{Introduction} 

The study of time-varying phenomena has led 
to some of the most important discoveries in astronomy. The ``new'' stars of
1572 and 1604 were shown to be beyond the moon and so the notion that the
heavens were unchanging was discarded \citep{NOVA1572}. Observations of variable
stars have led to discoveries of eclipsing binaries
\citep[e.g. Algol, ][]{algol}, which are the best systems for measuring the
masses of stars \citep{algolmass}; pulsating stars, which give the best estimates for
distances to nearby galaxies \citep{ceph1} and thereby the rest of the cosmic
scale; and cataclysmic variables, which give insights into the physics of
accretion discs and degenerate matter \citep[e.g.][]{CVs}. Observations of the motion of
objects such as planets and comets led to the laws of gravity \citep{Gravity}
and later parallax observations fixed the distance scales to the stars
\citep{Pllx}.

The word ``synoptic'' has been used frequently to describe wide-field,
multi-epoch surveys, designed to find rare variable sources \citep[e.g. the
Rossi X-Ray Transient Explorer, RXTE,][]{RXTE}. However, until the last decade,
optical variability studies were not very large-scale: the largest catalogue 
was the General Catalogue of Variable Stars \citep[ GCVS]{Khp98}. With the
advent of  new wide-field imaging cameras on survey telescopes, large surveys 
of moving and/or photometrically variable objects have become possible 
(\citealp{Pzk97}; \citealp[Northern Sky Variability Survey, NSVS]{Wznk04}; \citealp[All Sky Automated Survey, ASAS]{ASAS}). One very early
variability survey was the $25$ sq. deg. quasar variability survey using the UK Schmidt Telescope
\citep{Fld287}. This was produced using photographic plates, over a period of 
20 years, but the errors on the plates limited the survey to objects that 
varied with 0.2 mag or greater. In the last decade surveys have included
wide shallow surveys such as NSVS, and Super Wide Angle Survey for Planets
\citep[SuperWASP]{SWASP}, which are low resolution ($\sim14\arcsec$ pixels) and
therefore become confusion limited (at bright magnitudes) in the galactic plane.
NSVS is a systematic survey of variability of bright stars ($8<V<15.5$) in the northern
hemisphere, whereas SuperWASP is observing the transits of bright stars
($8<V<15$) by extra-solar planets. There are also deeper, narrower surveys,
such as the MACHO microlensing survey \citep{MACHO}, the Monitor planet transit
survey \citep{Mon} and the SDSS stripe 82 programme \citep{Ssr07}. The first
two  of these surveys observed $>10$ sq. deg. with hundreds of epochs, whereas 
the SDSS data covers $\sim300$ sq. deg. with $\sim80$ epochs. 
 
Some very recent surveys include very wide-field, medium-resolution 
($\sim2\arcsec$ pixels) transient surveys using new wide-field imagers on
old telescopes, e.g. Palomar Quest Sky Survey \citep{PQSS}, Catalina Real-Time
Transient Survey \citep{CRTS} and the Palomar Transient Factory \citep{PTF}. These are experiments to test some of the technology, particularly the Virtual 
Observatory event streams \citep{PQSSVO} necessary for the next generation of
high-resolution all-sky transient surveys and to find unusual
transients and variables.
 
CCD technology has improved to the point where all -sky,
high-resolution (sub-arcsec seeing, $\sim0.2\arcsec$ pixels) synoptic surveys
are possible. Surveys such as The Panoramic Survey Telescope \& Rapid
Response System \citep[Pan-STARRS][]{PSTR} have recently started operating
(early 2009) and in a few years, more ambitious projects such as the Large Synoptic Survey Telescope \citep[LSST; ][]{LSST,LSST08} and Gaia \citep{Gaia} will commence. Pan-STARRS and
LSST will hunt for near-earth asteroids, but will also do a wide range of science
such as finding and classifying variable stars and AGN; finding transients,
such as supernovae, gamma-ray bursts and micro-lenses, which can be quickly
reported and followed up by other telescopes; galaxy evolution studies, and 
large scale structure studies by taking advantage of the wide-deep images 
produced by stacking the individual exposures. Gaia will observe $10^9$ stars 80
times over 5 years to measure very accurate parallaxes (hence distances) and 
proper motions, vastly improving our knowledge of the
structure and dynamics of the Milky Way. LSST will observe $2\times10^{10}$
objects 1000 times over 10 years, covering 20,000 sq. deg.
 
Before the UK Infra-red Telescope Wide Field Camera (UKIRT-WFCAM) and
Canada-France-Hawai'i Telescope Wide-field Infra-red Camera (CFHT-WIRCAM), which
both have four $2k\times2k$ pixel detectors, there were no near
infrared instruments capable of doing high-resolution, wide-field surveys. 
The UKIRT Deep Infrared Sky Survey\footnote{http://www.ukidss.org} (UKIDSS) is
a series of five surveys undertaken by UKIRT-WFCAM. Three of these surveys are 
wide and shallow, with only one or two repeat observations in the same filter. 
The UKIDSS Deep Extragalactic Survey (DXS) and Ultra Deep Survey (UDS) have multiple 
observations of the same pointing in the same filter, to increase the magnitude
depth to find the most distant galaxies. The WFCAM standard star observations 
also observe the same fields through the same filters multiple times. However
these surveys are not true synoptic surveys since the cadences - the frequency 
of observations - are not designed for the discovery or study of variable
objects. This will not have any effect on the statistical analysis we describe in
\S\ref{sec:stats} but does make it more difficult to analyse the light-curves 
using Fourier analysis. These datasets, along with numerous smaller projects, 
led by Principal Investigators (PI) outside the main surveys, are suitable for 
multi-epoch analysis and benefit from the new pipeline and database tables 
described  in this paper.

While there are some multi-epoch data taken by WFCAM, the Visible
and Infra-red Survey Telescope for Astronomy \citep[VISTA-VIRCAM]{VISTA} will be
the  first near-IR instrument with planned wide-field synoptic surveys, i.e.
where the observing interval has been chosen to target particular types of
variables. There are three planned synoptic surveys amongst the VISTA Public
Surveys\footnote{http://www.vista.ac.uk/}: VISTA Variables in Via-Lactea (VVV),
a survey of the Galactic plane and bulge that will use RR-Lyrae and Cepheid
stars to measure distances to Galactic components; VISTA Magellanic Survey
(VMS), a survey of the Magellanic Clouds using variable stars as distance
indicators again; VISTA Deep Extragalactic Observations (VIDEO), which is
primarily a deep survey but has an observing strategy which will look for 
supernovae. These will be the first large synoptic surveys in the near-IR,
and much of the past work on infra-red variable stars has been concerned with
observing known optical variables in the near-IR, so these surveys may discover
many new types of variables.

The VISTA Data Flow System (VDFS) is responsible for processing and archiving
the data from UKIRT-WFCAM and VISTA-VIRCAM. The responsibilities are divided
between the Cambridge Astronomy Survey Unit (CASU), which does the nightly
processing and calibration and the Wide Field Astronomy Unit (WFAU, in
Edinburgh), which does the archiving. The data can be accessed through the
WFCAM Science Archive\footnote{http://surveys.roe.ac.uk/wsa/index.html} 
\citep[WSA,][]{Hmb08} and VISTA Science Archive (VSA).

This paper describes the philosophy, design and implementation of a relational
database science archive for synoptic data. The archive is designed to catalogue objects which are
varying both photometrically and astrometrically within the limits of the observations.
This model can be applied to data from a range of astronomical programmes that
are based on pointed observations. Scanning surveys such as SDSS and Gaia will
need to implement a slightly different design - the idea of breaking the
curation into sets of observed frames may not be so easily applicable in these
cases. 
 
In \S\ref{sec:datamodel} we describe the relationship between the different tables used
to archive synoptic data. In \S\ref{sec:curation} we describe the processes used to 
archive the data. In \S\ref{sec:stats} we describe the statistical methods that
analyse variability in the archive and in \S\ref{sec:DR5} we show some
examples of selecting variables in the UKIDSS-DXS Data Release 5 using the WSA
archive and show some useful analysis. We also highlight some existing problems that we
hope to correct in future releases. In \S\ref{sec:CAL} we show some objects
from the WFCAM standard star data, as an example of a correlated band pass data
set, including light curves of 3 standard stars in the Serpens Cloud Core.
In \S\ref{sec:VSA} we discuss additional issues that will be faced when
curating VIRCAM data, and in \S\ref{sec:otherPDB}, we discuss the differences
between multi-epoch archives such as that for the SDSS Stripe 82 data or the
NSVS public database and the WSA. Finally we summarise the work we have done and
suggest some improvements for the future.

The first release of variability data using the model described in this paper is
the UKIDSS Data Release 5, released on April 6th 2009. The previous releases
did not include the new synoptic tables described in \S\ref{sec:datamodel}.
Future releases of WSA or VSA data will extend this model or improve the 
attributes already available. Any modifications will be noted on the archive
webpages in the release history\footnote{http://surveys.roe.ac.uk/wsa/releasehistory.html}.

\section{Overall Data Model}
\label{sec:datamodel}

Our data model has been developed to enable users to find a wide range of
different types of variable in large data sets. These different
data sets and different science goals of WSA/VSA users have necessitated a 
very general approach. Some of the different science usages are listed below:

\begin{itemize}
  \item Search for low-mass brown-dwarf stars through their proper motions.
  \item Search for transiting extra-solar planets around M-stars.
  \item Search for RR-Lyrae and Cepheid pulsating variable stars.
  \item Search for supernovae.
  \item Find new faint infra-red standard stars.
\end{itemize}

These different items have put different constraints on the model. If we are
to look for moving objects, we cannot just use list-driven photometry --- where
fluxes are measured for a list of source positions in each observation,
regardless of whether there is detection in that observation at that point --- to measure the
fluxes of objects in each observation, but instead we have to link the 
observations together using an astrometric model. Transiting planets, 
eclipsing binaries and supernovae may not be detectable on all frames, so it is
important to keep track of all observations whether there
is a detection or not. Pulsating stars have asymmetric light-curves, so higher
order statistics, such as the skew, can be important indicators. Not only that,
but the variations are often highly correlated between filters. Finally it is important to understand the
noise characteristics of the data, if variables and non-variables are to be
distinguished. It should be noted that searches for transient objects
requiring prompt follow up such as supernovae, gamma-ray bursts and microlenses
are impractical through the archive, since data appears in the archive at least 6 weeks after observation
so that they can be processed and calibrated correctly beforehand. Transients
with large amplitudes do not need this level of calibration to be
noticed, and so a transient pipeline should be run at the telescope. The
archive is more suitable for long-term variables, low-amplitude variables and
slowly moving objects, which need multiple observations and the best 
calibration for their discovery and classification.
 
The heterogeneity of the data is another important issue: some datasets having
multiple filters and hundreds of epochs and others having one filter and two epochs means that the
pipeline has to be robust and serve many purposes. A few observations of a
star or galaxy may not be any use in determining whether it is a Cepheid
variable, but they can determine whether it is moving or not. 

The WSA is described in detail in \cite{Hmb08}. That paper discusses
production of deep stacks, simple recalibration, source merging and
neighbour tables, all of which are used in the production of the archive
for variable sources. In its discussion of synoptic data it mentions an early, 
very crude data model for curating the synoptic data and references
\citep[][hereafter Paper 1]{Crs07} for an advanced version. At the time of
writing, the data model for synoptic tables was only partially completed and work on
the pipeline had not yet been started.

In this section, we describe our new model, which develops and expands 
on the model in Paper 1. Since Paper 1, we have 
changed the philosophy, added astrometric statistics, added in noise
modelling and built a working pipeline to archive the synoptic data. In this
and later sections, we use the following conventions: 

\begin{itemize} 
  \item \verb+TableName+ indicates an archive table, which can be found in WSA
  Schema
  Browser\footnote{http://surveys.roe.ac.uk/wsa/www/wsa$\_$browser.html}.
  Tables which only contain data for a specific programme will be prefixed by a programme ID string. For instance, we refer to the \verb+Source+ table throughout this. In the UKIDSS-DXS programme, this becomes
  \verb+dxsSource+, and in the WFCAM Standard Star programme this becomes
  \verb+calSource+. Some tables such as \verb+Multiframe+ contain data
  from all programmes and are not prefixed in the archive.
  \item {\bf attributeName} indicates an attribute within an archive table,
  such as {\bf sourceID}, the unique identifier of a source in a
  \verb+Source+ table.
\end{itemize}

The procedures for multi-epoch surveys as described in \cite{Hmb08} are:
\begin{itemize}
  \item Quality control for each observation, deprecating poor quality frames.
  This is partly automated and partly done by survey teams checking the
  science frames.
  \item Quality bit flagging of catalogue data. A set of automated procedures
  that give warnings for objects in the catalogues that are too close to the
  edge of a frame, are saturated, have bad pixels, or are affected by
  electronic cross-talk. More issues will be flagged in the future.
  \item Stacking of individual epoch observations into deep stacks to detect
  faint objects. 
  \item Extraction of catalogues from deep stacks.
  \item Ingestion of deep stacks and catalogues into archive. 
  \item Updating the provenance of new deep stacks. This links a deep stack
  frame to all the frames that it is composed of. 
  \item Updating the quality bit flags of new deep catalogues. 
  \item Merging the deepest catalogues in each filter to produce 
  the \verb+Source+ table of unique sources. This associates different filter
  data by position and takes into account overlapping sets of frames.
  \item Creation of neighbour tables between \verb+Source+ and 
  \verb+Detection+, \verb+Source+ and itself and \verb+Source+ and external
  catalogues. 
\end{itemize}

These procedures mainly dealt with producing deep images and catalogues, but 
the neighbour table between \verb+Source+ and \verb+Detection+ allowed users to
compare the deep data to individual epochs. To make it easier to find and
categorise variable objects, we have developed the following new procedures:

\begin{itemize}
  \item Recalibration of intermediate stack detector zeropoints and deprecation
  of any frames with large zeropoint changes, since a large change indicates an
  error.
  \item Production of a merged bandpass catalogue at specific epochs for
  datasets with correlated bandpasses (see \S\ref{sec:corr}).
  \item Matching of the reseamed \verb+Source+ table to each observation.
  Reseaming the \verb+Source+ table finds objects in the table that are in 
  the table twice and prioritises them so that a unique list can be selected.
  \item Calculation of astrometric and photometric variability statistics.
  \item Calculation of the noise properties of data within each pointing.
  \item Classification of sources based on variability statistics.
\end{itemize}

The processing of all of individual epochs is done independently of each
other, with the exception of the calibration of the deep stack zeropoints,
which then feeds back into the recalibration of the individual epoch
zeropoints, and the calculation of variability statistics. The first two 
procedures must occur before the neighbour tables are produced \citep{Hmb08},
but the last three must occur afterwards. These new procedures require five
new tables:

\begin{itemize}
  \item \verb+SynopticMergeLog+: This has the frame merging information for
  different filter observations taken at (almost) the same time in a correlated
  pass band survey, see \S\ref{sec:corr}.
  \item \verb+SynopticSource+: This has the merged catalogue data from frames 
  in the \verb+SynopticMergeLog+ table, with many of the same attributes as the
  \verb+Source+ table.
  \item \verb+SourceX[Detection,SynopticSource]BestMatch+: This is the table of matches between individual sources in the \verb+Source+ and the nearest object in each
  observation frame, from the \verb+Detection+ table (for uncorrelated
  observations) {\bf OR} \verb+SynopticSource+ table (for correlated
  observations). Any dataset can only have either one, not both. This table will
  be called the best match (BM) hereafter.
  \item \verb+Variability+: This includes astrometric and photometric statistics
  from the different observations of each source, as well as classifications.
  \item \verb+VarFrameSetInfo+: This includes the noise properties of each
  frame set.
\end{itemize}

\subsection{Uncorrelated Observations}
\label{sec:uncorr}

Most multi-epoch data sets in the WSA were either taken through a single
filter or the observations in several filters are uncorrelated in time (e.g. DXS, UDS).
The \verb+SourceXDetectionBestMatch+ table is quite different from
the \verb+SourceXDetection+ neighbour table \citep{Crs07} since it has only one
match per observation frame and includes rows with default values for frames 
where there was no detection. The default values are usually very large
negative numbers that are well outside the range of sensible values
\citep[see][ for details]{Hmb08} and are therefore easily recognisable as a
non-detection. This is created using a matching algorithm which finds
the nearest match. We choose not to select by magnitude as well 
as position since some variable objects, which we are interested in may vary
(in magnitude) by several magnitudes and we do not want to bias our
observations. Some objects move measurably, though, but real motions are
typically composed of a proper motion (linear over small angles) and a  
parallax due to the Earth's motion around the Sun, which follows 
an ellipse where all the parameters apart from the size of the ellipse are 
determined by the coordinates of the object and time of year. The size of the
ellipse is determined by the distance to the object. For objects further than
$\sim20$ parsec, the parallax ellipse will be too small to see with WFCAM or
VISTA data. Our intention is to match objects based purely on their motion, incorporating a
linear proper-motion and a parallax. Since most objects will have no measurable
motion, or a motion that is very small, we split the matching process into two
parts. The first step is an initial match based on nearest match only, which we
have already implemented. The second step will rematch sources, which have
inconsistencies or whose measurements show motion, using a model that includes
motion. This
second step has not been implemented and will need to wait until we start fitting a model to the astrometric error.
Inconsistencies can occur when objects are incorrectly
deblended. This is likely to occur in dense regions of the Galaxy in
particular. Running list-driven photometry can help to determine whether the
deblending is correct, but list-driven photometry by itself would give no
astrometric information. We may incorporate list-driven photometry in the
future (see \S~\ref{sec:summary}), but we need to make sure that it can be run
efficiently, so that it doesn't place too many overheads onto our pipeline.

Using the neighbour table instead of the best-match table would produce
lightcurves that have more than one detection at some times and do not have important information
about missing data. That might occur in observations of eclipsing binaries, or
a failed supernova \citep{Kck08}.
 
\begin{figure*}
\includegraphics[width=140mm,angle=0]{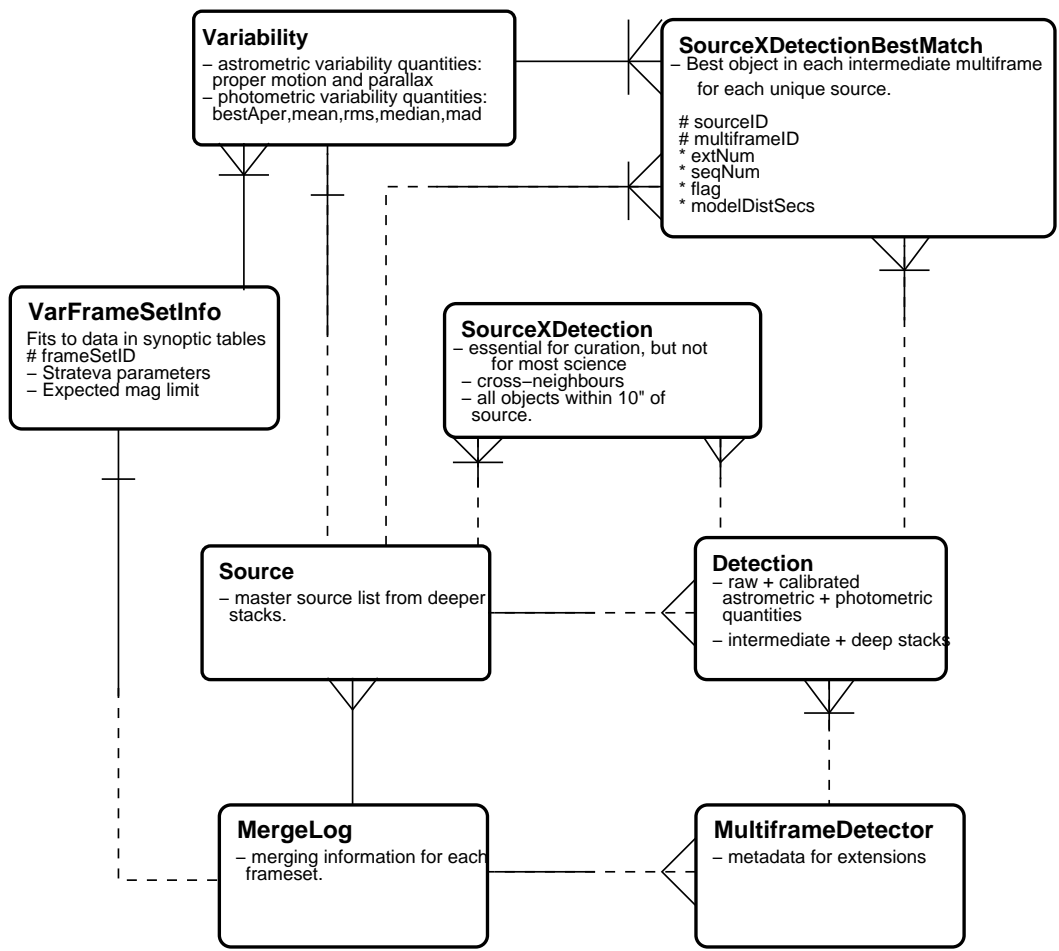}
\caption{Entity-Relation Model for synoptic data in the WSA. Within a box,
a \# indicates an attribute which is in the primary key , $*$ indicates
other attributes and $-$ indicates general description. The connections are a
solid line if there is a one-to-one relation, a fork connects the second box if
there are many rows in the second box joined to one row in the first box: e.g.
a frame in {\bf MultiframeDetector} contains many detections in
{\bf Detection}. If there is a dotted line, then some rows in that table are
not connected to rows in the other table: e.g. {\bf MultiframeDetector} has
rows associated with calibration images as well as science frames, but only the
science frames have detections associated with them. A short line perpendicular
to the joining line indicates that the tables are linked through the main
identifiers for those tables: the ``primary keys''.}
\label{fig:synERM}
\end{figure*}

Fig~\ref{fig:synERM} shows the new entity-relation model (ERM) for synoptic 
data in the WSA. The ERM shows how each of the tables relate to each
other. \cite{Hmb08} gives ERMs describing other features of the WSA. 

The most important step in archiving synoptic data is to produce a catalogue 
of unique sources, which is significantly deeper than a single epoch
observation. This is already available in the reseamed \verb+Source+
table, which contains measurements for each source from the deepest catalogues 
available in each filter. The procedures used to create the \verb+Source+ and
neighbour tables are described in detail in \cite{Hmb08}, so we will just
reiterate the salient points. In sparse regions, out of
the plane of the Galaxy, it is most advantageous to use all available good
quality data to create the deepest stacks possible, since these are also useful in faint object 
programmes. However, in crowded  regions in the Galactic plane, it may be 
advisable to only use a small number of intermediate stacks to avoid being 
confusion limited. This can be specified by the  Principal Investigator (or
survey team) in large surveys, although by default all good frames are 
stacked. If a restricted number is specified, we select the intermediate stacks
with the best seeing to get the highest resolution image. The \verb+Source+ 
table is reseamed so that any sources which are recorded multiple times  (i.e.
objects that are in two overlapping deep stacks) in the table are prioritised 
so that there is one primary source (from the frame set with most or the best 
observations) and one or more secondary sources. Using the {\bf priOrSec} flag
it is possible to select an unique list, or just the objects away from 
overlaps or just the objects within overlaps.

Once the neighbour tables have been produced, the \verb+SourceXDetection+ table
is used as a starting point for producing our best match table
(\verb+SourceXDetectionBestMatch+). This table is
designed so that it  can only contain one match to each source from each 
intermediate frame. There are two other important attributes in this table: a 
flag ({\bf flag}), which can indicate additional useful information to the user
and a separation distance ({\bf modelDistSecs}), which gives the separation
between the observation and the expected position. The expected position
can allow for motion. The flag indicates one of two cases. The first 
case occurs if the same intermediate frame object is linked to two sources. 
This can occur if the two source are blended in one frame but not in others 
(due to poor seeing or motion), or an object appears in some frames but not
others  (e.g. a supernova). In frames in which it does not appear the
neighbouring  object (e.g. the host galaxy) may be linked to the source
instead. In all these cases, the photometry is incorrect for one or both
sources, so it is important to note these occurrences. In this case the {\bf
flag} attribute is set to $1$. 

In the second case, the {\bf flag} is set to $2$ if there is no detection (a default row),
but the position is close enough to the edge of the frame that it would not 
have been detected in all the constituent observations that went into the frame.
Each individual epoch frame is made up of several ``normal'' frames that have 
slightly different pointings and are then ``dithered'' together to remove
artifacts in the image. In this case, the object is said to be within a dither
offset of the edge, where the exposure time decreases and therefore the noise
increases. If an object was not observed, then the most likely cause is the
rapid change in noise characteristics, rather than intrinsic variability in 
the object, so it is important to flag this fact. Detections which are within a
dither offset of the edge, are already flagged in the \verb+Detection+ table.

In Paper 1, the variability attributes were placed in the \verb+Source+ table,
but we decided to put them in a separate table for several reasons. The first
reason is philosophical: the \verb+Source+ table is the unique list of sources 
containing the merged catalogues extracted from the deep stacks in all the
different passbands in the survey, whereas the
\verb+Variability+ table contains the statistical information from multiple
short exposure time observations. \verb+Source+ may contain passbands where there was only a single pointing 
(for additional colour information), which are not necessary in the 
\verb+Variability+ table. The \verb+Source+ table contains many sources seen in
the deep stacks that are too faint to be detected on any of the short exposure stacks. 
Separating \verb+Source+ and \verb+Variability+ is good for curation: if the 
variability data has to be recreated (a more sophisticated motion or noise
model, recalibration of individual exposures, new statistical measurements etc), then the
\verb+Source+ table is unaffected. However, recreating the \verb+Source+ table
necessitates recreating the \verb+Variability+ table because the IDs of each
source would change. 

The \verb+Variability+ table contains information about astrometric
variability: the best fit proper motion and parallax, see \S\ref{sec:astrometry} for details.
It also gives information on the cadence --- the typical interval between
observations --- for each source, \S\ref{sec:observstats}. The main statistics
include simple photometric variability statistics in each band
\S\ref{sec:photometry}. The classification in each passband and overall is 
calculated. Careful use of many properties taken together can rapidly reduce the number of returns in a Structured Query Language (SQL)
query, so the user only has to look through the lightcurves of a small number of
possible sources. The cadence information, for instance, allows the user to
determine whether the data have the right sampling frequency for the science in question.

The final table \verb+VarFrameSetInfo+ records overall data, such as the fit to
the RMS as a function of magnitude and the expected magnitude limit for each
pointing (frame set). These are important for understanding the limits of the
data, and calculating whether an object is likely to be variable. It also 
records which type of astrometric fit was applied to the frame set in question (e.g. static, proper motion etc).
Processing on a frame set basis increases flexibility and simplifies parallelism
which improves speed of processing.

\subsection{Correlated Observations}
\label{sec:corr}

The WFCAM standard star observations (and some VISTA programmes) have data
which include repeated sets of observations of the same pointing taken in
several filters, where the filters are observed together in a batch over a much
shorter period of time than the interval between observation batches or the
time-scale of variability that we consider. In these cases, we say that the
pass-bands are correlated and the different observations are close enough together that they are at the same epoch. In the standard star observations
(hereafter CAL - short for calibration), a field is observed through the 5
broad-band filters one after another --- all within about 10 minutes ---  every
hour or two, although the same field is only repeated on a daily basis. 
Occasionally fields are also observed through the narrow band filters.

The data model in \S\ref{sec:uncorr} dealt with single filter data sets or
multiple filter data sets, where observations in different filters are not
synchronised (e.g. UKIDSS-DXS). However, if the observations in each filter are
correlated, then a more efficient method is to merge the different filter
observations for each epoch into a single table (\verb+SynopticSource+) and
match this to the \verb+Source+ table thereby reducing the size
of the best match table and more easily producing colour light-curves.

\begin{figure*}
\includegraphics[width=140mm,angle=0]{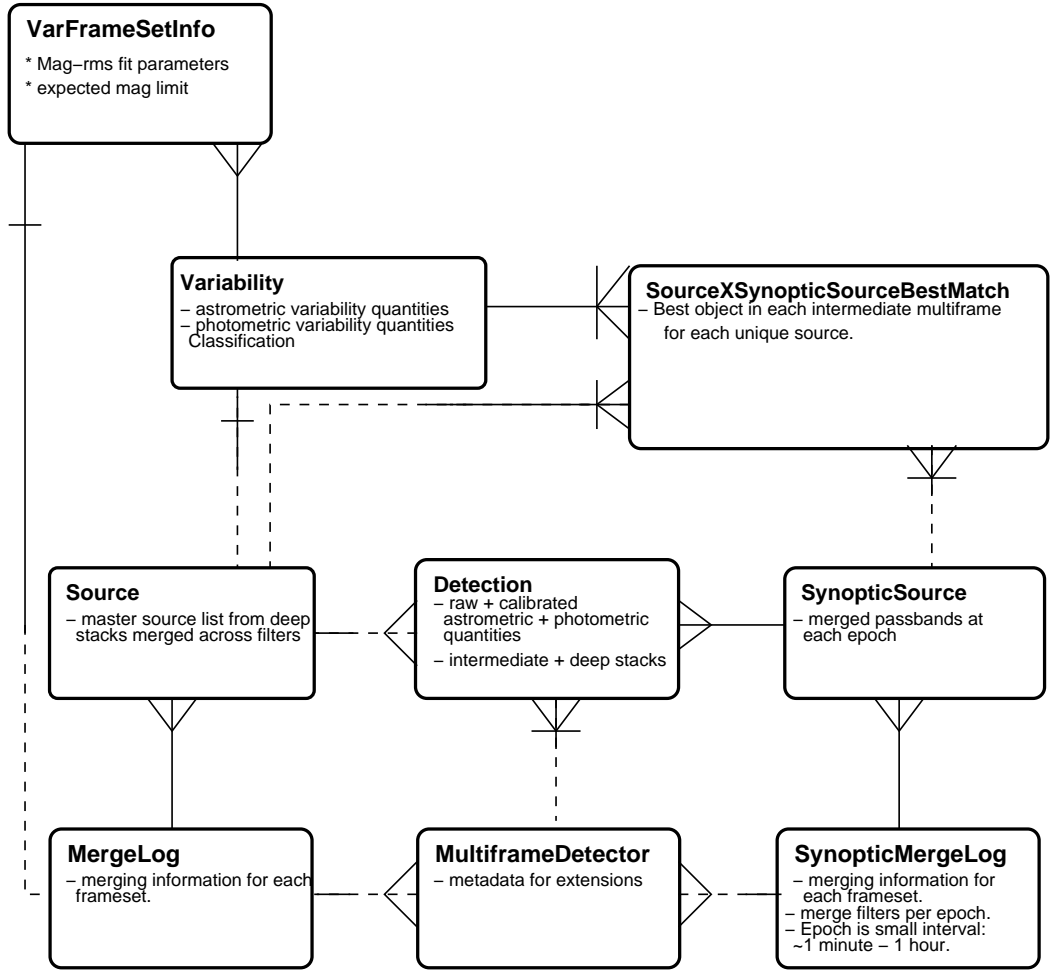}
\caption{Entity-Relation Model for correlated pass-band synoptic data in the
WSA. See Fig~\ref{fig:synERM} for details.}
\label{fig:synERMcol}
\end{figure*}

Fig~\ref{fig:synERMcol} shows the ERM for multiple pass-band data. Using this
model, data sets, such as the CAL observations are more usefully
processed. Band-pass merging for each set of observations to form a \verb+SynopticMergeLog+ table and a \verb+SynopticSource+
table has two advantages. The first is that the colour information at any epoch 
can be quickly looked up and variations in colour (i.e. whether variability is 
correlated between pass-bands) quickly found. This is extremely useful
information for variable classification
\citep[e.g.][]{Hu07,dwt06,HEH06}. 
Microlensing variations show no variations in colour; pulsating
stars and some eclipsing binaries show periodic colour variations with the same 
period as the magnitude variations; noise and cosmic rays are uncorrelated. 

The second advantage is that the cross-match table
\verb+SourceXSynopticSourceBestMatch+ is significantly smaller in size than the
equivalent \verb+SourceXDetectionBestMatch+ table would be, because the \verb+SynopticSource+ table for the CAL programme is 5-6 times 
shorter in row size than the \verb+Detection+ table. The 
\verb+SourceXSynopticSource+ is correspondingly shorter too. This 
reduces the time for curation of the data and lookup requests to the archive as 
discussed in Paper 1. However, while curation of the best match table is 
sped up, there is the additional curation time of creating the
\verb+SynopticSource+ table in the first place. The main advantages are to 
archive users, who have easier access to the information and smaller (and
therefore faster) lookup tables as well as additional correlated attributes to search
on.

To reduce the size of the \verb+SynopticSource+, we have removed most of the
magnitudes that are available in the \verb+Detection+ table and only left five
fixed aperture magnitudes, since most variable objects are point sources. Even galaxies that vary in brightness tend to vary
due to an active galactic nucleus or a supernova explosion, which are both
small scale events and therefore point sources in these data sets\footnote{
A supernova will often be off-centre in a galaxy. This may show up as a
separate point source, or it may be blended into the same object, changing  the
centre slightly.} This may be seen as a poor astrometric match as well as a
poor photometric match. Therefore Petrosian, Kron, Hall and larger aperture
magnitudes (which are useful for extended sources) are unnecessary in this table. The two main methods of measuring point source fluxes are aperture 
photometry \citep[e.g.][]{Mon} or PSF photometry \citep[e.g.][]{DAOPHOT}. We
use seeing corrected aperture photometry, where light is measured in a small 
aperture (typically $\sim1\arcsec$ radius) which includes most of the light of
the galaxy, but is small enough that the chance of contamination is very low. 
The median correction is measured between this aperture and a much wider
aperture for point source objects and this is applied to correct for the light lost in the 
wings of the profile. PSF photometry fits a 2-D profile PSF to all point
sources in the image (or in parts of the image). PSF photometry automatically
removes contamination from other detected stars and typically does a better job
in very crowded stellar fields, but cannot take into account contamination from
galaxies. \cite{MOMF} points out that aperture photometry is better for
isolated  stars and PSF photometry for faint stars or stars in crowded regions,
and suggests a method that combines the advantages of both methods. Tests by
CASU  (private communication) suggest that list-driven aperture photometry 
performs as well as PSF photometry in crowded regions. The advantage of 
list-driven photometry is that it removes some of the uncertainty in the
centroid that can be a large source of error, but only by assuming that the 
objects have no proper-motion. This assumption does not always hold true.

In the \verb+SourceXSynopticSourceBestMatch+ table, if two rows have the same
single epoch detection (in the \verb+SynopticSource+ table) matched to two
different sources, then ${\bf flag}=1$, just as in \S\ref{sec:uncorr}. However,
non-detections within a dither offset of the edge of a frame are more difficult
to handle. The different frames for each filter may not lie exactly on top of 
one another, and it is important to keep the 
information for each filter. To flag this, we have adopted the 
following convention: ${\bf flag}=\sum_f\,2^f$, where f is the {\bf filterID}.
Thus if a survey is observed in $Y$ (${\bf filterID}=2$), $J$ (${\bf
filterID}=3$), $K$ (${\bf filterID}=5$) and there are non-detections in each 
of these filters, but only $Y$ and $K$ are within one dither offset of the edge,
then ${\bf flag} =2^2+2^5 = 36$.

The \verb+Variability+ is calculated in the same way, except that once the
photometry statistics are calculated in each filter separately, the
Welch-Stetson statistic \citep{WS93} is calculated for each pair of broad band
filters.

The full current schema of the new tables can be found on the WSA Schema
Browser.

\section{Curation of the data}
\label{sec:curation}

In this section we give an overview of the data processing that goes into 
creating the archive product. The curation of the synoptic tables is an 
automated process once the survey requirements have been specified in the 
following curation tables, which are themselves setup automatically using
the metadata from the science frames in each programme \citep[see][]{Cll09}:

\begin{itemize}
\item \verb+RequiredSynoptic+, which states whether the programme is correlated
and the correlation time scale for the programme. The correlation time scale is
the maximum time delay between the first and last filter in any given ``epoch''.
\item \verb+RequiredFilters+, which lists the filters used in each programme
and which filters are synoptic. In some programmes some filters may be
synoptic and others observed a definite small number of times (e.g. UKIDSS
LAS, GCS and GPS all have 2 repeats for some filters). 
\item \verb+RequiredStack+, which lists the different pointings for
deep stacks in the survey, gives information on producing stacks and the required extraction parameters for cataloguing.
\item \verb+RequiredMosaic+, which lists the different pointings for mosaics in
the survey and size of the mosaics, gives information on the software to be used
and the required extraction parameters for cataloguing.
\item \verb+RequiredNeighbours+, which lists the neighbour tables to be produced for the survey
and the maximum radius for matching.
\end{itemize}

\subsection{Production of deep stacks and catalogues}

The production of deep stacks or mosaics uses information in the
\verb+RequiredStack+ or \verb+RequiredMosaic+ curation tables to produce 
the correct stacks or mosaics. In general, we produce deep stacks,
rather than mosaics, but the pipeline can handle both. Catalogues are extracted
from these deep images, again using the extraction parameters in these tables, which depend on the amount of
micro-stepping used when creating the stack. Source extraction is done using
the VDFS extractor (Irwin et al., in preparation) used to extract individual
epoch data or in a few cases, such as the UDS, using SE{\tiny XTRACTOR} \citep{Btn96}.

After extraction, as part of the curation process, the deep stack catalogues
are calibrated. Since the WFCAM intermediate stacks have already been
calibrated against the Two-Micron All Sky Survey \citep[][2MASS]{2MASS}, and the
zeropoints are accurate to better than 0.02 mag \citep{Hdg09}, the simplest method of
calibration is to use a clipped median of all the intermediate stacks, or a
random selection of them to save processing time if there are very many. The deep
stack products are then ingested into the archive, the \verb+Provenance+ table
is updated to link these new deep images with the component intermediate
stacks. Quality bit flags are calculated for the deep image catalogues.
Finally, source merging is run to create the master \verb+Source+ table. This
is controlled by the \verb+RequiredFilters+ table which contains the different filters, the number of passes in each filter and whether the filter is synoptic.

\subsection{Individual epoch observations}

The individual epoch observations (intermediate stacks) are recalibrated to 
give the best relative photometry. The recalibration is done separately for
each detector. The main pipeline calibration for all WFCAM data compares WFCAM
data to 2MASS data \citep{Hdg09} but does not have enough 2MASS stars in each
frame to do a detector by detector calibration, so uses  a month of
data to measure the mean offset between detectors. Since we are comparing short WFCAM exposures to deep
stacks, we have many more stars per frame, so we can get a much more accurate
relative calibration. We recalibrate the data by finding the average difference
in magnitude of bright stellar sources in the relevant deep stack and each intermediate stack
and modifying the intermediate stack zeropoint by this amount. Since the
zeropoints should already be accurate to $\leq0.02$ mag from comparison to
2MASS, any differences in magnitude more than $\sim0.05$ mag is a clear sign of an error. We set the
{\bf deprecated} flag to 110 in \verb+MultiframeDetector+ on these frames. These
frames go into the data release unlike other deprecated frames, since the frame
may already be a component of the deep stacks. These frames will then be
removed  from the next release\footnote{There are no DXS frames in the Data 
Release 5 with {\bf deprecated}$ =110$, because these frames  were found while
testing the synoptic pipeline and were deprecate before processing of the DR5
commenced.}. The change in zeropoint of a detector frame should be less than the
formal error on the zeropoints derived from comparison with 2MASS. In \S\ref{sec:recal} we show the improvements to the accuracy of
the data from this simple recalibration. 

The new zeropoints replace
old values in both the archive tables and the archived FITS files.
The old zeropoints are recorded in the history lines of the FITS file and in
the \verb+PreviousMFDZP+ table with a version number. The timestamp for the 
version number is in the \verb+PhotCalVers+ table. With this setup, users can
keep track of the changes we make and results in older publications can be 
checked and compared to current results.

\begin{figure}
\includegraphics[width=84mm,angle=0]{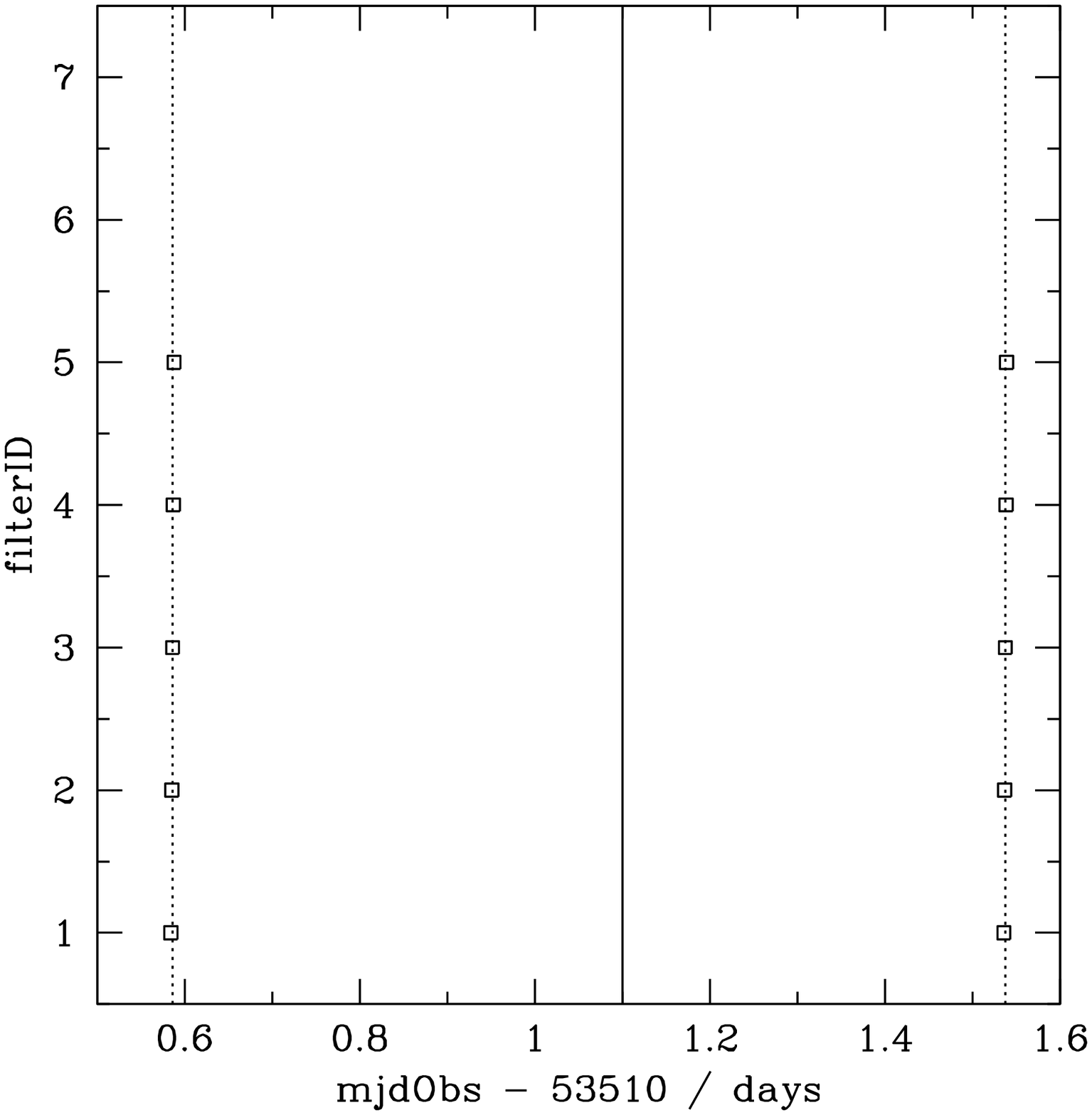}
\includegraphics[width=84mm,angle=0]{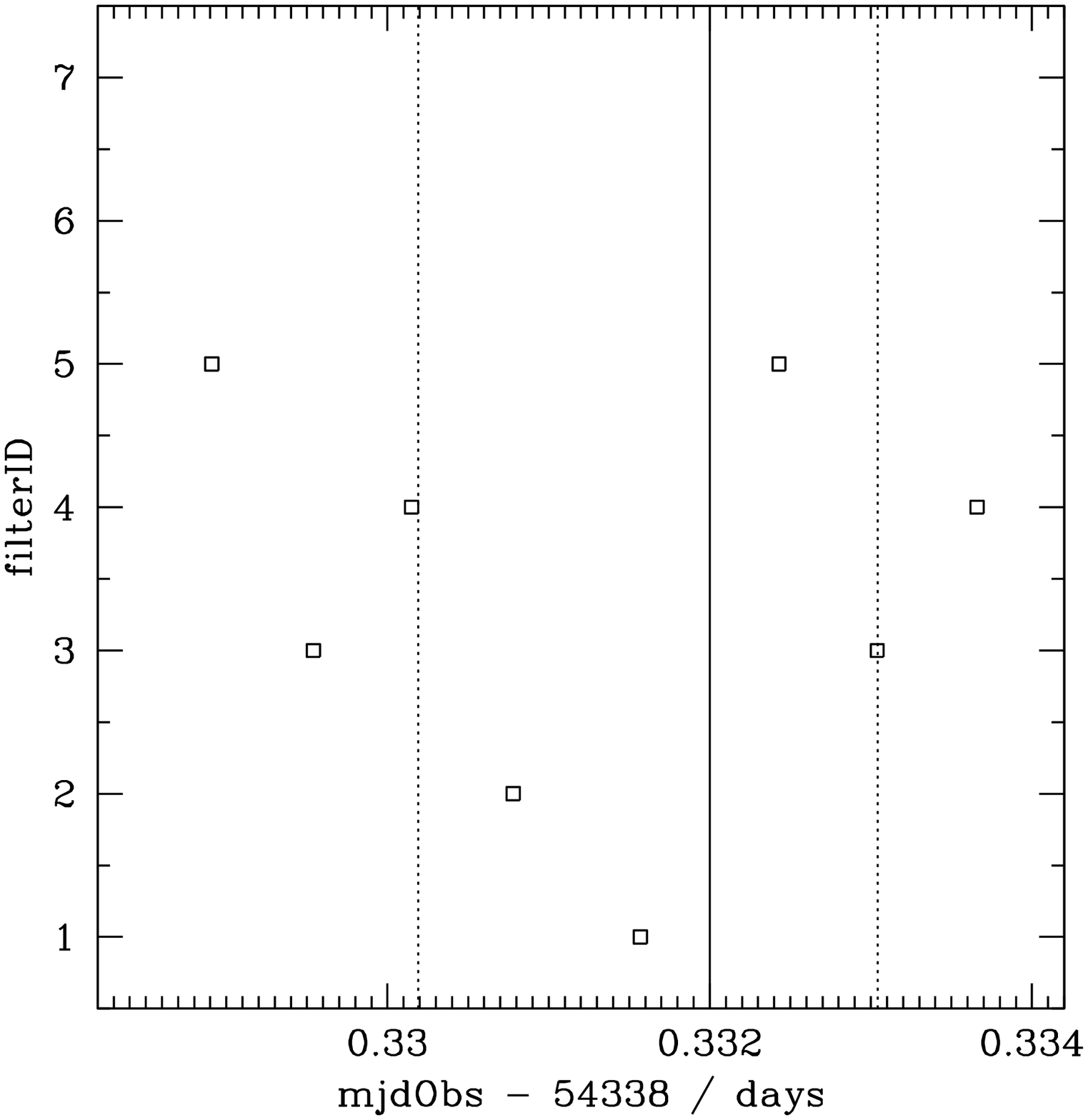}
\caption{Plot of observation time vs filterID for two ``epochs'' in the
WFCAM standard star programme. This is a correlated filter programme. The upper
plot shows the normal occurrence: two sets of broad-band filter observations
separated by one day with the observations in each set taken within the 15
minutes specified. The lower plot shows an abnormal case: 8 observations in 5
different filters taken close together. Only $5\%$ of intervals between epochs
are less than 15 minutes and are classed as abnormal. The other intervals are
at least 30 minutes long. In this case the observations are split into two
epochs of 5 and 3 observations. The mean in each epoch is shown by the dotted line and the solid line separates the two epochs.}
\label{fig:synEpochs}
\end{figure}

If the survey has correlated bandpasses, then the \verb+SynopticSource+ table
is created. This is created in the same way as the \verb+Source+ table,
except frame sets are created with a specific timespan designated in the
\verb+RequiredSynoptic+ table: the band merging criterion. If the criterion is
15 minutes, as in the case of the CAL programme, then only frames of different
filters that are observed within 15 minutes of each other are used to make a frame set at that
epoch - one row in the \verb+SynopticMergeLog+ table. If there are frames from
the same filter within 15 minutes of each other (see Fig~\ref{fig:synEpochs}),
then they are split into two epochs. Frames of different filters 
that are more than 15 minutes apart are also split into two epochs. 

\subsection{Joining the Source table to the intermediate data}

\begin{figure}
\includegraphics[width=84mm,angle=0]{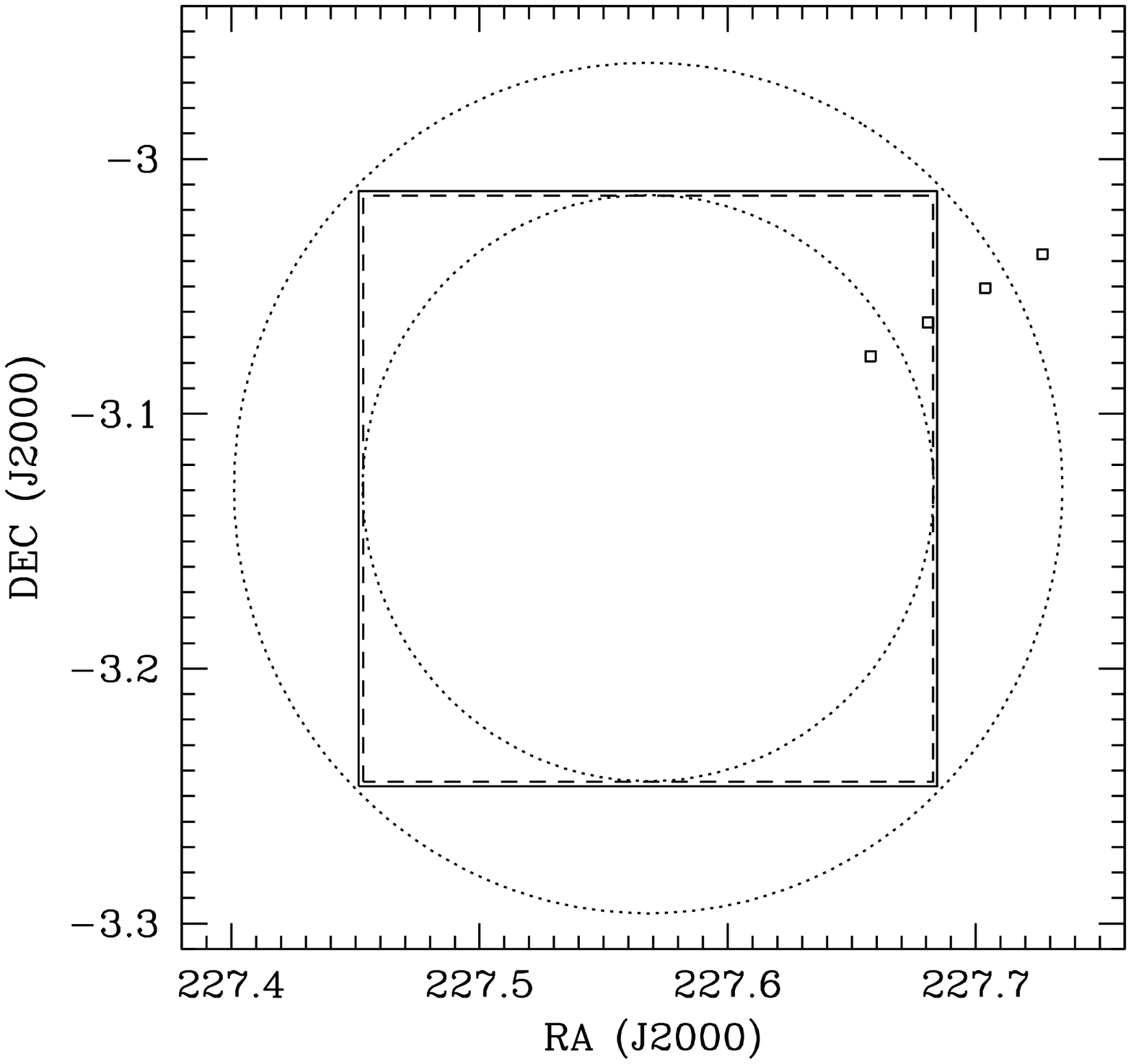}
\caption{Plot demonstrating the algorithm used to test whether an object should
have been observed. We define two radii (shown by the dotted circles) from the
centre of each detector frame. The inner one is inside the dither offset region
(marked by a dashed line) and the outer one is at the furthest extent of the
image. If the object is closer to the centre of the frame than the inner
radius, then it should definitely have been observed. If it is further than the
outer radius then it definitely was not observed. Any missing objects within the
two radii are tested more carefully. The four points demonstrate this. The inner
most point is closer to the centre than the inner radius, and the outer most
point is further than the outer radius, so these two points can be dealt with
in the first stage: the first is a missing detection and the second wasn't
observed. The two middle points need further testing. The inner most of these
two has been observed and the outer has not.}
\label{fig:checkmissWSA}
\end{figure}

The \verb+SourceXDetection+ or \verb+SourceXSynopticSource+ neighbour table is
created along with the other neighbour tables, with a matching radius specified in
\verb+RequiredNeighbours+. Currently we use a radius of $10\arcsec$, since this
includes most detections of stars with a measurable proper motion over a
timespan of 5-10 years. Users are warned that the neighbour table may well
include multiple matches of the same observation within this radius for a 
source
or indeed no matches. Only detections in the intermediate stacks in
\verb+Detection+ table are matched to the sources in the \verb+Source+ table.
This is the starting point for the creation of the best match table. The best
match table and the \verb+Variability+ table only include objects in the
\verb+Source+ table which are primary sources {\bf priOrSec}$= 0$ or {\bf
priOrSec}$=${\bf frameSetID}. The 
nearest detection in each frame is taken as the best match, if there is a 
match within $0.5\arcsec$ ($\sim10\times$ the typical astrometric error). If
there is no such detection on a detector frame that covers the position, then 
a default value is entered, allowing the user to know that an observation was
made but no object observed. The process of finding whether a particular source
should have been observed is time-consuming and is an important factor in
scaling this pipeline up to very large datasets. At present we split this 
process into two steps. In step 1, we calculate the great-circle distance from the source to the centre
of each missing frame. We accept an object as missing (i.e. should be within
the frame) if the distance is less than a minimum radius that is the shortest
distance from the frame centre to a dither offset from the frame edge, see
Fig~\ref{fig:checkmissWSA}. We reject all objects which are further out than
the maximum distance from the frame centre to the frame edge (the image extent). 
The sources which lie between the two radii have to be treated more carefully. This is step 2. The expected x and y positions are calculated using the equatorial
positions and the world coordinate system information from the frame. This
allows us to very accurately tell if the object is on the frame or not, and
whether it is within the dither offset. Step 1 is quick and step 2 is slow, 
so it is important to minimise the number of objects which require step 2 processing, see \S\ref{sec:missObjVSA}.

\subsection{Variability table curation.}

Once the \verb+SourceXDetectionBestMatch+ table has been created, the 
\verb+Variability+ table can be populated. The astrometric statistics are
calculated first, using data from all bands together, even those filters with
only one observation. Next, photometric data from each filter is analysed
separately. We calculate the numbers of good, flagged and missing observations,
the cadence information, and the best aperture to use for a given
source, similar to the method used by the Monitor project \citep{Mon}. The best
aperture is selected to have good signal-to-noise while avoiding contamination
by nearby objects, which can vary with seeing. 
   
Once the best aperture has been selected, we calculate the rest of the
photometric statistics for each source. Then we calculate the
intrinsic variation in the data by fitting a function to the noise that a
non-variable point source is expected to have. We then calculate the 
additional noise --- the intrinsic variation that the object has on top  of
this noise --- and use these measurements to classify whether an object is 
variable in this band.
 
When we have a \verb+SynopticSource+ table, we also calculate the
Welch-Stetson statistic: a measure of the correlation between two bands for
pairs of filters. We always use the same aperture magnitude in this case
($1\arcsec$ radius; {\bf aperMag3}) since using different aperture sizes, even
with aperture corrections for lost light in the wings of the point spread 
function (PSF), adds in
additional noise. Finally we use the number of good detections and the
ratio of the intrinsic-RMS to the expected-RMS in each filter to give a final 
classification of whether the object is variable or not.

\section{Analysis of Variability}
\label{sec:stats}
In this section we give the methods used to calculate the properties in the
variability table. In all cases, only the data that are not rejected by quality
control as possibly being unreliable are used. This reduces the total number of
real variable sources that can be discovered, but allows for greater confidence
in the remaining sample.

\subsection{Astrometry}
\label{sec:astrometry}

We calculate the mean right ascension {\bf ra} and declination {\bf dec} and the
errors ({\bf sigRa} and {\bf sigDec}) in the tangential coordinates. We define the direction of {\bf sigRa} as the
tangential coordinate that is perpendicular to both the Cartesian z-axis and 
the direction of the object from the Cartesian origin, $\underline{r}$. The
direction of {\bf sigDec} is defined as perpendicular
to the ``{\bf sigRa}'' direction and $\underline{r}$. These are calculated
through standard tangent plane astrometry. Currently, we assume the simplest
model for matching our objects between observations --- no motion --- but we have left place-holders 
for proper-motion and parallax parameters in the relevant tables. We also give 
the number of good frames ({\bf nFrames}) that go into the
astrometric calculation and the $\chi^2_{\nu}$ ({\bf chi2}) statistic for
astrometric fit to the multi-epoch data. Until we have evaluated a proper 
noise model for the astrometry (see \S\ref{sec:summary}), this will always have a value of $1$. 
 
\subsection{Observation statistics}
\label{sec:observstats}

We produce a number of statistics for observations through a single filter. The
first are to do with the number of observations in the band. We give the 
number of good observations ({\bf nGoodObs}), the number of flagged 
observations ({\bf nFlaggedObs}), where ${\bf ppErrBits}>0$, and the number of 
missing observations ({\bf nMissingObs}), where {\bf seqNum} is default. 
{\bf nMissingObs} is the number of frames that the object was not detected on.
It is good observations alone that contribute to the main variability 
statistics.
Users of the WSA and VSA may worry about incompleteness rate due to the
missing observations. Always, there is a decision to be made between
reliability and completeness. We have decided to tend towards
increased reliability in the classifications and statistics that we use,
but users can group data for each source through the best-match table and 
calculate statistics on data which has been flagged as having possible 
photometric errors if they think that these observations are
useful for their science and can even select observations with particular
error-bit flags. Using the UKIDSS-DXS data, we calculate the fraction of
incomplete observations in a particular filter (f):

\begin{equation}
f_{\rm incomp}(f)=\frac{\sum_i {\rm nFlaggedObs_i(f)}}{\sum_i {\rm
nGoodObs_i(f)+nFlaggedObs_i(f)}},
\end{equation}

\noindent where ${\rm nGoodObs_i(f)}$ and ${\rm nFlaggedObs_i(f)}$ are the
number of good and flagged observations of the $i^{th}$ source observed through
the filter (f) and the sum is over all sources in the programme. We don't
include the number of missing observations because they depend on the depth of the deep
frame compared to the depth of the individual observations. In the DXS, 
$f_{\rm incomp}(J)=0.26$ and $f_{\rm incomp}(K)=0.25$. The number flagged depends on the density of
sources. In dense regions, there will be more deblended observations and more
objects contaminated by cross-talk, so these numbers may be different in other
programmes.

We include four parameters which describe the cadence --- the interval between
observations. These are the {\bf minCadence} (minimum interval between any two
consecutive observations in this band), the {\bf medCadence} (median interval 
between any two consecutive observations in this band), the {\bf maxCadence} 
(the maximum interval between any two consecutive observations in this band) 
and the total period: the difference between the date of the final 
observation and the first observation.

\subsection{Photometric statistics}
\label{sec:photometry}
\begin{figure}
\includegraphics[width=84mm,angle=0]{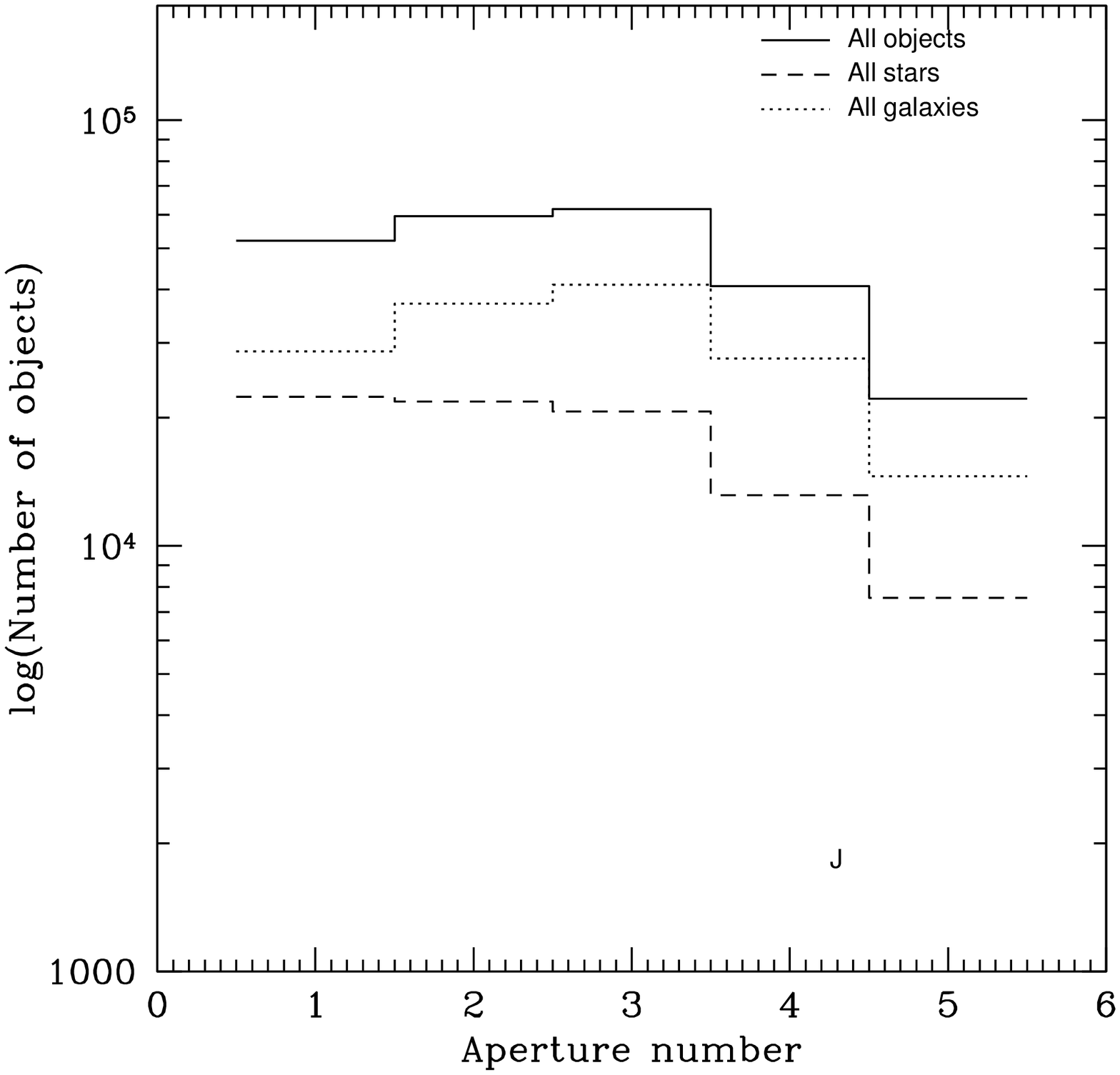}
\caption{Histogram of the best aperture for all J-band UKIDSS-DXS objects and
separately stars and galaxies.}
\label{fig:bestAper}
\end{figure}

For the good observations, we calculate the median absolute deviation (MAD) 
of the magnitude and the median magnitude for the first five aperture 
magnitudes. The best aperture is the aperture with the 
minimum MAD for apertures of diameter: $0.5, 0.7,
1.0, 1.4~\&~2.0 \arcsec$ ({\bf aperMag1} - {\bf aperMag5}). The distribution 
of best apertures is shown in Fig~\ref{fig:bestAper}. 
 
Using data measurements in the best aperture, we calculate the mean ($\mu$), 
standard deviation ($\sigma$) and skew ($\gamma$), defined as,
 
\begin{equation}
\gamma=\frac{n^2\mu_3}{(n-1)(n-2)\sigma^3},
\label{eq:skew}
\end{equation}

\noindent where $\mu_3$ is the third moment of the distribution,

\begin{equation}
\mu_3=\frac{1}{n}\displaystyle\sum_{i=1}^{n}(m_i-\mu)^3, 
\label{eq:mu3}
\end{equation}

\noindent $n$ is the number of observations and $m_i$ is the magnitude of
the i$^{th}$ observation. The skew can only be calculated for sources with 3
or more observations. The skew tells how symmetric the distribution of 
magnitudes around the mean is. For less than 3 observations, the skew is given
a default value. For faint objects near the detection limit of the epoch
images, these quantities are biased since detections only occur in 
images where the flux is
scattered to brighter values. This affects all photometric statistics. This
bias can be removed by using list-driven photometry, see \S\ref{sec:summary},
where the light in an aperture is measured, whether there is enough flux for an
independent detection or not.

Next we calculate the expected RMS, $<\zeta(m)>$, in the frame set. The expected
RMS is the RMS for a non-variable point source. Selecting only those sources
which are classified as star-like objects and are sources in only one set of
deep stacks (i.e. have ${\bf priOrSec}=0$) to avoid overlaps (see
Appendix~\ref{app:overlaps}), we order the data by mean magnitude over a range
of eight magnitudes with the faintest source half a magnitude brighter than the 
expected magnitude limit of the intermediate stacks. The data are split into 
bins of equal numbers of objects, with each bin having $\sim100$ objects, or a
minimum 10 bins. In each bin the median and MAD of the standard deviation are
calculated as well as the median of the mean magnitude. We then use a 
least-squares method to fit the best fit function to the noise model.
Currently we use the Strateva function \citep[see][for more
details]{Stv01,Ssr07} as a functional fit to the noise. The Strateva function,

\begin{equation}
<\zeta(m)>=a+b\,10^{0.4m}+c\,10^{0.8m},
\label{eq:strat} 
\end{equation}

\noindent gives the noise properties as a function of magnitude, where $m$ is
the magnitude and $a$, $b$ \& $c$ are the Strateva parameters. These parameters
are recorded in the \verb+VarFrameSetInfo+ table ({\bf aStrat} etc). In this
model, the noise tends to a minimum equal to the parameter, $a$, at bright
magnitudes. This is an empirical fit to the data, and has the advantage that it
can be fitted to the data as the pipeline is processing, with no prior
modelling of the noise. However, some datasets may have significant differences
in the exposure time and sky background, and therefore noise properties of each
individual epoch frame. An empirical fit can only give an average noise for the
whole set of epochs, whereas a noise model based on the underlying processes 
can weight each epoch correctly. Most datasets will use the same or very
similar exposure times for each epoch: given a fixed total integration time and
a fixed number of epochs the most efficient way to target as many objects in
each epoch is to divide the total exposure time equally among the observations.
For this reason, we will suffice, for now, with the empirical model.

A chi-squared statistic for the hypothesis of no variability can be
calculated. The model is the mean magnitude $\mu$  and the error is the 
expected magnitude $<\zeta(m)>$. The chi-squared per degree of freedom is given
by,

\begin{equation}
\chi^2_{\rm
ndof}=\frac{1}{n-1}\displaystyle\sum_{i=1}^{n}\frac{(m_i-<\mu)^2}{<\zeta(m)>^2}.
\label{eq:chi2}
\end{equation}

The probability of this source being variable can be calculated by integrating
the chi-squared distribution, 

\begin{equation}
p(\chi^2/\nu,\{m\},I)=\int_0^{\chi^2}\frac{y^{(\nu-1)}\exp({-0.5y})}{2^{\nu}\Gamma(\nu)}dy,
\label{eq:pchi2}
\end{equation}

\noindent where $\nu$ is half of the number of degrees of freedom.  

The intrinsic RMS, $\sigma_{\rm int}$, is the RMS intrinsic to the
source. Assuming that the flux errors that make up the expected RMS are
uncorrelated, and independent of the intrinsic variation, then the 
intrinsic variation is

\begin{equation}
\sigma_{\rm int}=(\sigma^2-<\zeta(m)>^2)^\frac{1}{2}.
\label{eq:intsig}
\end{equation}

Objects are classified as variable in a filter if $p(\chi^2_{\rm ndof})\geq0.96$
and $(\sigma_{\rm int}/<\zeta(m)>)\geq3$. i.e. the probability of it being
a variable is greater than $96\%$ and the standard deviation is at least 3
times the expected noise for this magnitude. We calculate a final variability
classification based on data from all the filters. An object is a variable
if it matches the criteria,

\begin{equation}
\Sigma_s=\frac{\sum_f w_f\frac{\Sigma_{\rm int,f}}{<\zeta(m(f))>}}{\sum_f
w_f}\geq3,
\label{eq:varClass}
\end{equation}

\noindent where $\Sigma_s$ is a weighted ratio of the standard deviation to the
expected noise summed over all filters (f). The weighting factor $w_f$ in each
filter is based on the number of observations.

\begin{equation}
w_f=\frac{N_{\rm Obs,f}-N_{\rm min}}{N_{\rm Obs,max}-N_{\rm min}}.
\label{weightFac}
\end{equation}

$N_{\rm Obs,f}$ is the number of good observations of that
source in filter $f$. $N_{\rm min}$ is the minimum allowable number of
observations for variability classification (5), and $N_{\rm Obs,max}$ is the
maximum number of observations of that source in any filter. We illustrate
this with an example source: UDXS J105644.55+572233.4, see Fig~\ref{fig:lc2}.
This object has 25 good observations in $J$ and 38 in $K$, $\sigma_{int,J}=0.037$ and
$\sigma_{int,K}=0.061$, $<\zeta(J)>=0.008$ and $<\zeta(K)>=0.008$. The
weighting factor in $K$ is 1, since $K$ contains the maximum number of
observations. The weighting factor in $J$ is 0.606. In this case the source
varies at more than 3 times the noise in each filter ($4.6\times$ in $J$ and
$7.6\times$ in $K$), so the weighted ratio is 6.5. In this case, both ratios
were greater than the limit, but if a source had a ratio less than the limit in one filter where there were many observations and greater in one where there were
few (or vice-versa) then the filter with most observations is given the most
weight. If there are less than or equal to five observations, then the filter
has no weight: i.e. only objects with greater than or equal to five good
observations in one filter can be classified as variable. This methodology
uses a simple prior --- the relative number of good observations --- as a
weighting function for each filter, but does not use full Bayesian analysis currently. The
classification may be improved in the future, to correctly use Bayesian methods and to provide a wider range of classifications that point towards different types of variable.

\subsubsection{Correlated observation programmes}

We produce the same single pass-band statistics as above for correlated programmes. In addition, for 
each pair of broadband filters, ordered by wavelength, we calculate the
Welch-Stetson statistic \citep{WS93}, 

\begin{equation}
I_{\rm
WS}=\sqrt{\frac{1}{n(n-1)}}\displaystyle\sum_{i=0}^n(\delta\,b_i\delta\,v_i),
\label{eq:iWS}
\end{equation}

\noindent where $\delta\,b_i$ and $\delta\,v_i$ are the weighted 
differences between the i$^{th}$ observed magnitude and the weighted mean 
magnitude in the two filters. If the differences correlate or anti-correlate 
then $|I_{\rm WS}|$ is large. If they are random then $I_{\rm WS}\sim0$.

\section{UKIDSS Deep Extragalactic Survey: Data Release 5}
\label{sec:DR5}

The UKIDSS DXS is a deep, multi-epoch survey intended to study galaxy
and galaxy-cluster evolution at intermediate redshifts. The depth is built up
from $\sim20$ individual epochs taken at various times when fields were
visable and the observing conditions best met the DXS requirements. In
UKIDSS-DR5, the DXS covers $\sim14.8$ sq. deg. The magnitude limits for
each individual epoch are $J\le21.12$ mag and $K\le19.74$ mag. The DXS
observations are in four main regions, the Lockman Hole ($l\sim148$ deg,
b$\sim52$ deg), the XMM-LSS (l$\sim171$ deg, b$\sim-58$ deg), the 
European Large Area ISO Survey - North 1 field (ELAIS-N1; l$\sim85$ deg,
b$\sim45$ deg) and the Visible Multi-Object Spectrograph 4 field (VIMOS
4; l$\sim63$ deg, b$\sim-44$ deg). For detailed information about the
UKIDSS-DXS, see \cite{Dye06} and \cite{UKIDSS-DR1}. {\bf The DXS is the first WFCAM programme to be released having been processed with the new synoptic pipeline. It is large and varied enough to test most aspects of the pipeline and give a range of interesting results.}
b$\sim52$ deg), the XMM-LSS (l$\sim171$ deg, b$\sim-58$ deg), the ELAIS-N1 (l$\sim85$ deg, b$\sim45$ deg) and the VIMOS 4 (l$\sim63$ deg, b$\sim-44$ deg). For detailed information about the UKIDSS-DXS, see \cite{Dye06} and \cite{UKIDSS-DR1}. The DXS is the first WFCAM programme to be released
having been processed with the new synoptic pipeline. It is large and
varied enough to test most aspects of the pipeline and give a range of
interesting results. 

\begin{figure}
\includegraphics[width=84mm,angle=0]{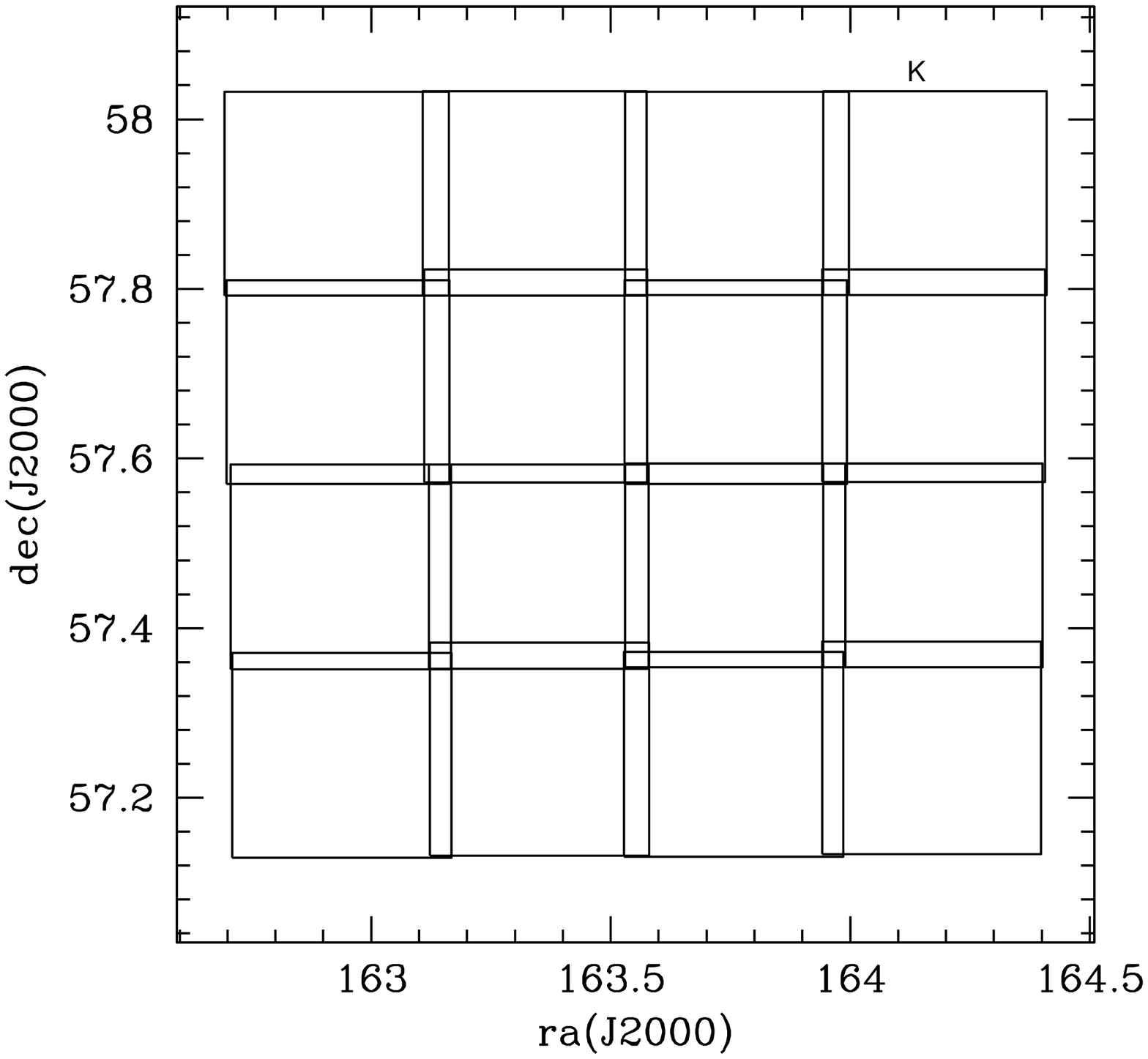}
\caption{Plots of the overlap of the 16 K-band deep image extensions in the 
UKIDSS-DXS DR5. These are in the Lockman Hole region, stacks LH010\_0,
LH011\_0, LH010\_1 \& LH011\_1.}
\label{fig:pointingK}
\end{figure}

The results up to the end of \S\ref{sec:recal} are
from the first four pointings (the first eight products in {\bf RequiredStack}
of the UKIDSS-DXS using Data Release 5). We use this subset only to make the 
figures easily readable to avoid confusion. The four pointings are made from
eight deep stacks (four J and four K) and these are merged into 16 frame-sets.
The overlap of the deep stacks are shown in Fig~\ref{fig:pointingK} for the K band.

\begin{figure}
\includegraphics[width=84mm,angle=0]{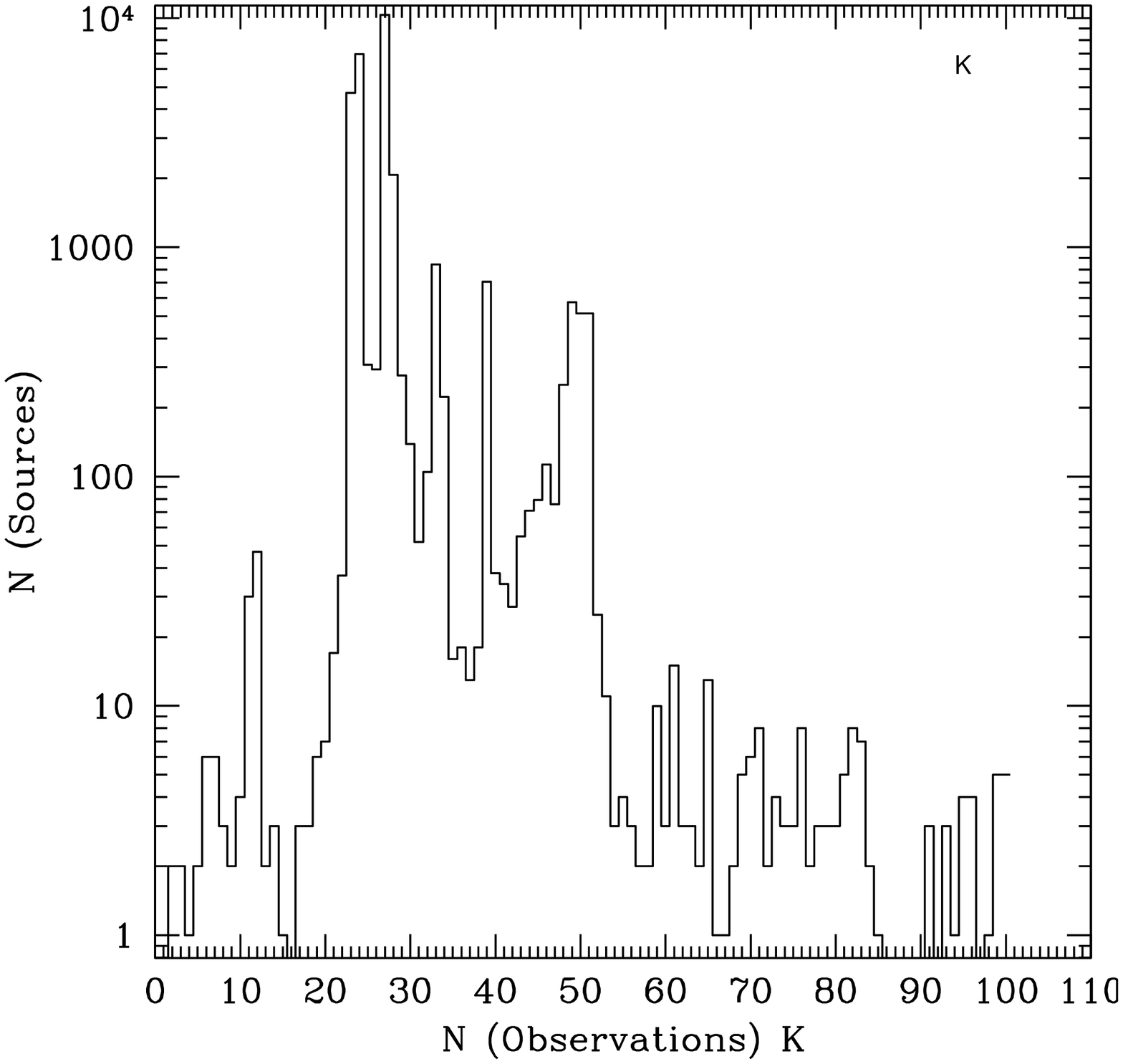}
\caption{The histogram of the number of observations for each source in the 
DXS K-band. The number of observations is the sum of the number of good 
observations {\bf nGoodObs}, the number of flagged observations {\bf 
nFlaggedObs} and the number of missing observations {\bf nMissingObs}, in
the {\bf Variability} table. The 
main peak at 27 observations is the number of K-band epochs 
that go into each K-band deep stack. The second peak at $\sim52$ observations
occurs when two deep stacks overlap and the further peaks at $\sim100$
observations  are when four deep stacks overlap. The peaks are from overlaps
with fields outside this main area. The overlap regions in each case cover a smaller
area, so there are fewer sources. The number of sources outside the 
peaks occur because each intermediate stacks is slightly offset relative
to the others.}
\label{fig:nPointSouK}
\end{figure}

The histogram of the number of observations is shown for the K-band in 
Fig~\ref{fig:nPointSouK}. This plot demonstrates that the modal number of
epochs is 27 in the K-band which corresponds to the number of epochs in two of
the pointings. There are 23 and 24 epochs in the other two pointings. The
number of epochs in the J-band is 14 or 15. There are also sources with more
than than 27 observations, particularly around 50. These are sources where
two pointings overlap. There can be up to 100 observations for a source, where
4 pointings overlap, as can be seen in Fig~\ref{fig:pointingK}. 

\subsection{Effects of Internal Recalibration}
\label{sec:recal}

\begin{figure}
\includegraphics[width=84mm,angle=0]{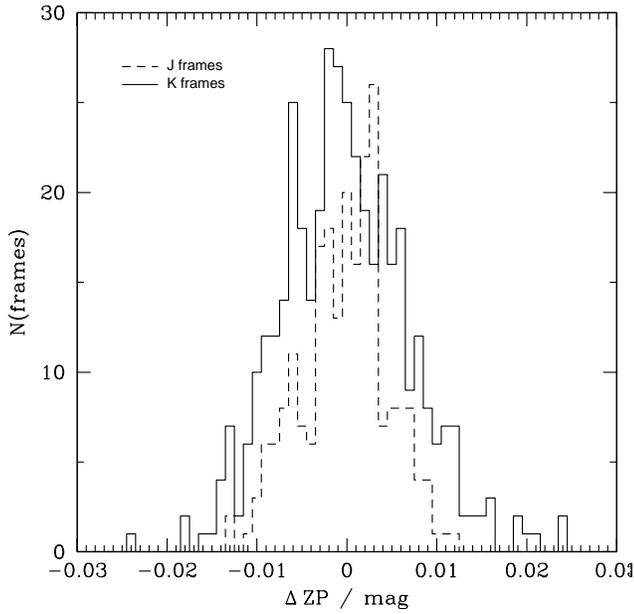}
\caption{Histogram of the difference in zeropoint when DXS frames are
recalibrated. The average shift in each case is close to zero
($-0.0002\pm0.0048$ mag in J and $-0.0002\pm0.0071$ mag in K), so there is no
systematic shift in the photometry, just a reduction in the variation between
the frames.}
\label{fig:histdfZPs}
\end{figure}

\begin{figure} 
\includegraphics[width=84mm,angle=0]{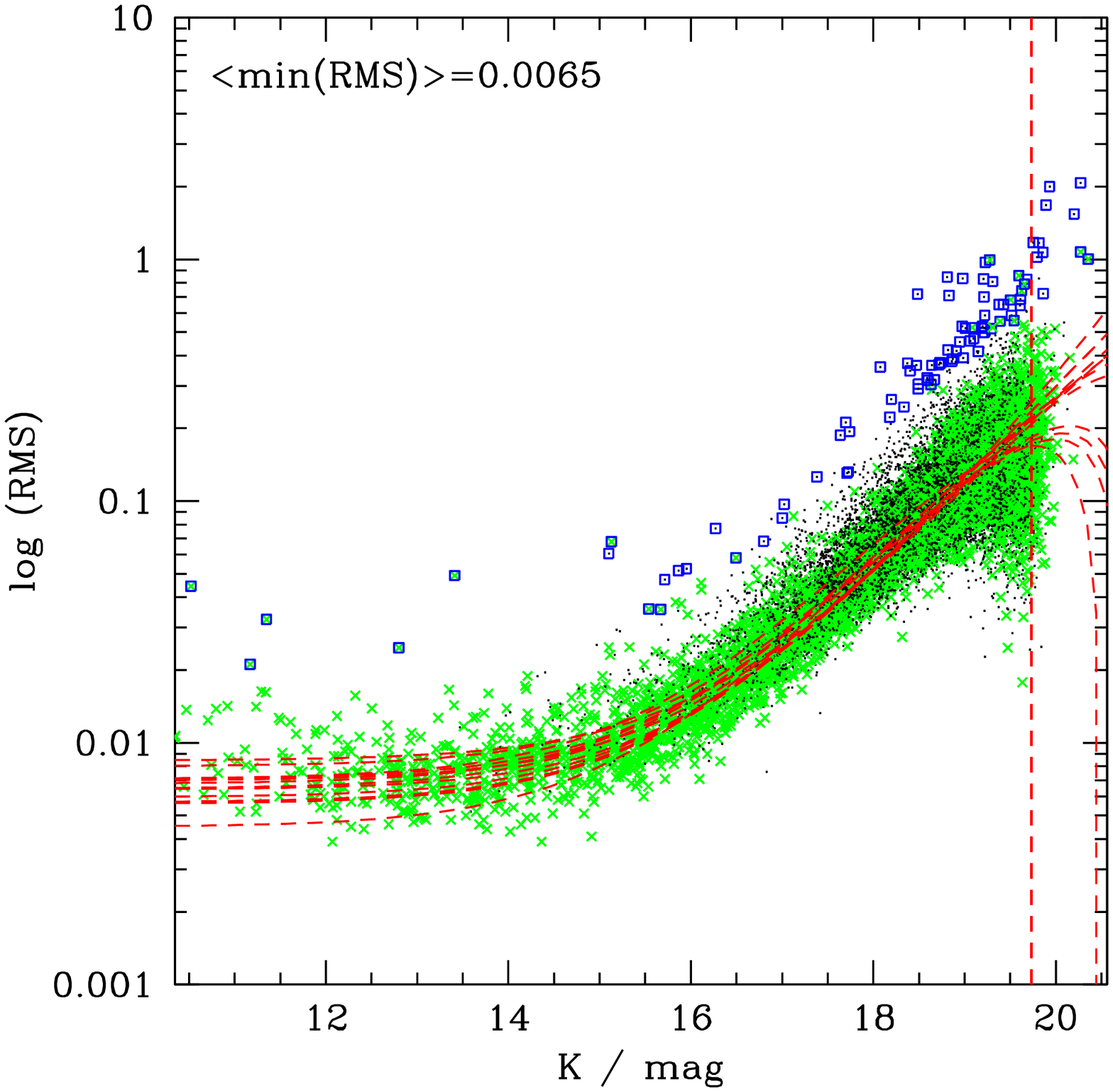}
\caption{RMS versus magnitude plot for K-band data before recalibration.
The black dots show all the data, the green crosses are objects classified as
stars in the {\bf Source} table and the blue squares are objects classified as
variable. The red dashed vertical line is the expected magnitude limit for
the intermediate stacks and the dashed curves are the best fit Strateva
curves to the minimum RMS as a function of magnitude for each frame-set. 
Each line represents the empirical fit for the noise in a different pointing. 
The mean of the values {\bf aStrat}, which represents the minimum RMS for a bright
star is 0.0065 mag.}
\label{fig:magRmsK}
\end{figure}

\begin{figure} 
\includegraphics[width=84mm,angle=0]{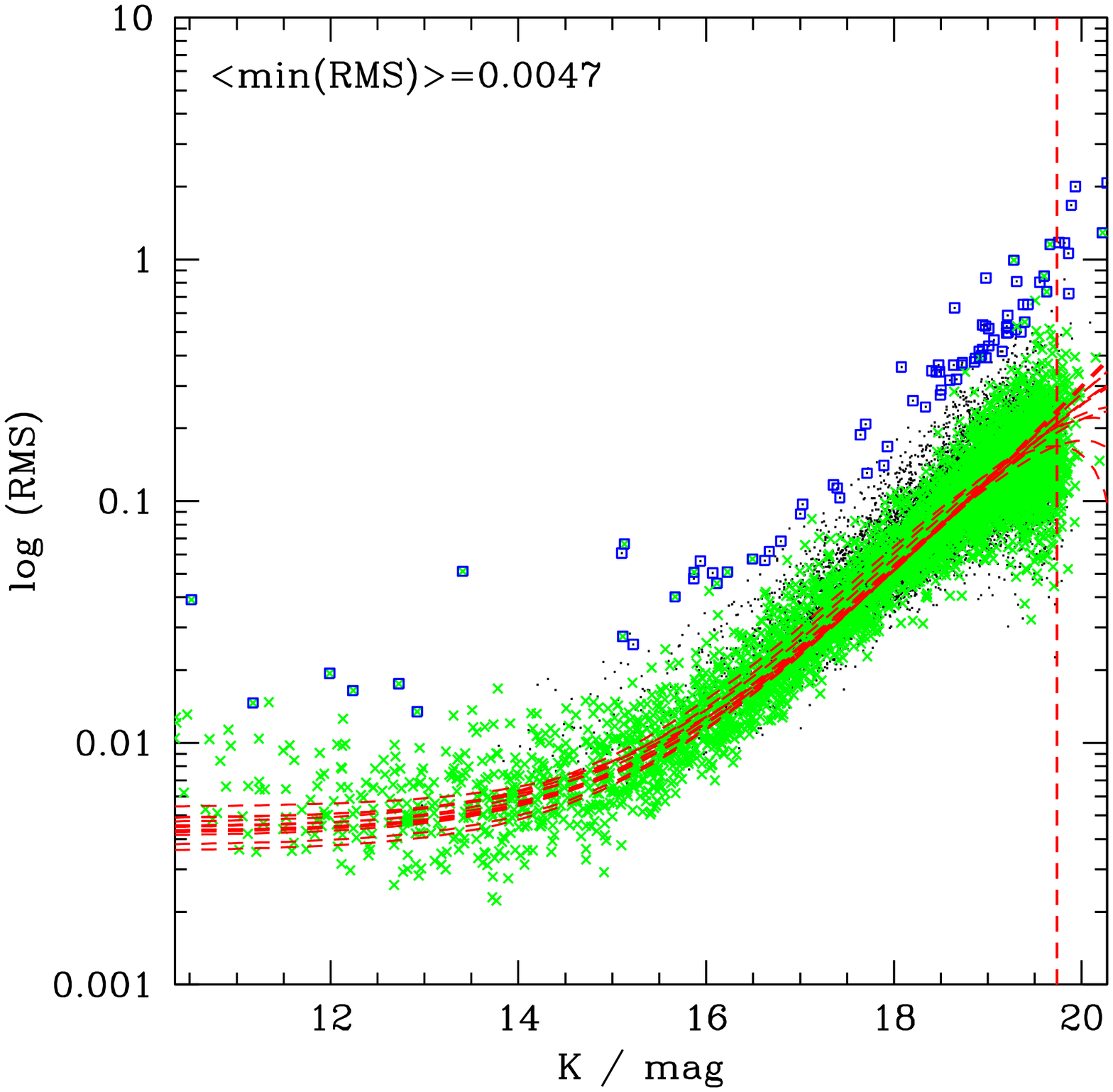}
\caption{RMS versus magnitude plot for recalibrated K-band data.
See Fig~\ref{fig:magRmsK} for details. The mean of the values 
{\bf aStrat}, which represents the minimum RMS for a bright star is 0.0047 
mag.}
\label{fig:magRmsKrec}
\end{figure}

Fig~\ref{fig:histdfZPs} shows the histogram of the difference in zeropoints for
intermediate stacks before and after recalibration. 
Recalibration of individual epochs makes a significant improvement in quality
of the variability statistics and classification, as
can be seen by comparing the ``before'' and ``after'' magnitude-RMS plots:
Fig.~\ref{fig:magRmsK} and Fig.~\ref{fig:magRmsKrec}. These plots show the RMS as a function of magnitude and are useful for diagnosing the noise properties of a frame or dataset and for finding variables. The red-dashed
lines show the fit to the minimum RMS for each frame. Generally speaking there
is good agreement between the stellar locus and the noise model, particularly at
the faint end. The additional divergence at the bright end reflects the fewer data points. The noise flattens at the bright end
when the random, ``white'' noise ceases to dominate and correlated ``red''
noise \citep[see][]{Mon} becomes significant, as seen in Fig~\ref{fig:magRmsK}.
In the better calibrated data, Fig~\ref{fig:magRmsKrec}, it is
noticeable that the noise increases for the very brightest objects, which is
not reflected in the noise model. This may be due to saturation effects. We have
marked the objects classified as variables by blue boxes. In the recalibrated version, the
typical minimum-RMS is $0.0047$ mag rather than $0.0065$ mag, meaning that the
noise across all frames for a bright object is $\sim0.002$ mag lower. If we are
confident that $3\sigma-$detections are good, then we can detect variables
with amplitudes of $0.013$ mag rather than $0.020$ mag. This is reflected in
the larger number of blue squares in Fig~\ref{fig:magRmsKrec}.

This is just a very simple recalibration using a change in zeropoint.
More complicated changes, fitting for spatial variations in both the astrometry
and photometry are possible too. The recalibration only 
affects frames within a single pointing and we have not made any effort to 
recalibrate across pointings using overlaps, since the  number of objects that
can be used is much fewer. Only $\sim3\%$ of the objects in a frame are in the overlaps, see Appendix~\ref{app:overlaps}. Very
good relative calibration can be achieved this way, but to get much better absolute calibration macro-stepping of the 
detectors is necessary to remove all instrumental effects. This involves
observing the same large group of stars multiple times with different parts of
the same detector and different detectors.

\begin{figure}
\includegraphics[width=84mm,angle=0]{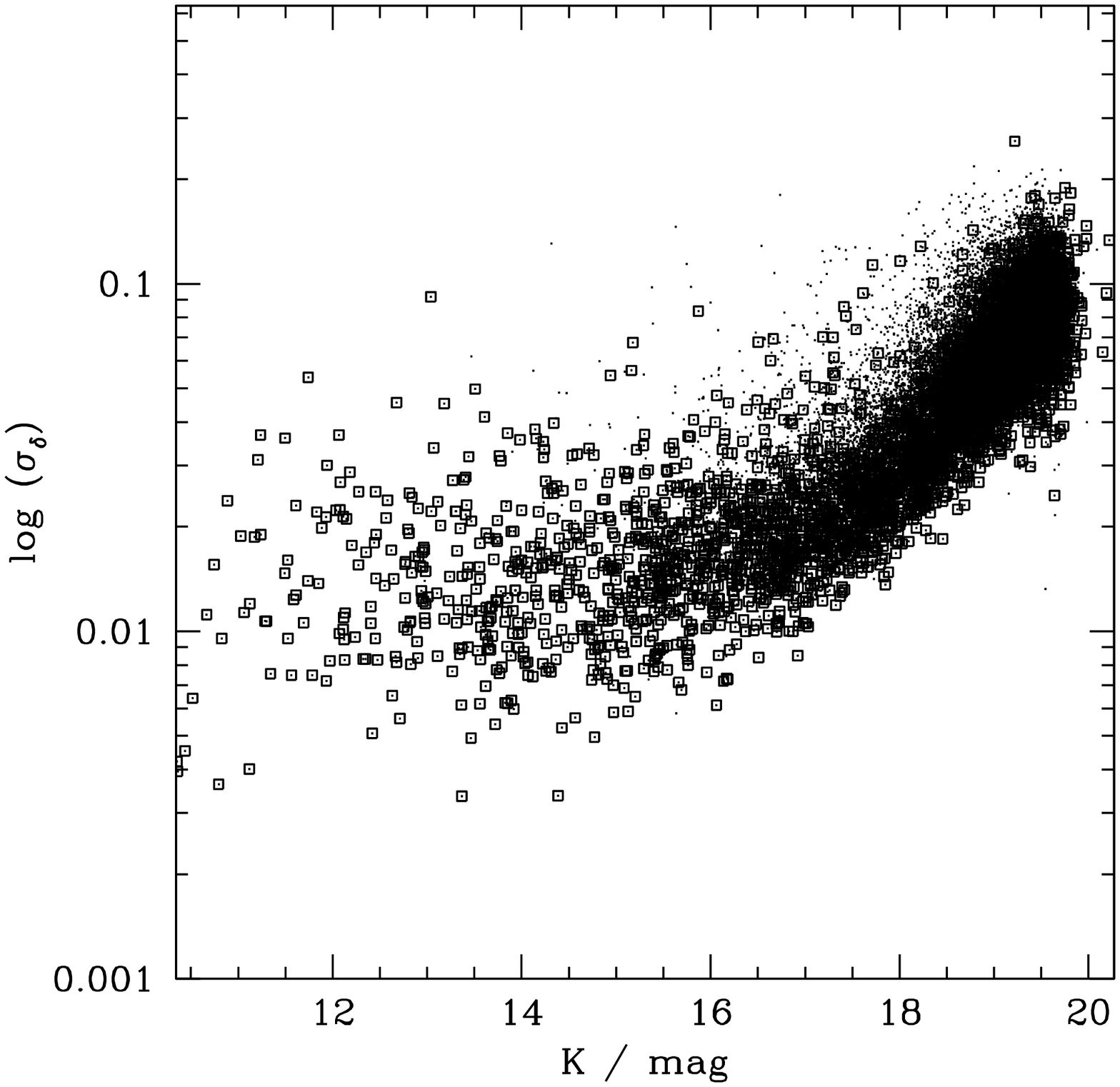}
\caption{Astrometric error versus magnitude plot in K-band for UKIDSS DXS data.
The dots are galaxies and the squares are stars. For objects with $K<16$, the
typical error is $0.015\arcsec$, and then gets larger for fainter objects.}
\label{fig:magSigDecK}
\end{figure}

The astrometric error, {\bf sigDec} ($\sigma_{\delta}$) is shown as a function
of K-band magnitude in Fig~\ref{fig:magSigDecK} and shows a similar variation
with magnitude as the photometric error. In the future, we will fit the
magnitude-astrometric error in the same way as we fit the magnitude-photometric
error in \S\ref{sec:photometry}, see Fig~\ref{fig:magRmsK}.

\subsection{Variable Objects}

To find interesting variable objects, we select sources which
are classed as variable, have mean magnitudes that are at least 3 magnitudes
brighter than the expected magnitude limit in each band and not default
and have more than 20 good observations in the K-band and more than
12 in the J-band. These last criteria are used since a typical DXS stack has 25
K epochs (see Fig~\ref{fig:nPointSouK}) or 15 J epochs and we want to be close
to the maximum to be able to see structure in the light curves.
With this selection we found 40 objects. There are 3686 objects (variables
and non-variables) which match these criteria apart from the variability
classification. We looked through the light curves of all of these and selected
the most interesting.

\begin{figure}
\includegraphics[width=84mm,angle=0]{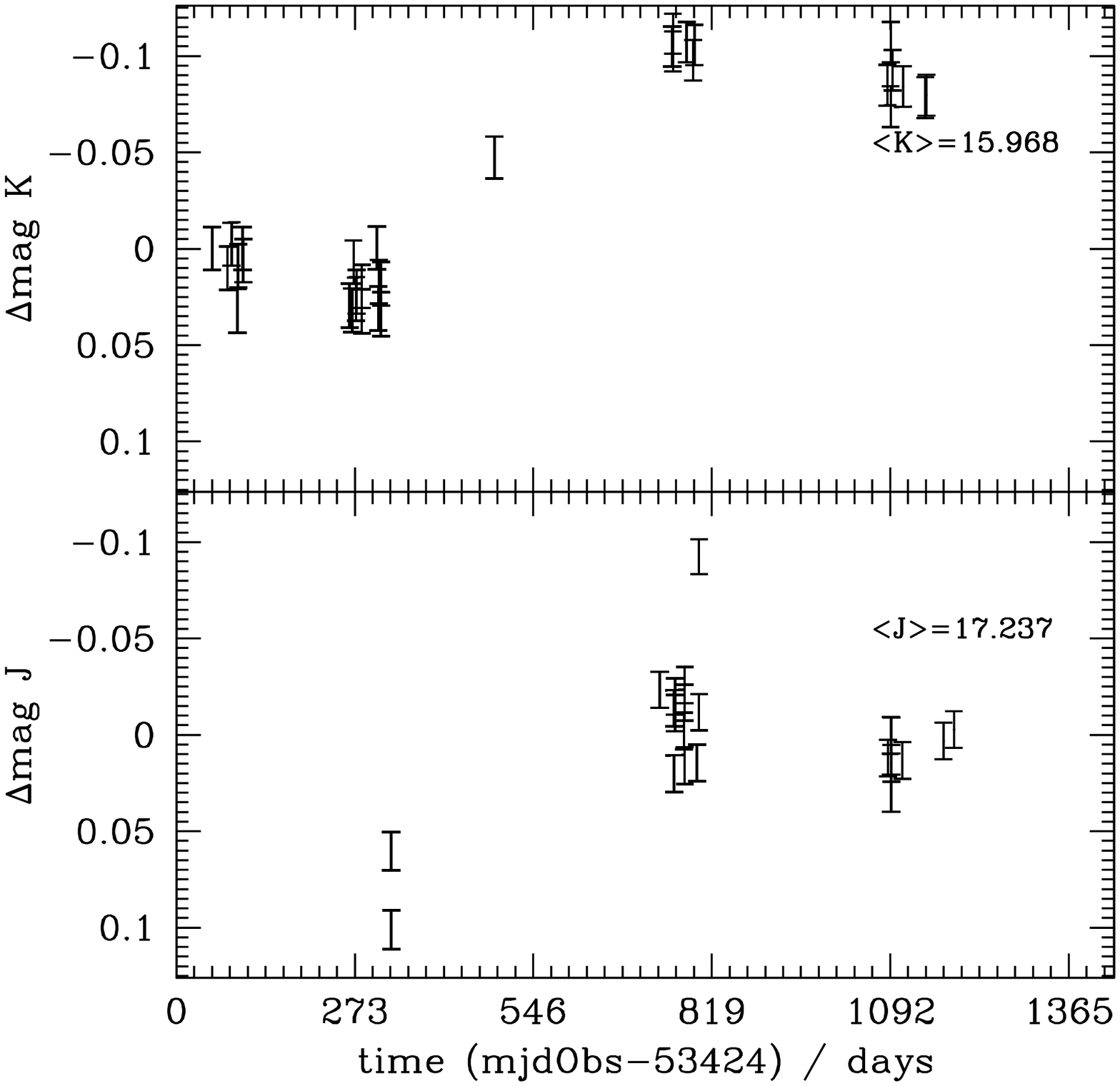}
\caption{Light curve for a variable galaxy, UDXS J105639.43+575721.6 
Difference in magnitude around the median is plotted vs time in days from the
first observation. Points with error bars show
good observations. All the observations in this case were good. The statistics
are calculated from the good observations only. The median magnitude of the observation is given. The light curve shows a clear minimum and maximum in the K-band, with an amplitude of 
$\sim0.1$ mag. There is only the maximum in the J-band. There are not enough 
data to determine whether this is a periodic variable or not. This object is
quite red: $(J-K)=1.27$ mag and has the profile of an extended source.}
\label{fig:lc1}
\end{figure}

\begin{figure}
\includegraphics[width=84mm,angle=0]{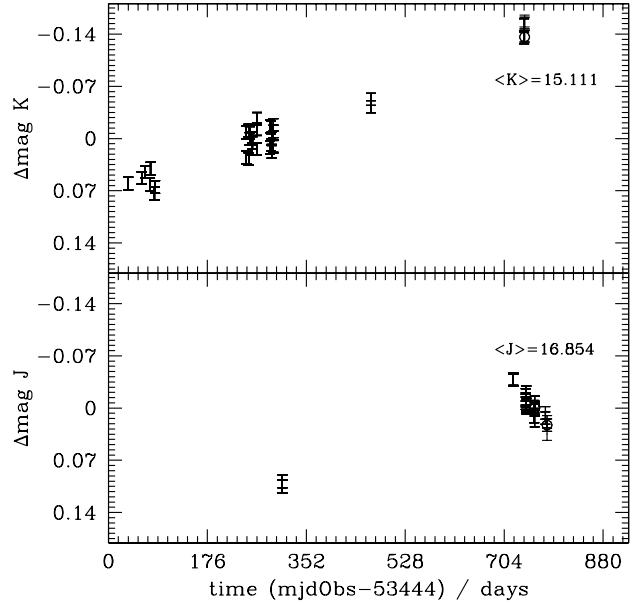}
\caption{Light curve for a variable galaxy, UDXS J105644.55+572233.4. 
Points with error bars show good observations and circles {\it without} error
bars show flagged observations. The light curve shows a linear increase in
brightness in the K-band, with an increase of $\sim0.2$ mag over $\sim700$ days. In the J-band there is a decrease in brightness, but there
are very few points taken at the same time. This object is quite red:
$(J-K)=1.74$ mag and has the profile of an extended source.}
\label{fig:lc2}
\end{figure}

\begin{figure}
\includegraphics[width=84mm,angle=0]{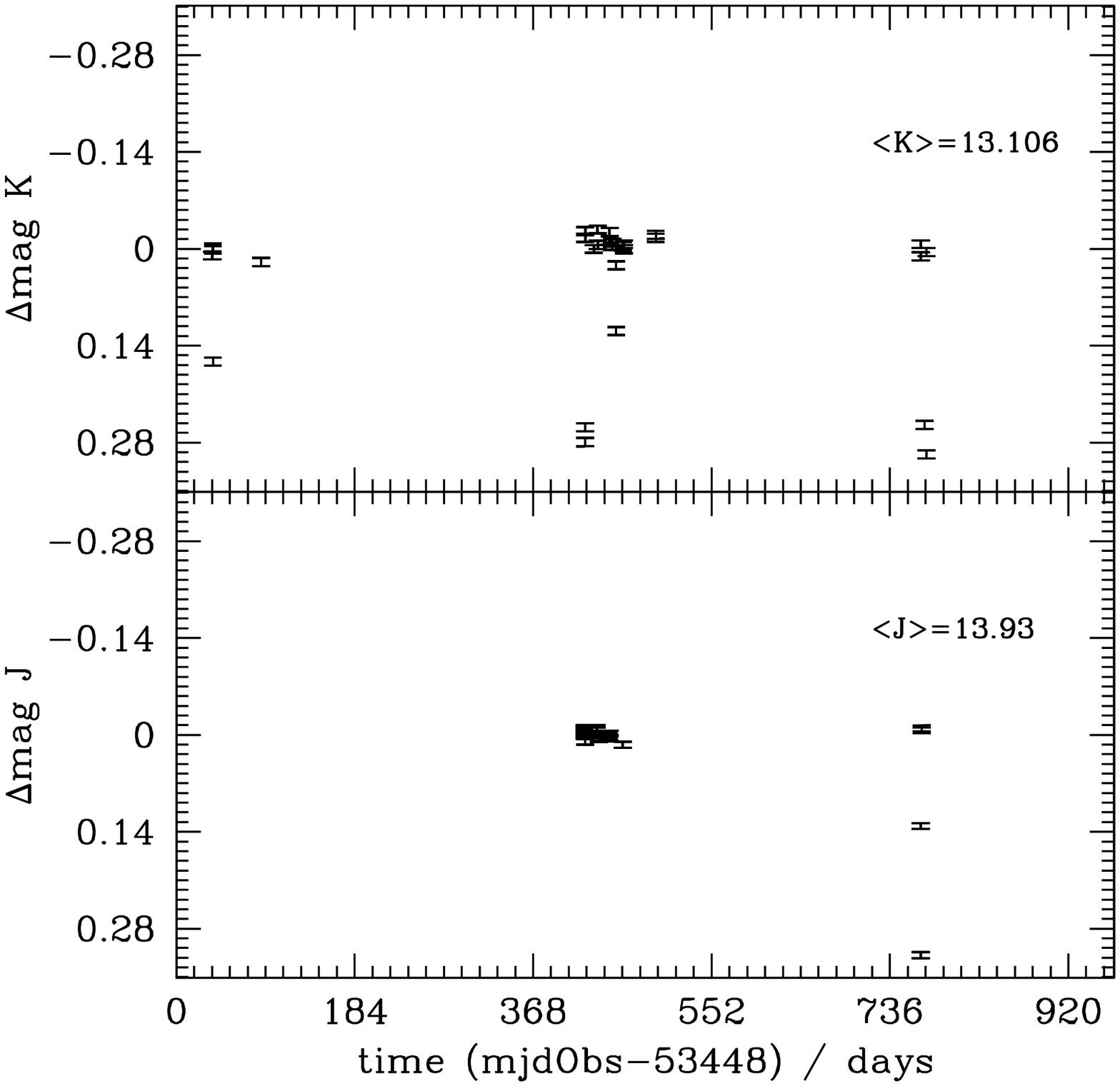}
\caption{Light curve for a variable star, UDXS J160650.11+544924.5. 
Points with error bars show good observations. The light curve is mainly
flat in both bands but several dips of $\sim0.2$ mag. More closely spaced
observations could determine whether this is real (possibly an eclipsing
binary) or not.}
\label{fig:lc3}
\end{figure}

To use SQL to generate lightcurves for a specific source, see
Appendix~\ref{app:examp} or the SQL cookbook on the WSA
interface\footnote{http://surveys.roe.ac.uk/wsa/sqlcookbook.html\#LightCurve}.
We give examples of a variety of variables in Fig~\ref{fig:lc1} ---
\ref{fig:lc3}. Fig~\ref{fig:lc1} \& \ref{fig:lc2} show two variables that are also classified as galaxies by the star-galaxy classifier and are also much
redder than typical stars (in low extinction regions). The colours suggest that
these are extragalactic objects, Active Galactic Nuclei (AGN), or heavily
reddened stars, such as Asymptotic Giant Branch (AGB) stars \citep{AGB} which
produce dust in their outer layers. These two objects are
classified as extended sources in the deep images, but a slowly moving star may
appear elliptical in a combination of images. We look at the distribution of
the star-galaxy separation statistic {\bf classStat} for the individual
observations of these two objects. UDXSJ105639.43+575721.6 has $\overline{\rm
classStat}_K=16.5\pm4.7$ and $\overline{\rm classStat}_J=14.5\pm2.5$. UDXSJ105644.55+572233.4 has 
$\overline{\rm classStat}_K=6.3\pm1.6$ and $\overline{\rm
classStat}_J=8.1\pm2.0$. A point-source object is expected to have $-3.0\le{\rm
classStat}\le2.0$, so these two objects are certainly extended sources and are
likely to be AGN. The first shows an undulating variation in both $J$ and $K$
bands, whereas the second shows are linear increase in brightness in $K$ over 700 days and
a subsequent decrease in brightness in $J$. Fig~\ref{fig:lc3} shows a star
(based on colours and star-galaxy separation) that
dims by more than 0.2 mag on several occasions in both $J$ and $K$. Follow up
observations may prove this to be an eclipsing binary and determine the period.

In addition to finding many real variables such as the example above, we 
also found some cases of poor calibration between adjacent overlapping frames,
see Appendix~\ref{app:overlaps}. To avoid regions with overlaps, it is best to
set ${\bf priOrSec}=0$ in the \verb+Source+ table. 

\begin{figure}
\includegraphics[width=84mm,angle=0]{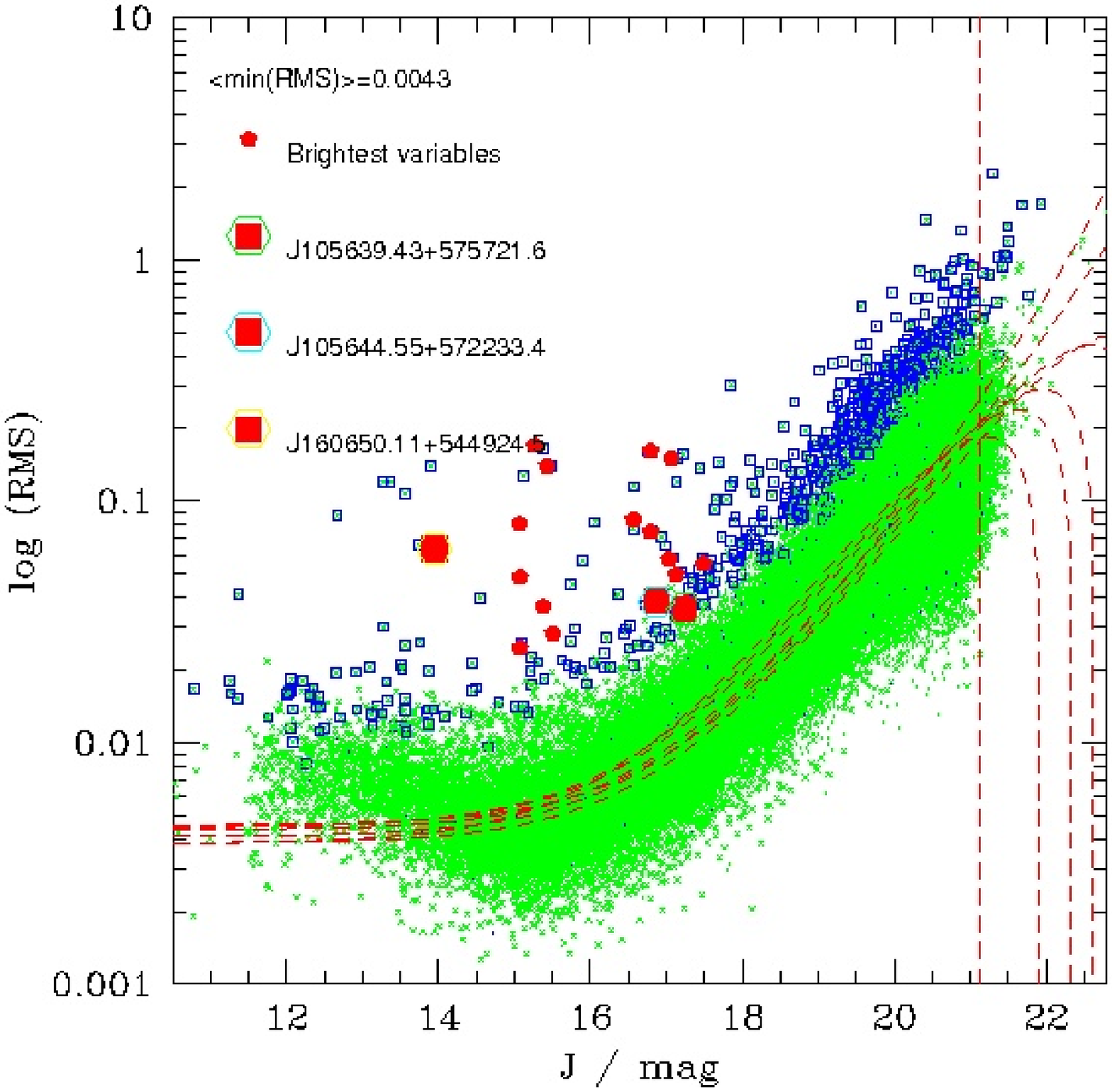}
\caption{Magnitude versus RMS plot for $J$-band data. See Fig~\ref{fig:magRmsK}
for details. The mean of the values {\bf aStrat}, which represents the minimum
RMS for a bright star is 0.0043 mag. The small red pentagons are the
brightest good variables described in Table~\ref{tab:brightDXS} and the 3 large
squares with coloured circles are the 3 objects of interest in
Figs~\ref{fig:lc1} --- ~\ref{fig:lc3}. There is a significant deviation in the
RMS of the stellar population compared to the noise model for $J<14$
mag.}
\label{fig:magRmsJfin}
\end{figure}

\begin{figure}
\includegraphics[width=84mm,angle=0]{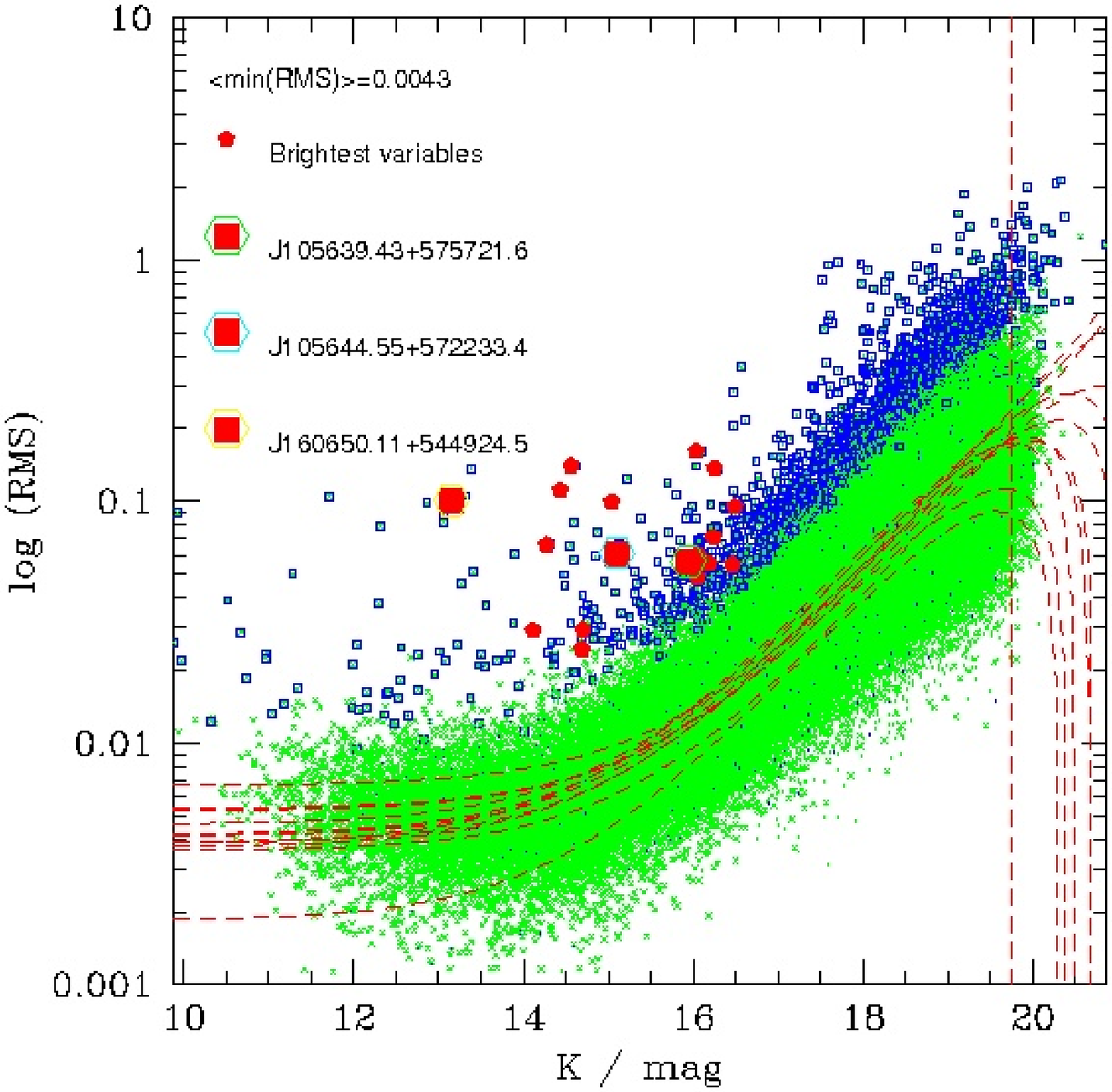}
\caption{Magnitude versus RMS plot for $K$-band data. See
Fig~\ref{fig:magRmsJfin} for details. The mean of the values {\bf aStrat}, which
represents the minimum RMS for a bright star is 0.0043 mag.}
\label{fig:magRmsKfin}
\end{figure}

Figs~\ref{fig:magRmsJfin} \& ~\ref{fig:magRmsKfin} shows the magnitude-RMS
plots for the whole of the UKIDSS-DXS Data Release 5 recalibrated intermediate data.  Objects
in the overlap regions have also been removed, apart from the 3 objects with
interesting lightcurves shown in Figs~\ref{fig:lc1} --- \ref{fig:lc3}. These
three objects are in overlap regions, but the offsets across overlaps are
minimal.

Fig~\ref{fig:magRmsJfin} shows a very noticeable increase in noise at the bright
end ($J\leq14$ mag), from the locus of the stellar population. There is not
such a strong increase in the K-band. This noise has not been 
adequately modelled by the Strateva function and so the noise that goes into
the variability calculations is under-estimated for $J<14$ mag, leading to
excessive classifications of variable stars. This additional noise may be caused
by a non-linearity or a saturation effect.

Table~\ref{tab:brightDXS} lists the brightest 
variables in the DXS which are not in these overlap regions and which have
$J\geq14$ mag to avoid the effects of an incomplete noise model. We found
15 sources that matched our new criteria: classified as variable in both filters
and having at least 10 good detections in each filter, out of a population of
11,957 sources that matched all the criteria apart from the variability 
criteria.
These 15 objects are away from overlap regions and have magnitudes where the
noise model is well fit and are the very best candidates for real variables. 

These variables are plotted in
Figs~\ref{fig:magRmsJfin} \& ~\ref{fig:magRmsKfin}. Unfortunately most of the lightcurves are difficult to classify with only 15-20 points in 
each filter. To find and measure periodic variables, or eclipsing binaries, 
many more points would be needed. Some objects like supernovae can be usefully 
studied with this amount of data and these observations are very good for 
improving the calibration of the data. The objects in Table~\ref{tab:brightDXS}
can be followed up with more observations to properly determine their 
characteristics.

\begin{table*}
\begin{tabular}{lcccccc}
\hline
IAUName & \multicolumn{2}{|c|}{Mag} & \multicolumn{2}{|c|}{RMS} &
\multicolumn{2}{|c|}{Skew} \\ 
(UDXSJ) & J & K & J & K & J & K \\ 
\hline
{\it 160509.10+542920.4} & 15.1 & 14.1 & .024 & .029 & 1.97 & 1.31 \\
221809.44+002730.9 & 15.1 & 14.3 & .048 & .066 & +2.59 & +0.94 \\ 
222052.62-000205.9 & 15.1 & 14.4 & .080 & .111 & +0.55 & +0.87 \\ 
221706.04+002646.6 & 15.4 & 14.6 & .140 & .140 & +1.83 & +1.67 \\ 
161206.50+543814.1 & 15.4 & 14.7 & .036 & .024 & +1.62 & +1.05 \\ 
161244.51+541552.6 & 15.5 & 14.7 & .028 & .029 & +0.05 & +0.17 \\ 
222139.61+004959.7 & 15.3 & 15.0 & .170 & .100 & -0.01 & +0.17 \\ 
222049.30+003544.1 & 16.8 & 16.0 & .074 & .161 & +2.35 & +1.73 \\
{\it 222106.51+004846.7} & 17.3 & 16.0 & .036 & .046 & +2.31 & -0.20 \\
{\it 160519.04+542059.9} & 17.1 & 16.1 & .049 & .058 & +1.93 & +0.33 \\
222215.25+010049.5 & 16.6 & 16.2 & .084 & .053 & +0.75 & +1.11 \\
221953.96+001007.5 & 17.0 & 16.2 & .057 & .070 & +3.29 & +2.39 \\
221714.27+003346.5 & 17.1 & 16.2 & .150 & .136 & +1.41 & +1.36 \\
{\it 160504.88+543602.0} & 17.5 & 16.5 & .054 & .052 & +1.52 & +1.53 \\
222117.36+010517.2 & 16.8 & 16.5 & .161 & .093 & +0.09 & +0.49 \\
\hline
\end{tabular}
\caption{Table of reliable variables, at least three magnitudes brighter
than the per observation magnitude limit, in the DXS which are classified 
as variables in both filters and have more than 10 observations in each band and
are not observed across an overlap. We have also removed any with $J\leq14$ mag, 
since the noise properties of these are not well fit by the Strateva function
at the bright end. These are mostly stars, but the entries marked in
italics are classified as galaxies.}
\label{tab:brightDXS} 
\end{table*}

Table~\ref{tab:variables} is a table of the number of bright objects in each
filter, as a function of object type (star, galaxy, noise, probable star) and variability (variable V,
or non-variable NV), for objects outside the overlap regions. The proportion of
stars that are classified as likely variables (classified using observations in
that filter only: {\bf jvarClass} or {\bf kvarClass}) is $0.45\pm0.05\%$ in the
$J$-band ($14.0<J\leq18.2$ mag) and $0.55\pm0.05\%$ in the $K$-band
($11.5<K\leq16.7$ mag). The proportion of galaxies classified as variable is
$1.0\pm0.1\%$ in the $J$-band and $1.5\pm0.1\%$ in the $K$-band. While these
limits are three magnitudes brighter than the limiting magnitude, the noise 
has already started increasing at $J\sim16.5$ mag and $K\sim15$ mag, so the
lowest amplitude variables cannot be found.  If we do limit the magnitudes to 
these brighter levels, we find $0.6\pm0.1\%$  of stars are variable with 
$\Delta\,J\geq0.015$ mag ($3\sigma$) and $1.7\pm0.6\%$ of galaxies are
variable.  We find similar values in the $K$-band ($0.6\pm0.1\%$ of stars and
$2.3\pm0.6\%$ of galaxies are variable with $\Delta\,K\geq0.015$ mag). This 
estimate for the fraction of stellar variables is an underestimate, since
variables with a much longer period than the total interval between
observations will be excluded,  and so will objects, like some eclipsing binaries, which have very 
little variation most of  the time, but occasionally dip in brightness. If there
are not enough observations to get several eclipses then these objects will 
also be excluded, as will objects that only vary sporadically. For galaxies,
the  noise model is not quite right, because all the photometry has been 
corrected for light loss outside the aperture, assuming that the objects are 
point spread functions. For stars and distant galaxies, this is the correct 
approach, but some nearby galaxies will not be corrected properly and the 
differences between the correction used and the true correction is an
additional  source of noise. This noise is not taken into account, and so the 
number of variable galaxies may be over-estimated.
 
\begin{table}
\begin{tabular}{lcccc}
\hline
Filter & Object & Var Type & Var Type & Number \\
 & Type & (filter) & (overall) &  \\
\hline
J & Star & NV & NV & 16252\\
J & Star & NV & V & 12\\
J & Star & V & NV & 47\\
J & Star & V & V & 27 \\ 
J & pStar & NV & NV & 275 \\
J & pStar & NV & V & 4 \\
J & pStar & V & NV & 3 \\
J & pStar & V & V & 2 \\
J & Galaxy & NV & NV & 5577 \\
J & Galaxy & NV & V & 14 \\
J & Galaxy & V & NV & 50 \\
J & Galaxy & V & V & 10\\ 
J & Noise & NV & NV & 14 \\ \hline
K & Star & NV & NV & 24164 \\
K & Star & NV & V & 8 \\
K & Star & V & NV & 24 \\
K & Star & V & V & 110 \\
K & pStar & NV & NV & 192\\
K & pStar & V & NV & 3\\
K & pStar & V & V & 5\\
K & Galaxy & NV & NV & 12851 \\
K & Galaxy & NV & V & 2 \\
K & Galaxy & V & NV & 59\\
K & Galaxy & V & V & 137\\
K & Noise & NV & NV & 38 \\
K & Noise & V & V & 1 \\ \hline
\hline
\end{tabular}
\caption{Table of the classification of bright sources, at least 3 magnitudes
brighter than the per observation magnitude limit, in the DXS after recalibration and
fainter than the failure in the noise model.
The sources must have at least 5 good observations to be included (since
variables are only counted for objects with at least 5 good observations) and
${\bf priOrSec=0}$ to avoid bias from overlaps. }
\label{tab:variables} 
\end{table}

\begin{figure}
\includegraphics[width=84mm,angle=0]{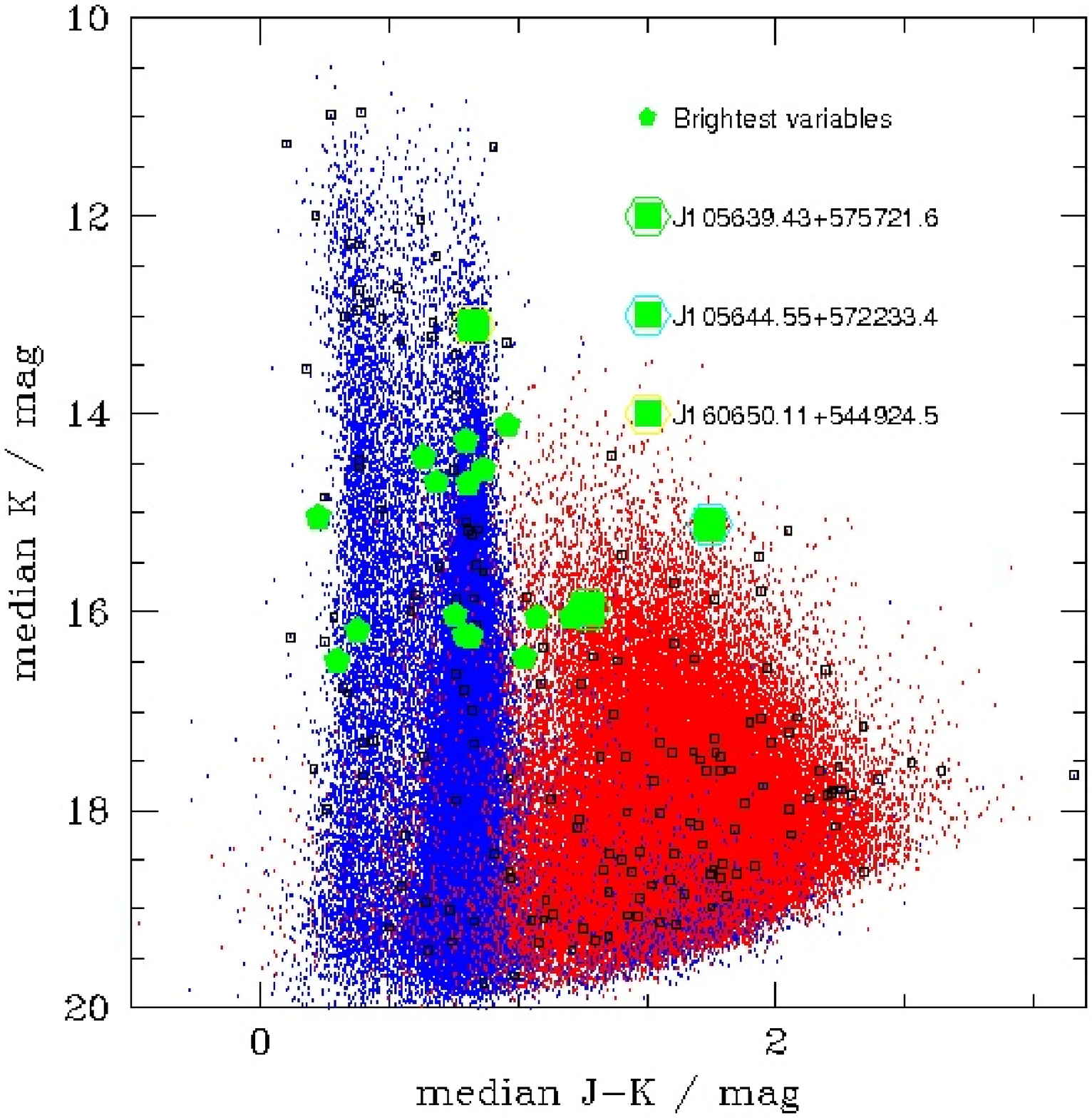}
\caption{Colour-magnitude plot for UKIDSS-DXS data. The blue points are
classified as stars and the red points are galaxies. The black squares are all
variables. The good bright variables are plotted as green pentagons and the
three interesting objects are marked by larger squares surrounded by a with different colour circle.}
\label{fig:colourMagDxs} 
\end{figure}

Next we look at the distribution of the variable stars
versus non-variables. If we plot the ($J-K$) vs $K$ colour magnitude plot
(Fig~\ref{fig:colourMagDxs}), we find three main groups of objects: 1) galaxies,
with $(J-K)>1$; 2) stars with $0.6<(J-K)\leq1$ and 3) stars with
$(J-K)\leq0.6$. We find that there are bright variables in each of these groups.

\begin{figure}
\includegraphics[width=84mm,angle=0]{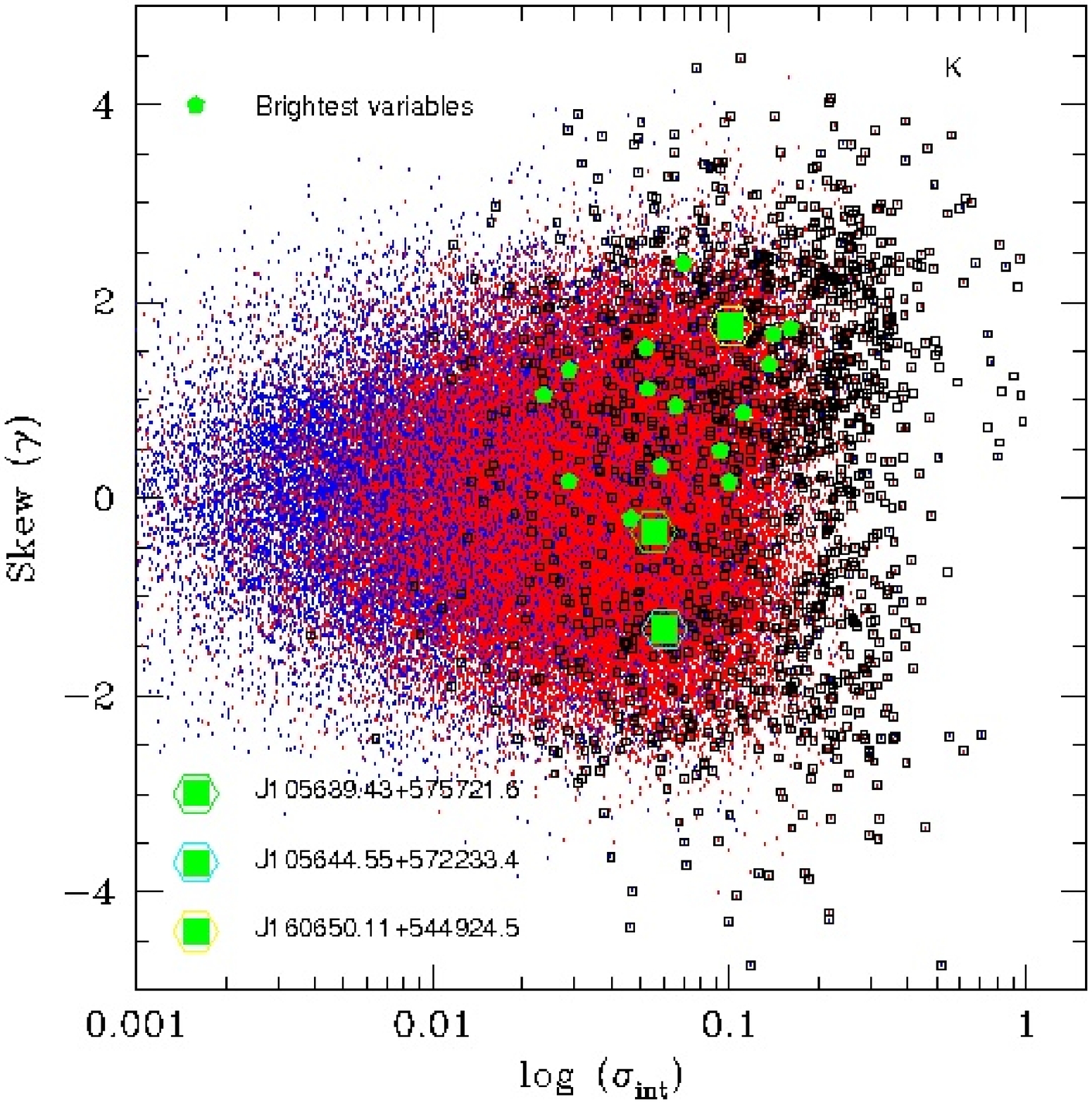}
\caption{K-band intrinsic RMS vs skew for the UKIDSS-DXS. The blue points are
classified as stars and the red points are galaxies. The black squares are all
variables. The 20 good bright
variables are plotted as green pentagons and the three interesting objects are
marked by larger squares with a colour circle around.}
\label{fig:RMSSkewK}
\end{figure}

In Fig~\ref{fig:RMSSkewK}, we look at the distribution of variables in the
intrinsic RMS versus skewness plane. Here we can see a definite bias towards
positive skew for the brightest variables, and for variables in general,
although the overall population of objects is quite symmetrical around a skew
of zero.

\section{Standard star data}
\label{sec:CAL}

While we have not yet released the WFCAM standard star data using this new
archive model, we have produced some test data with our pipeline. We did the 
tests on one standard
star field, the Serpens Cloud Core, chosen for the large number of individual  observations
$\sim100$, the high density of stars, and because it includes three standard
stars in the field, close to the cloud core. The standard star fields are 43 non-overlapping fields, although in
some cases two pointings have been done around the same field to put the known standard onto different
detectors. Since the Serpens Cloud Core is in a dense region of sky, liable to
be confusion limited, we have only used seven epoch frames in each deep
stack.

\begin{figure}
\includegraphics[width=72mm,height=70mm,angle=0]{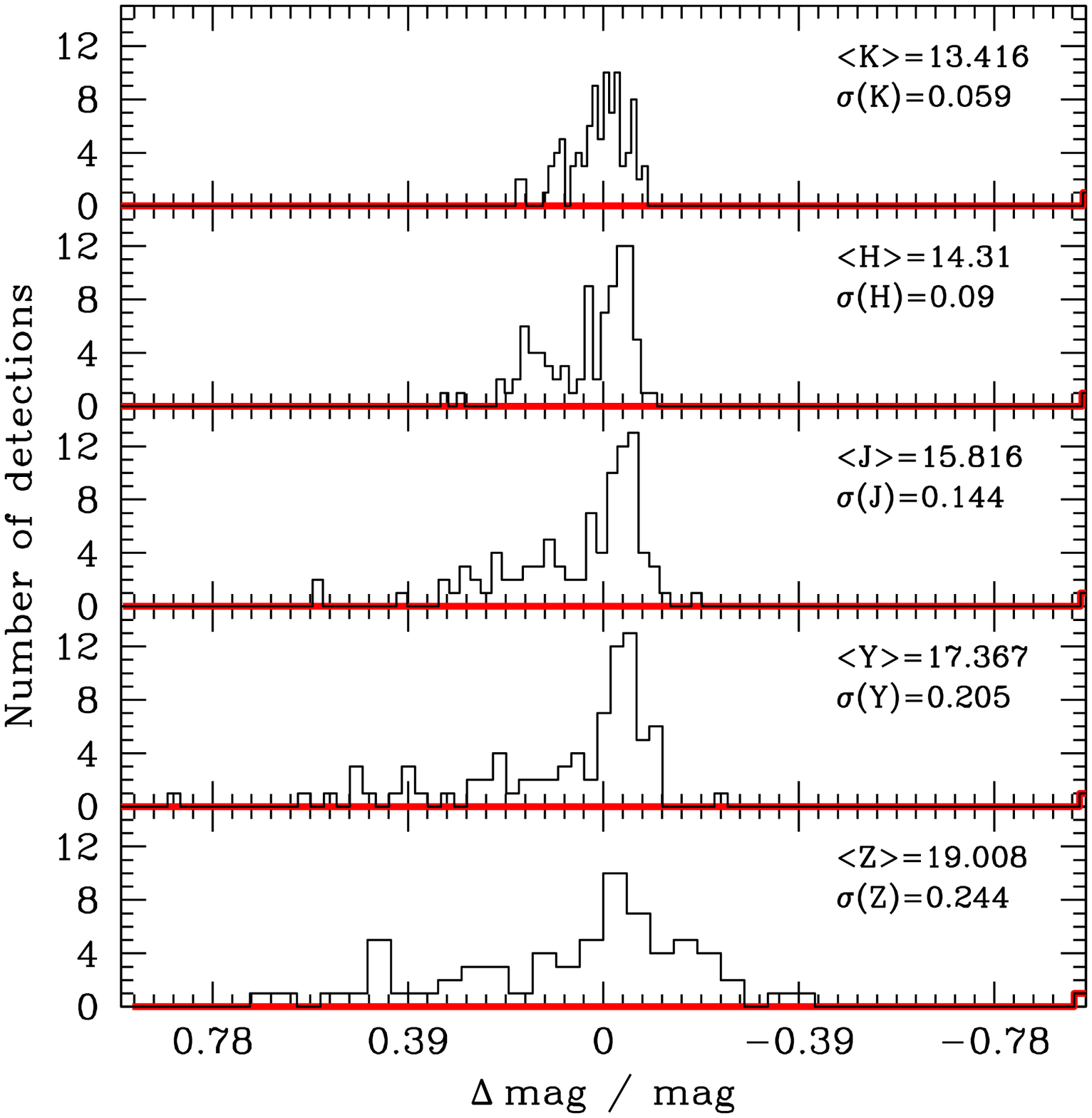}
\includegraphics[width=72mm,height=70mm,angle=0]{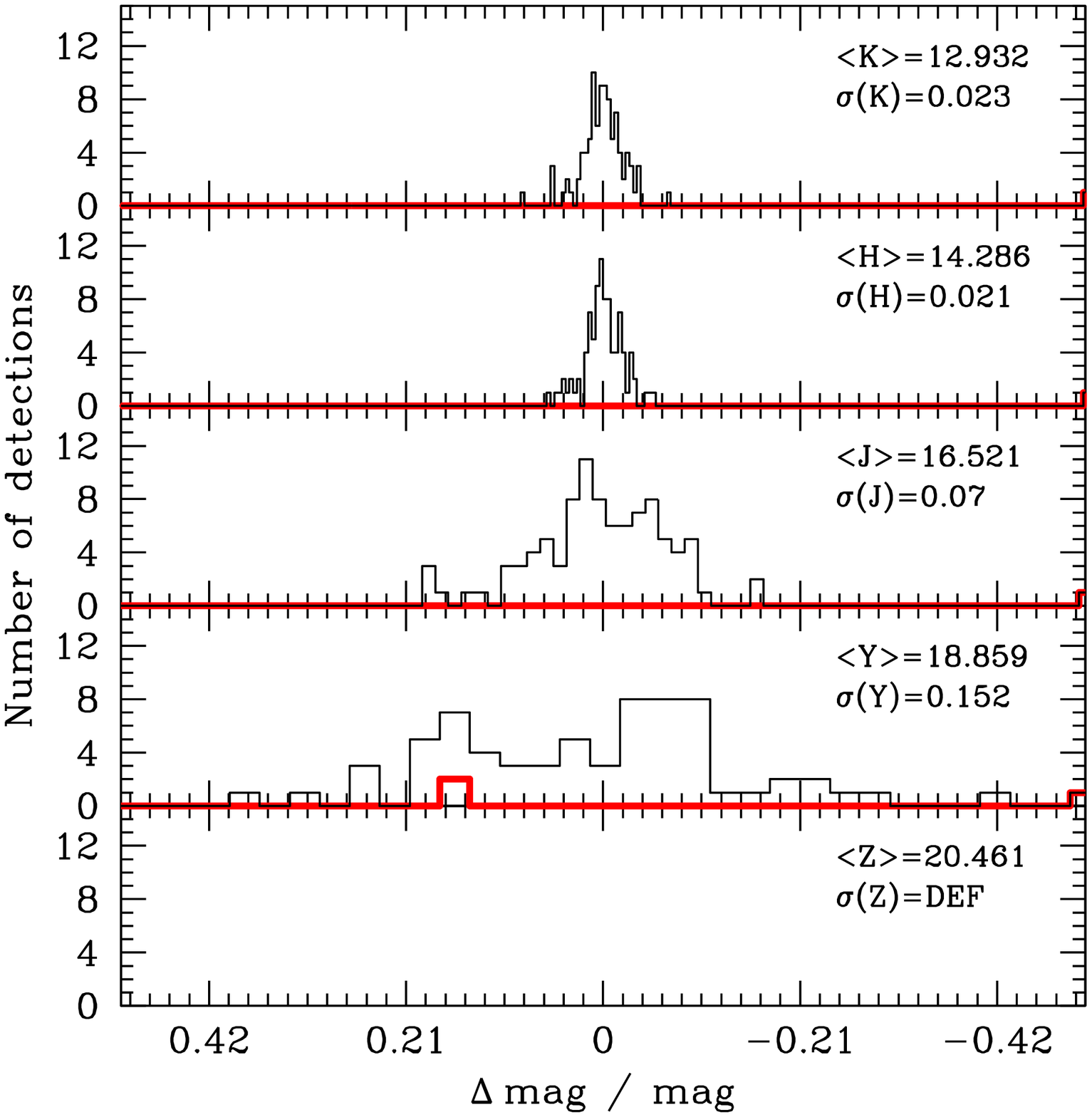}
\includegraphics[width=72mm,height=70mm,angle=0]{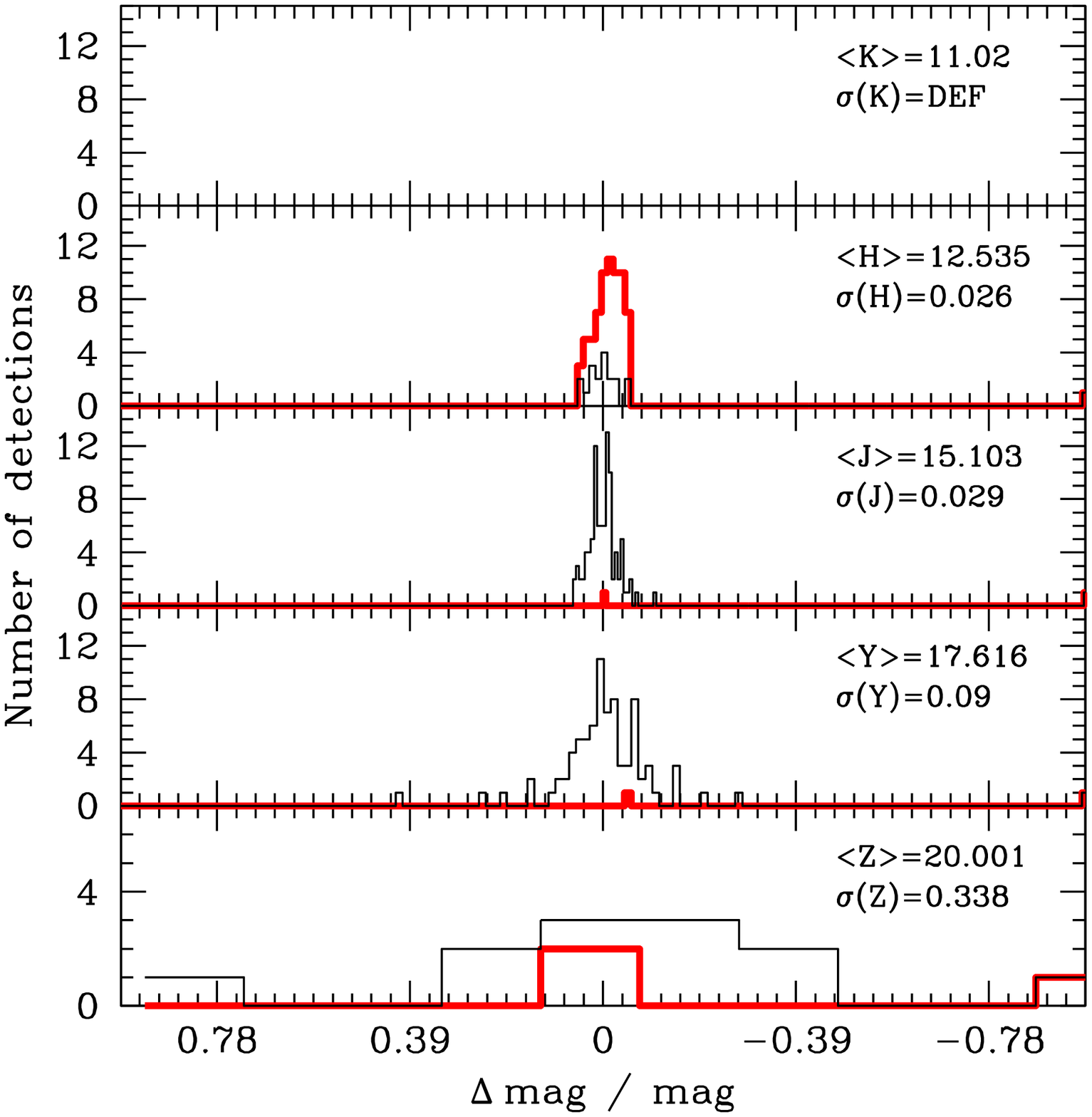}
\caption{Histogram of difference in magnitude from the median for standard
stars in the Serpens Cloud Core. Ser-EC51 (top), Ser-EC68 (middle) and Ser-EC84 bottom. 
The thin solid black histograms show the good observations and the thick solid
red histograms show 
observations that we have flagged as having photometric problems. 
These stars are all very red, so there is very little Z-band 
flux and the K-band is saturated in the case of Ser-EC84.}
\label{fig:lcS1}
\end{figure}

Since the observations are correlated, the same times are sampled in each
light curve, which makes it much easier to distinguish which features are real
variations. Fig~\ref{fig:lcS1} shows the histogram of the deviations in
magnitude from the mean for three UKIRT faint standard stars Ser-EC51,
Ser-EC68 and Ser-EC84 \citep{Hwd01}. These are all in the dense nebulosity of
the centre of the cloud core. Ser-EC68 and Ser-EC84 show very little variation 
although EC84 is saturated in $H$, too bright for good detections in $K$ and too 
faint in $Z$. EC68 is also too faint in $Z$. The extinction in the cloud core
means that very little radiation shorter in wavelength than $1\mu$m is
visible. Ser-EC51 shows some large deviations from the median, particularly to 
fainter magnitudes. The light curve for this object shows some coherent 
variations 400 days after the first observation and 820 days after the first 
observation. Ser-EC51 should not be considered as a useful standard.

\clearpage

\begin{figure}
\includegraphics[width=84mm,height=100mm,angle=0]{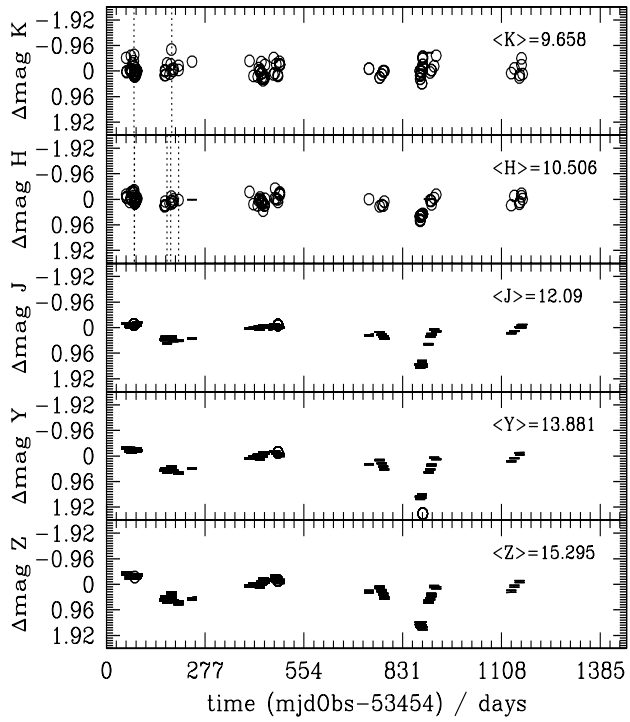}
\includegraphics[width=84mm,height=100mm,angle=0]{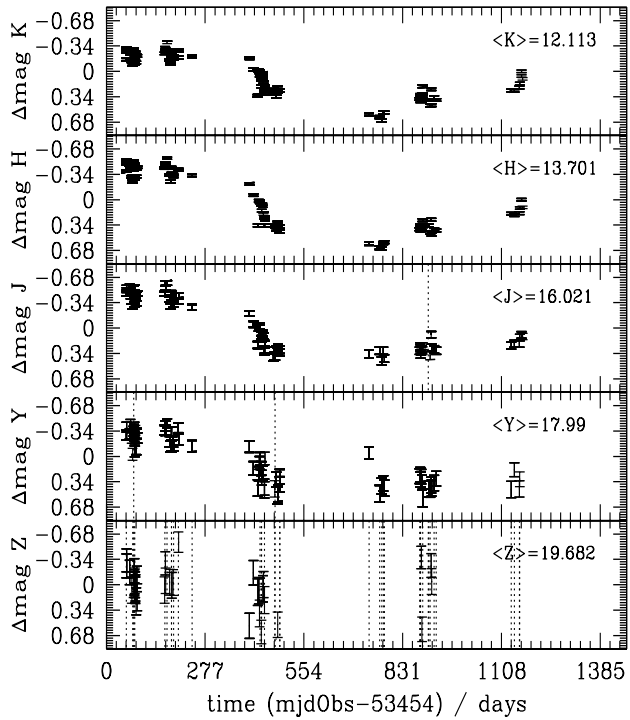}
\caption{Light curve for two variable stars in the Serpens Cloud Core field.
UCAL J182931.98+011842.5 (top) and UCAL J182955.19+011322.0 (bottom). Points
with errorbars indicate good detections, circles indicate flagged detections
and a dotted vertical line indicates a missing observation. The
variations are highly correlated between each band.}
\label{fig:lcCALv}
\end{figure}

Fig~\ref{fig:lcCALv} shows the light curves of two other variable stars, 
chosen because of their interesting features. The top one shows a star that 
undulates slowly over a few hundred days, by 0.4 mag, before more rapidly dimming, by 1.3 mag, and rapidly brightening again. This may be an eclipsing binary. The lower object shows a longer term
variation, with a rapid fading after 400 days, followed by a slow brightening. These were selected partly by using the Welch-Stetson statistics.

\begin{figure} 
\includegraphics[width=84mm,angle=0]{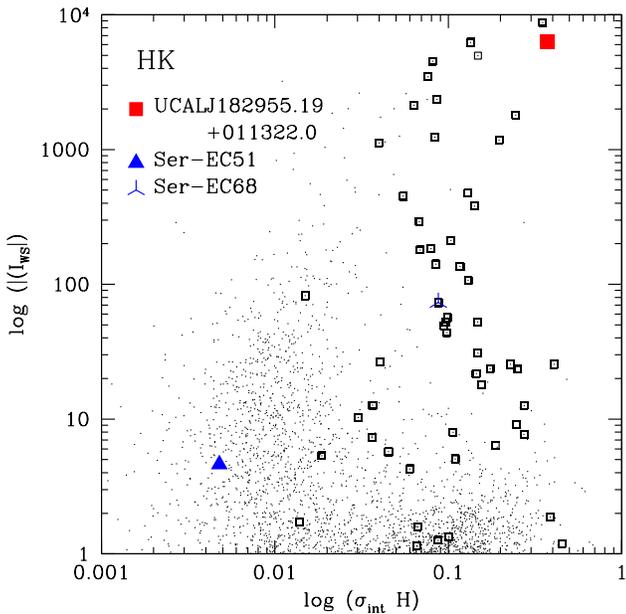}
\caption{Plot of $H$-band intrinsic RMS vs absolute value of Welch-Stetson
$|I_{WS,H,K}|$ for objects in the WFCAM Standard Star programme.  Objects
classified as variable have been marked by large squares. Three of the five
objects with light curves are shown too. The other two, Ser-EC84 and
UCALJ182931.98+011842.5 are saturated in $K$ and/or $H$ and so have a default
value for $I_{WS,H,K}$. Ser-EC68 has low values of both parameters and is
clearly non-variable, UCALJ182955.19+011322.0 has very large values of both and
is clearly variable. Ser-EC51, is in between the two, but does show some signs
of variation in the light curve.}
\label{fig:rmsWSHK}
\end{figure}

The correlated band data is very useful as most real fluctuations are
correlated (or anti-correlated) across many filters whereas most noise does not have a
filter dependent correlation. Having so many more observations per source than
the DXS also makes it easier to separate truly variable objects from ones with
a few spurious measurements. Fig~\ref{fig:rmsWSHK} shows that variable objects
tend to have large absolute values of the Welch-Stetson statistic, as well as large values of the intrinsic
RMS. 
   
\clearpage
   
\section{VISTA-VIRCAM}
\label{sec:VSA}
We will apply the same data model to the VISTA Science Archive data. There are
some features of VIRCAM which give additional problems. 

\begin{itemize}
  \item VIRCAM will usually have tiled images, with tiles made up
  of 6 observations, with the observations in the x-direction separated by
  $90\%$ of a detector width and those in the y-direction separated by $45\%$
  of a detector width. These tiles are much larger than WFCAM detectors and are
  likely to have more calibration issues with half detector overlaps, and 
  large distortion effects at the edges.
  \item The top and bottom of the tiles have a half detector width strip which
  only gets observed once, so cosmic ray and artifact removal is not possible.
  Some PIs may try to improve observing efficiency by stitching together
  these overlaps. Catalogued objects extracted from these stitched together
  overlaps will not have a single observation time, so they could not be
  used in any variability analysis and would have the additional problems of
  different noise and PSF and distortions in the original images.
  \item The focal plane can be independently rotated, so it is possible to have
  multiple rotations at the same pointing. 
\end{itemize}

Much of the synoptic pipeline design has these issues in mind, but further work and tests with
VIRCAM data will be necessary to fully solve them. As a result of the work on
the synoptic pipeline that is presented in this paper, we have done away with
the fixed number of filter passes used in shallow UKIDSS surveys and made all
multi-epoch data sets synoptic (e.g. VISTA-VIKING). This gives the advantages
of deep stacks, and internal recalibration, which surveys such as the
UKIDSS-LAS will not enjoy. Additionally, this extra flexibility means that the
schema does not have to be changed if an extra epoch is added in later.

\subsection{Large Data Volumes}

Frame sets from WFCAM are typically $0.05\deg^2$ 
(the area of one detector), but VISTA frame sets will be 
$\sim1.5\deg^2$ (the area of a VISTA tile). Thus the number of objects per frame set
will increase from a few thousand to more than 50,000 in a typical pointing and 
$\sim1$million in a dense region of the Galactic plane. The UKIDSS UDS contains
a single frame set of area $\sim1\deg^2$ with 100,000 objects and has several hundred
individual pointings. This has been successfully processed, so typical VISTA frame sets
should pose few additional problems. Eventually, we will have to process the
whole of the VVV: $\sim10^9$ sources, with $\sim100$ observations, producing a
best match table with $\sim10^{11}$ rows. The full processing of this must be
done in $\leq1$ month, if it is not going to significantly interfere with other
archive processing and if we are going to be able to rerun the task. Our current
processing speed is $\sim2$hrs for the UDS field (using an older server). The VVV data set will be $\sim3000\times$ larger. This problem is easy to parallelise, and factoring in Moore's law (the main variability part of the VVV survey will not
take place until year 3 of the surveys: 2012), a factor of $20\times$ the speed
can be found without any optimisation. This gives a total time of $\sim300$
hrs or $\sim2$ weeks on four or more machines. Optimising the code so that it
can run two or three times faster would allow the processing to be done on one
or two machines over a sensible time scale. 

\section{Comparison with other public databases with multi-epoch observations}
\label{sec:otherPDB}

\subsection{SDSS Stripe 82 database}
\label{sec:SDSS}
The SDSS Stripe 82 is a 300 sq. deg. strip which has been observed 80 times in
$u$, $g$, $r$, $i$, $z$ filters. The observations are taken in a drift scan
mode with objects observed through each filter consecutively. The filter
observations are therefore correlated within our criterion.  

The Stripe 82 data has its own database (http://cas.sdss.org/stripe82/en),
separate from other surveys. The database includes
notes\footnote{http://www.sdss.org/dr7/coverage/sndr7.html} about how to search
for all the detections of different objects. This uses the hierarchical
triangular mesh identifier to search by position. This has the same drawbacks as
the neighbour table approach in Paper 1:

\begin{itemize}
  \item It is not possible to know whether there are missing observations, which
  are important in a lot of transient searches.
  \item In dense regions you will be contaminated by neighbouring objects,
  which having different magnitudes will make the objects appear variable.
\end{itemize}
 
However, there are also a couple of useful variability tables: \verb+Stetson+,
which is like a simplified \verb+Variability+, containing a few photometric 
statistical values in each band and a continuous classification, and 
\verb+ProperMotions+ which contains the astrometric fit comparing the SDSS and
USNO-B catalogues. These do not include any noise model as yet, although
\cite{Ssr07} have used one on the Stripe 82 data. \cite{Ssr07} \& \cite{Brm08}
calculate many more statistics than are currently accessible through the main
archive database. 

\cite{Brm08} describe two useful tables, a Higher Level Catalogue (HLC) and a
Light Motion Curve Catalogue (LMCC) which are similar to our \verb+Variability+
and \verb+SourceXSynopticSourceBestMatch+, but these are only available as 
downloadable files which can be processed by an {\tiny IDL} programmes. It is
not possible to search through them using the SDSS query tools and then to match up with external catalogues.

While much work has been done on measuring variability in the SDSS Stripe 82
data, this work is in several separate publications and very little of this is
currently available in the main SQL query tool, so it is not easy for users to
search on variability statistics. In contrast we have designed our multi-epoch
pipeline and archive together, so users can access all our parameters and do more
detailed searches on a wide range of different types of variables.

\subsection{NSVS public database}
\label{sec:NSVS}

The NSVS public database\footnote{http://skydot.lanl.gov} \citep{Wznk04}
contains six tables. These are \verb+Field+, \verb+Frame+, \verb+Object+, 
\verb+Synonym+, \verb+Observation+ and \verb+Orphan+. The NSVS
observations  were all taken in a single, wide optical filter, which is closest
to the Johnson R filter of all the standard filters. The main variability table, \verb+Object+, contains $\sim2\times10^7$ sources and is 
similar in scope to our \verb+Variability+ table. It contains an ID, the median
and standard deviation of the right ascension and declination, the median of 
the magnitude and median and standard deviation of the differences in
magnitude of the from ``good'' points, the number of points, number of good
points and number of points with a certain flag type as well as the flags 
associated with the object. The \verb+Observation+ table is similar to a 
combination of our best match and \verb+Detection+ tables, listing all the 
individual observations linked to each object. It includes the position, 
magnitude, magnitude error and flags only. The \verb+Synonym+ table is 
equivalent to our \verb+SourceNeighbours+ table linking identical objects to 
each other. The \verb+Frame+ table is similar to our \verb+Multiframe+ table, 
describing each observation and the \verb+Field+ table is similar to 
\verb+RequiredStack+ describing the pointing information. The \verb+Orphan+ 
table is very interesting: it contains bright objects that aren't linked to any
object. These could be fast moving objects (solar-system objects) that have 
moved too much between observations to be linked to each other or objects that 
are very faint but flare up, only to be seen on a few frames.

The NSVS public database is similar to ours and light-curves can be easily
selected from it, like the WSA, but it lacks any modelling of the noise or 
any attempt at classifying variables so it is much more difficult to reduce 
the number of objects that you are searching through to a manageable number.
This is left for the user to do. However, searching thousands of light-curves
is no simple task and many users will be repeating the same type of 
analysis: checking the noise properties and eliminating objects that are 
too close to the noise limit, so it would be good to have these useful 
quantities in the database to search on.

A follow up paper
\citep{NSVS2}, describes how to use the database to select candidate red
variables, using light curve information from \verb+Observation+ and aggregate
data from \verb+Object+.

\section{Summary and Future Work}
\label{sec:summary}

We have designed, implemented and released a dynamic archive for analysing time
series data containing both astrometric and photometric variations which works
on a wide variety of data sets observed by either the UKIRT-WFCAM or VISTA-VIRCAM instruments.
The design of the archive can be used (with small modifications) on any
astronomical data from pointed observations. It is designed so that
the data can continue to be updated and improved, and additional refinements 
included without repeating every step again. We have included measurements of
the noise properties that are essential for determining whether the object is
variable and properties of the observation sequence, such as the
typical observation interval.

This paper describes the design of the synoptic pipeline in the WFCAM
Science Archive and is useful for the design of future archives for synoptic
surveys such as Pan-STARRS and LSST. This paper is also a useful handbook
for users of the WSA and VSA, so that they can use the facilities to do useful
science with variable objects.

The archive of synoptic data is also useful for global calibration of UKIDSS and
VISTA data, since the data has now been more efficiently matched and
non-variables can be selected. Some spatial variations within the detectors have
been seen in synoptic data. New standard stars can be selected from the data and previous standards tested more
thoroughly. 

We have made it possible, using these archive tools, to rapidly reduce a
massive database of millions or billions of sources into a few tens of
potentially interesting variable objects. The synoptic pipeline is still  being
developed and we expect to add in many of the following features in the future: 

\begin{itemize}
  \item Improved calibration
   \subitem --- a spatially dependent relative photometric and astrometric
   recalibration. This will be more sophisticated than the simple zeropoint
   shift at present.
   \subitem --- calibrate across the overlaps to correct for the discontinuity 
   in calibration across overlaps. 
  \item Improved noise model
   \subitem --- an additional component that takes into account the increase in
   astrometric noise at the bright end.
   \subitem --- fitting astrometric errors as a function of magnitude.
   \subitem --- moving from an empirical fit to a model dependent on filter,
   exposure time and sky brightness, so that sets of observations with 
   different exposure times can be weighted correctly. This will involve a lot 
   of careful analysis of the noise in different frames, across many programmes
   first.
   \subitem --- estimation of effects of red (correlated) noise. 
  \item Moving objects
   \subitem --- fitting the astrometric data for proper motions (and parallax).
   \subitem --- a table of fast moving objects: i.e. objects that are too far
   apart to be stacked together in deep stacks, but are good detections at individual epochs. 
  \item List-driven photometry. This gives un-biased statistics for faint
  objects and could improve the RMS for non-moving sources by removing
  errors caused by inconsistent centring of the object.
  \item Difference imaging. This may be necessary to find variables in very
  crowded regions.
  \item More sophisticated cadence statistics. The median cadence gives a
  useful idea of the typical interval between observations, but the mode (or
  modes in a multi-peaked distribution) would be more useful. The minimum and
  maximum give the range of time between observations, but these can be skewed
  by an unusual observation, so taking the 10th percentile and 90th percentile of 
  time between observations would give a more robust measure of the range of
  observations.
  \item Additional photometric statistics, such as the fraction of points
  more than 3 standard-deviations from the mean magnitude, the kurtosis of the
  $\Delta$mag distribution and statistics on the star-galaxy separation. The 
  first two of these constrain the properties of the light-curve and the last
  gives better morphological information.
  \item Fourier analysis to get periodicities and the light curve shape. This
  could be part of our data analysis services, so users request particular objects, rather than a
  wholesale processing of all data.
  \item More sophisticated variable classification and classification of
  different types of variable. This will use more of the photometric
  statistics such as the skewness, the Welch-Stetson statistic as well as
  astrometric information to produce a wider range of classes that will
  increasingly help users to isolate a group of objects that interest them,
  such as eclipsing binaries, periodic variables, high proper motion stars,
  cataclysmic variables.
\end{itemize} 

We have started investigating and planning some of the improvements, and will
continue to do so as WFCAM and VISTA continue operations. All changes will be
documented on the archive websites. 

\subsection{Moving Objects}

\begin{figure}
\includegraphics[width=84mm,angle=0]{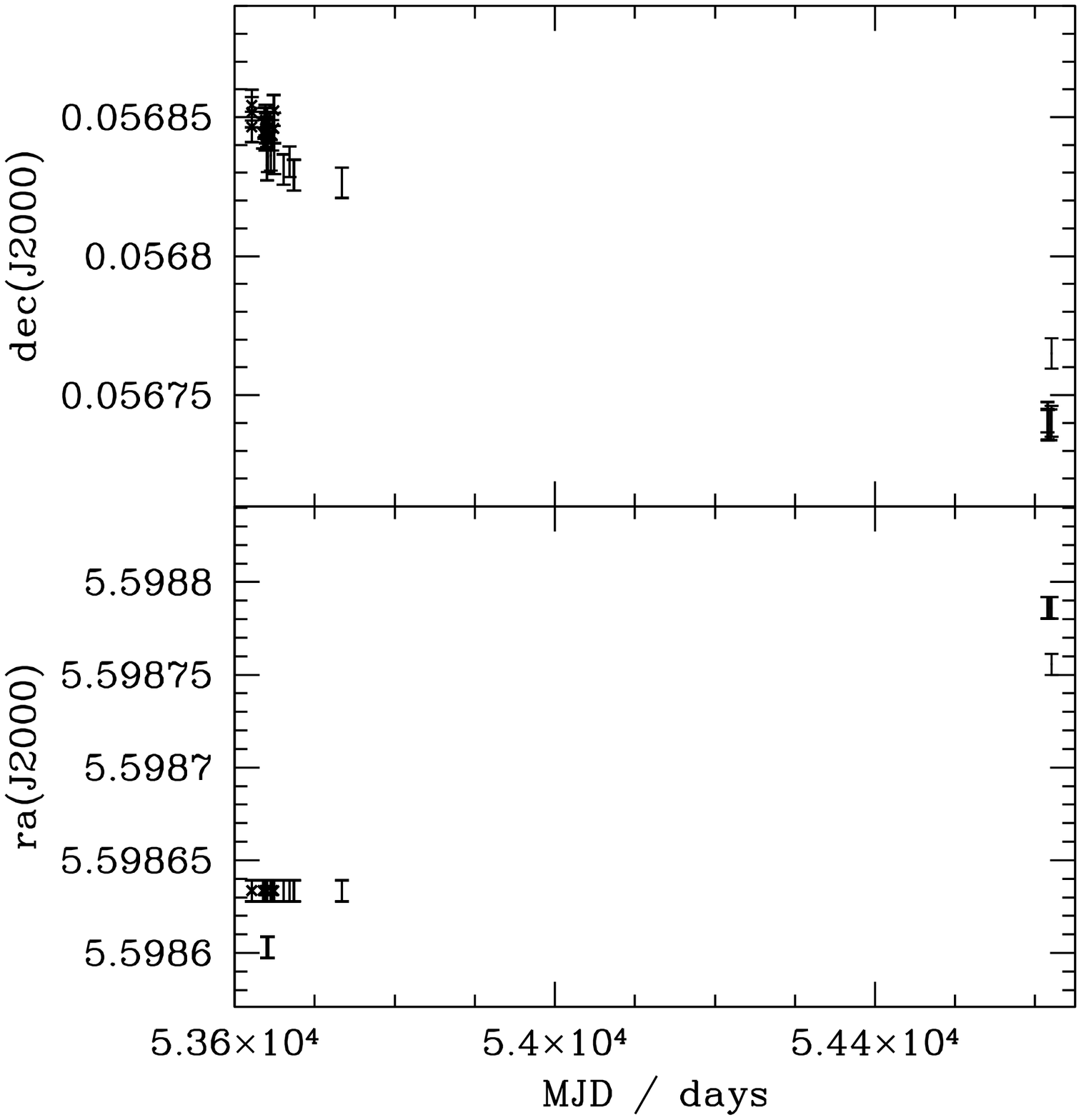}
\includegraphics[width=84mm,angle=0]{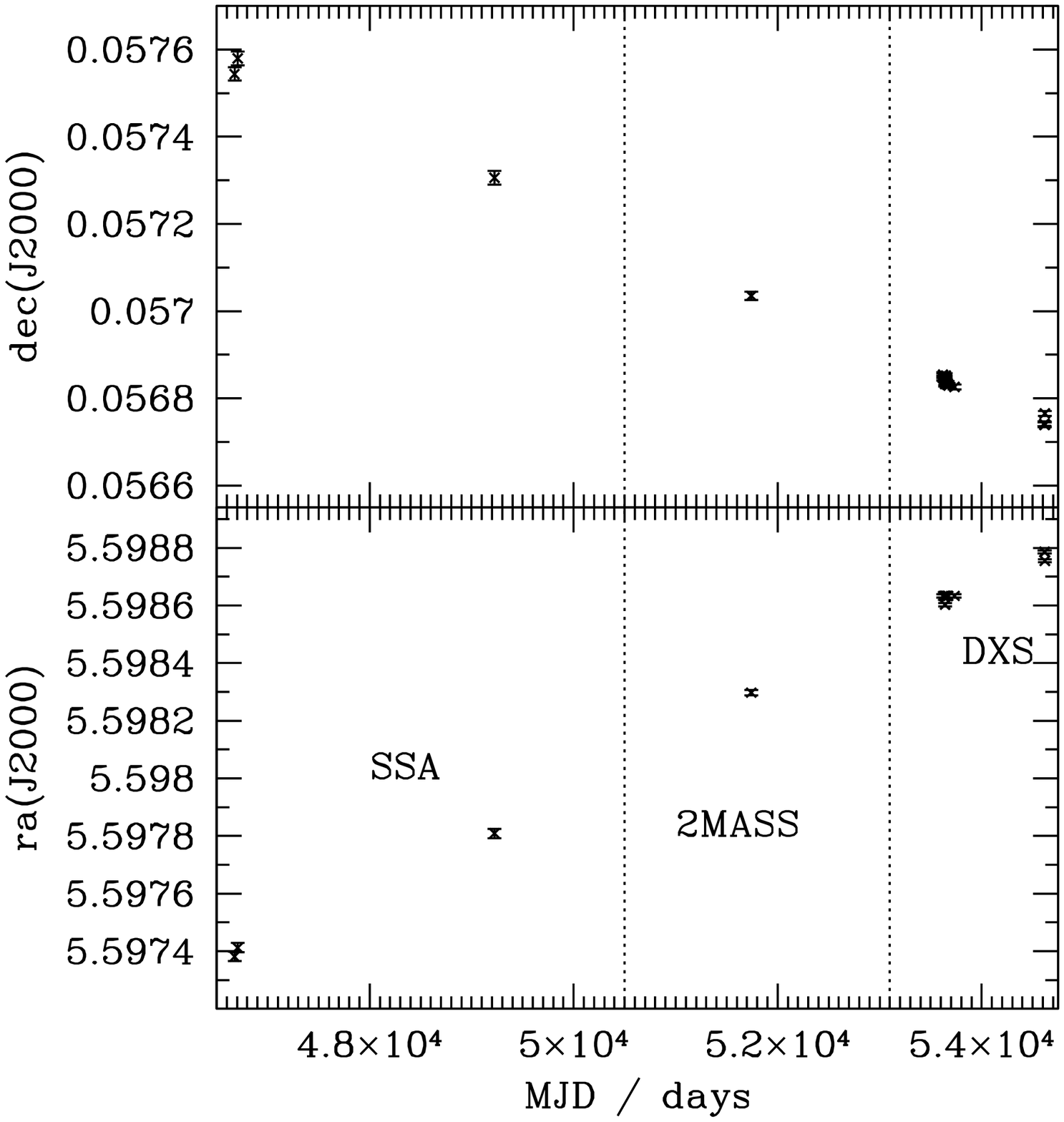}
\caption{Time varying position of high proper-motion star UDXS
J222223.70+000324.3. The top plot (a) shows the UKIDSS-DXS observations, which
are clustered around two epochs. The dots with errorbars are the J-band
observations and the crosses with errorbars are the K-band ones. The lower plot
(b) shows the same star, but older observations from 2MASS and the
Super-Cosmos Science Archive (SSA) have been added in. It has a very clear
constant proper motion. In both plots RA=RA$-330$ deg for presentation purposes.}
\label{fig:pm}
\end{figure}

One of the future improvements is to add a proper-motion calculation into the
astrometric statistics. The astrometric part of the
pipeline has not progressed far enough for it to be easy to find moving
objects, such as the high proper-motion star UDXS J222223.70+000324.3, see
Fig~\ref{fig:pm}. This was found through a comparison with 2MASS instead. This
object has $\mu_{\alpha}\sim0.23\arcsec$year$^{-1}$ and 
$\mu_{\delta}\sim0.13\arcsec$year$^{-1}$. The DXS observations fall into two
batches  (see Fig~\ref{fig:pm}a), and only the later smaller batch are linked to
the source through the best-match table. The other detections are too far apart
and so this source has many missing observations even though it is bright 
($J=12.6$ mag, $K=11.8$ mag). A faster moving object, and one where the 
observations are more evenly sampled may be missed in the deep images
altogether,  because the stacking algorithm uses median clipping to remove
false  detections.

The object can be tracked back through the neighbour table to 2MASS, and
through the Supercosmos Science Archive \citep{Hmb04} to plates taken over 20
years ago, see Fig~\ref{fig:pm}b. 

This is a particularly high proper motion star, and most of those that users are
interested in will be moving much more slowly. However, it is a useful example
of the problems that are faced when automating a search for moving objects.
Because of examples like this, we increased the maximum distance in the 
\verb+SourceXDetection+ neighbour table for UKIDSS-DR5 from
$1\arcsec$ to $10\arcsec$.

In many cases, if the complete data set is taken over a very short time
--- much less than a year --- very few stars have moved enough to be
detectable. Therefore, in later variants of the pipeline, it would be sensible
to incorporate a test which calculates the minimum proper motion rate,

\begin{equation}
\mu_{\rm min}=\frac{\sigma_{\rm astrm}}{T_{\rm max}},
\label{eq:pmrate}
\end{equation}

\noindent where $T_{\rm max}$ is the maximum observation time between 
observations for any source in a dataset and $\sigma_{\rm astrm}$ is the
typical astrometric error for bright sources and $\mu_{\rm min}$ is the minimum
proper motion rate. If $\mu_{\rm min}$ is less than a rate which will find
many scientifically interesting candidates ($\mu\sim0.05\arcsec\,$year$^{-1}$),
then the proper motion will be evaluated for all objects in the dataset,
otherwise no proper motions will be calculated. The error in the DXS astrometry
is $\sigma_{\rm astrm}\sim0.015\arcsec$, see Fig~\ref{fig:magSigDecK}, and the
period of observations in any pointing is $T_{\rm max}\sim2$ years, giving a
minimum proper motion of $\mu_{\rm min}\sim75$mas year$^{-1}$.

The parallax calculation may be
invoked similarly. 

Solar system objects have much larger motions, so running
the proper motion code on shorter timescale datasets could pick out
these objects. On long timescale datasets, these will show up as detections
unmatched to any source, since the stacking clips out detections that are not
repeated at the same position in other frames that contribute to a deep stack.

\subsection{Improved test for missing objects}
\label{sec:missObjVSA}

\begin{figure}
\includegraphics[width=84mm,angle=0]{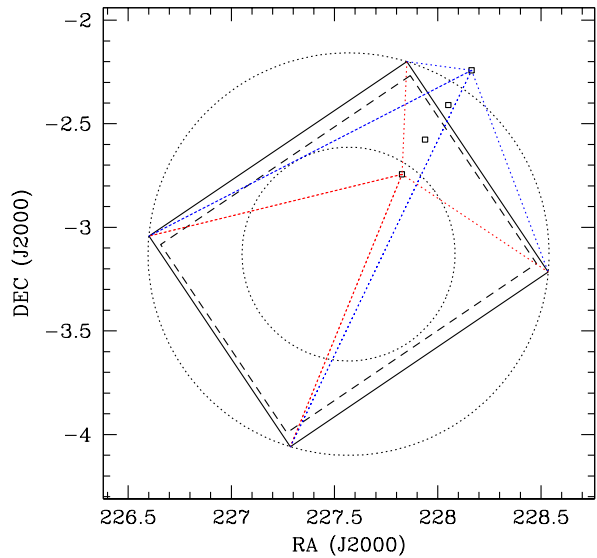}
\caption{The difficulties of finding whether an object should have been
observed on a VIRCAM tile. A tile at orientated at 34 degrees is shown as an
example. For rectangular tiles, the radius method used in
Fig~\ref{fig:checkmissWSA} is very inefficient. Calculating distances from the
four corners and the areas of the triangles formed by the object and the
corners is slower than the radius method, but much faster than calculating the x
and y position. The area of the triangles formed by the red
dotted lines and the edges of the tile is the same as the area of the tile. The
area of the triangles formed by the blue dotted lines and the edges is greater
than the area of the tile. Objects inside and outside the tile can easily be
distinguished. Only a very small number of objects very close to the edge will
need to have their x and y positions calculated.}
\label{fig:checkmissVSA}
\end{figure}

In Fig~\ref{fig:checkmissWSA} we showed the current method of testing whether a
source should have been observed. Fig~\ref{fig:checkmissVSA} shows that this
method will be very inefficient on VSA rectangular tiles: the area between the
two radii is extremely large and so a lot of sources must undergo the slow
step 2 of determining whether the object was in the pointing and should have
been observed. A more efficient routine needs to reduce the number of objects
for which step 2 is necessary for determination. The modified step 1 must still be very
efficient. Using a simple RA and DEC selection is no good, because VISTA tiles
can be rotated. Instead we use a method in which the distance from each of the
four corners is calculated, and from those calculations, the areas of the four 
triangles from each edge of the tile to the source are calculated. If the 
areas of the four triangles is equal to the area of the rectangle, then the 
source is within the tile, else it is outside. The area of the rectangle and
triangles will be somewhat distorted by the image projection and so the area
calculated for the triangles may not be equal to the area of the rectangle,
even when it is inside the rectangle. Some uncertainty may exist until the area
of the triangles is quite a bit larger than the area of the rectangle. This may
mean increasing the area that the more careful second stage is used in, but
this area should still be significantly smaller than the area between the
circles.

An alternative method is to use the half-space method \citep{fprint}. In
this method, each edge of the frame is an edge on a plane that intersects the sphere
of the sky. The plane is represent by the orthonormal vector and an offset from
the centre of the sphere. It is easy to tell which side of the plane a point is
through its dot-product. Repeating this for each edge of the frame tells you
whether the object is in or out of the frame. The plane can be calculated using
cross-products of the coordinates. These only have to be done once for each
frame, and the dot-products that need to be calculated four times for each
point are computationally very simple and quick. This also has the advantage
that it is clear-cut whether an object is either side of each line. We will 
test these two methods to see which gives the best performance.

\section{Acknowledgements} 

NJGC would like to thank Nicolas Lodieu and Richard Jameson for
their inspiration at the 2007 UKIDSS meeting in Munich. NJGC would also like to
thank Leslie Hebb for useful early discussions on analysis of variables and
Tam{\'a}s Budav{\'a}ri for more recent discussions on footprints. We would like
to thank the anonymous referee for useful suggestions and a quick report.

Financial resources for VISTA/WFCAM Science Archive development were 
provided by the UK Science and Technology Facilities Council
(STFC; formerly the Particle Physics and Astronomy Research Council)
via e--Science project funding for the VISTA Data Flow 
System and Rolling Grants. Operations resources, including provision for
hardware, are provided in the main by the STFC, but we would also like to 
acknowledge significant contributions to hardware funding from University  of
Edinburgh sources.  

\bibliographystyle{mn2e}
\bibliography{wsasynoptic} 

\appendix
  
\section{Example Queries}
\label{app:examp}

Throughout these queries \verb+xxx+ replaces the programme name: e.g.
\verb+dxs+, \verb+uds+, \verb+cal+ for the UKIDSS-DXS, UKIDSS-UDS and WFCAM
standard star observation programmes respectively. To select all reliable
variable stars which are 3 magnitudes brighter than the expected magnitude limit in the K-band.\\[5mm]
\noindent
\begin{verbatim}
SELECT v.sourceID,v.kMeanMag,v.kMagRms 
FROM   xxxSource AS s,xxxVariability AS v,
       xxxVarFrameSetInfo AS i
WHERE  s.sourceID=v.sourceID AND
       v.frameSetID=i.frameSetID AND
       s.mergedClass=-1 AND v.variableClass=1
       AND v.kMeanMag<(i.kexpML-3.) AND v.kMeanMag>0.
       AND s.priOrSec=0
\end{verbatim}
\noindent
The first two lines of the ``WHERE'' statement links the 3 tables used in this
SQL statement. If these tables are not properly linked then you have what is called
a Cartesian join: every line in one is joined to every line in the others,
which takes a long time to return and does not return what is wanted. It is
best to link tables through the primary key (see the Schema 
Browser for the primary key and indices for each table), if possible, since the
primary key is unique and indexed, making a particularly quick return. The
\verb+Source+ and \verb+Variability+ tables have {\bf sourceID} as the primary
key, and \verb+VarFrameSetInfo+ has {\bf frameSetID} as the primary key. The
other selections are straight-forward: ${\bf s.mergedClass}=-1$ selects objects
classified as stars, ${\bf v.variableClass}=1$ selects objects classified as
variable, ${\bf s.priOrSec}=0$ selects objects that are away from overlaps.
${\bf v.kMeanMag}<({\bf i.kexpML}-3.)$ selects objects which are 3 magnitudes
brighter than the expected magnitude limit for that frame. The ${\bf
kMeanMag}>0$ removes any objects with default magnitudes. Additional terms 
for the number of good matches, the skewness, the median interval, or the  RMS
can also be applied as necessary.

To select a light curve from the archive, use the following, if the band passes
are uncorrelated (the best match table is a \verb+SourceXDetectionBestMatch+):\\[5mm]

\noindent
\begin{verbatim}
SELECT m.mjdObs,d.aperMag3,d.aperMag3Err, 
       d.ppErrBits,x.flag 
FROM   xxxSourceXDetectionBestMatch AS x, 
       xxxDetection AS d,Multiframe AS m 
WHERE  x.sourceID=NNN AND x.multiframeID= 
       d.multiframeID AND x.extNum=d.extNum AND 
       x.seqNum=d.seqNum AND x.multiframeID=
       m.multiframeID AND d.filterID=5 
ORDER BY m.mjdObs 
\end{verbatim}
\noindent 
This selects the full light curve for {\bf sourceID=NNN}.
Both the \verb+SourceXDetectionBestMatch+ and \verb+Detection+ tables contain
objects from all filters used in the programme, so it is important to have a
selection on {\bf filterID}. The time term is a modified Julian date, which is
found in the \verb+Multiframe+ table. Ordering by {\bf mjdObs} puts the results
in time order. The last 2 output terms are useful for determining how good the
light curve is. The {\bf ppErrBits} flag tells the user if the object has been
flagged for potential poor photometry. The best match table flag is also
useful: if {\bf flag=1} then the object has been matched to two sources,
suggesting either motion or deblending; if {\bf flag=2} then the lack of
detection is probably due to the object being within a dither offset of the
edge of the frame, and not a source that has taken a drastic dip in brightness. 
   
If the band passes are correlated, so that a \verb+SynopticSource+ table is
produced, then the following produces a similar light curve:\\[5mm]
\noindent
\begin{verbatim}
SELECT ml.meanMjdObs,e.kaperMag3,e.kaperMag3Err
       e.kppErrBits,x.flag
FROM   xxxSourceXSynopticSourceBestMatch AS x,
       xxxSynopticSource AS e,
       xxxSynopticMergeLog AS ml
WHERE  x.sourceID=NNN AND x.synFrameSetID=
       e.synFrameSetID AND x.synSeqNum=e.synSeqNum 
       AND e.synFrameSetID=ml.synFrameSetID 
ORDER BY ml.meanMjdObs
\end{verbatim}
\noindent  
For correlated bandpasses, light curves can be produced for several filters
together:\\[5mm]
\noindent 
\begin{verbatim}
SELECT ml.meanMjdObs,e.japerMag3,e.japerMag3Err,
       e.jppErrBits,e.kaperMag3,e.kaperMag3Err,
       e.kppErrBits,x.flag
FROM   xxxSourceXSynopticSourceBestMatch AS x,
       xxxSynopticSource AS e,
       xxxSynopticMergeLog AS ml
WHERE  x.sourceID=NNNNNNNNNN AND x.synFrameSetID=
       e.synFrameSetID AND x.synSeqNum=e.synSeqNum
       AND e.synFrameSetID=ml.synFrameSetID
ORDER BY ml.meanMjdObs
\end{verbatim}

\section{Overlaps}
\label{app:overlaps}

\begin{figure}
\includegraphics[width=84mm,angle=0]{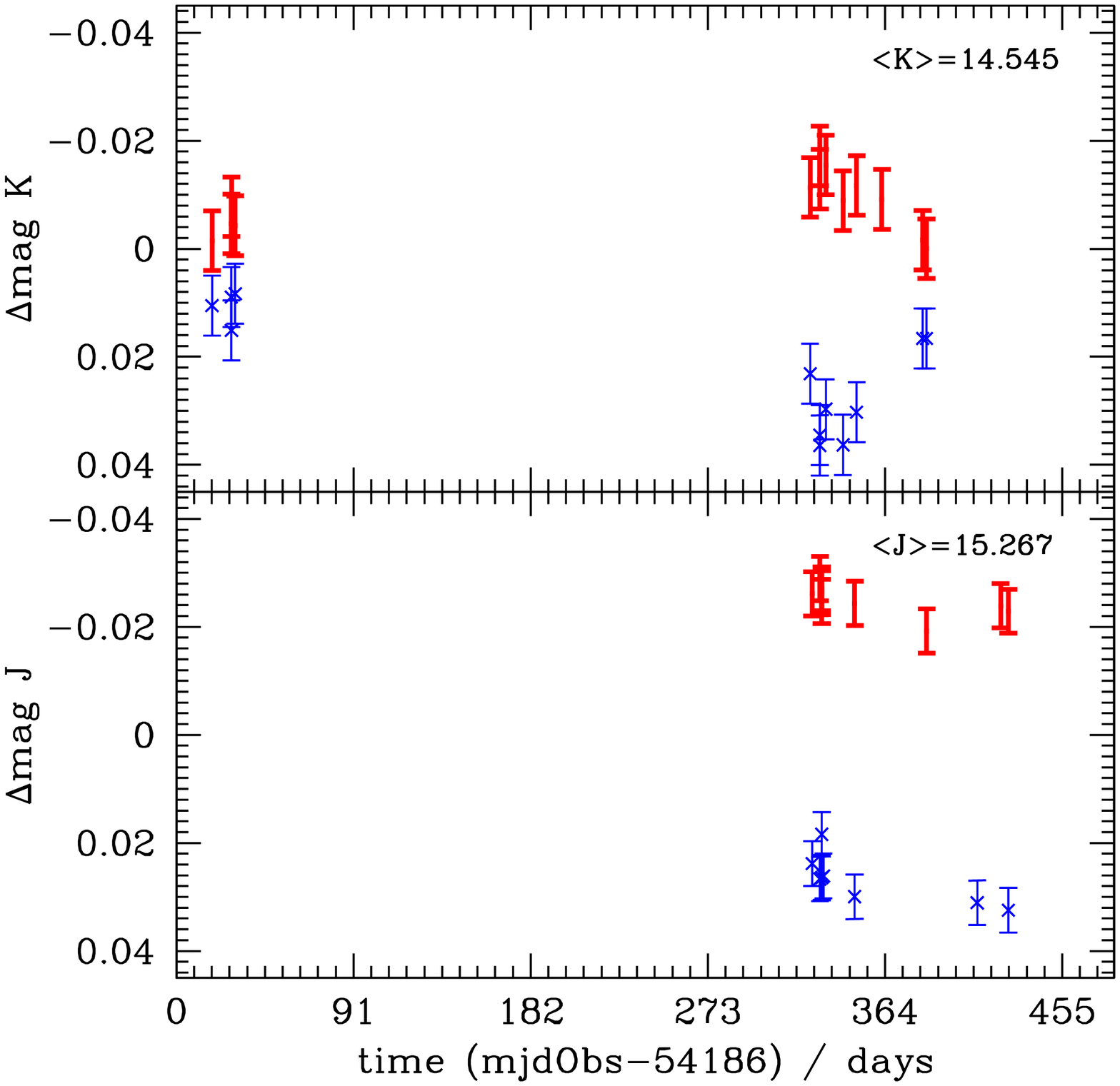}
\caption{Light curve for object, UDXS J105553.80+583930.7. The points that have ${\bf extNum}=4$ and $y<500.$ are marked by points with thick error bars and those with ${\bf extNum=4}$ and $y>3500.$
are marked by crosses with thin error bars. It is immediately apparent that the majority of
the variation is dependent on the position on the field, and is not intrinsic.}
\label{fig:lc4}
\end{figure}

While there are many real variables, there are also a great many which are
false. Consider the object in Fig~\ref{fig:lc4}. This object seems to have two
separate light-curves. Unfortunately this is not a high frequency periodic variable, but
indicates that there are still some spatial systematics that have not been
removed in the calibration \citep{Hdg09}. The variation is strongly correlated
with position on the field, since this object is in a region overlapping two
pointings.

\subsection{Describing overlaps}
\begin{figure}
\includegraphics[width=84mm,angle=0]{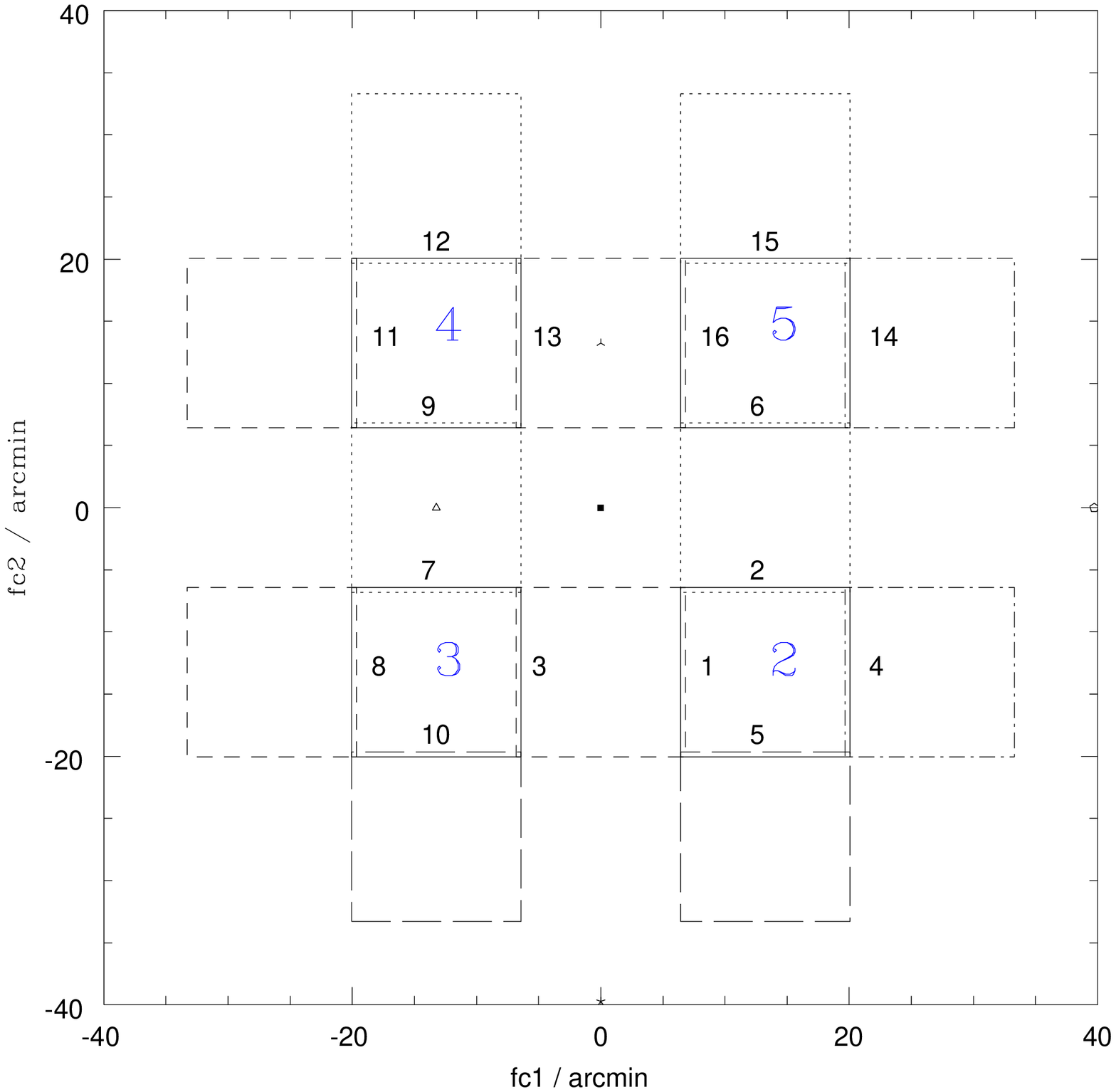}
\caption{Typical arrangements of pointings in the DXS (and most WFCAM
datasets). The black solid lines show the pointing in question. The dot,
dashed, long-dashed and dot-dashed lines show the neighbouring pointings with
overlaps along a detector edge. The extension numbers are given in large blue
numerals. Each overlap is numbered by a smaller black number. These numbers are
the indices in Table~\ref{tab:overlaps}.}
\label{fig:typoverlap}
\end{figure}

\begin{figure}
\includegraphics[width=84mm,angle=0]{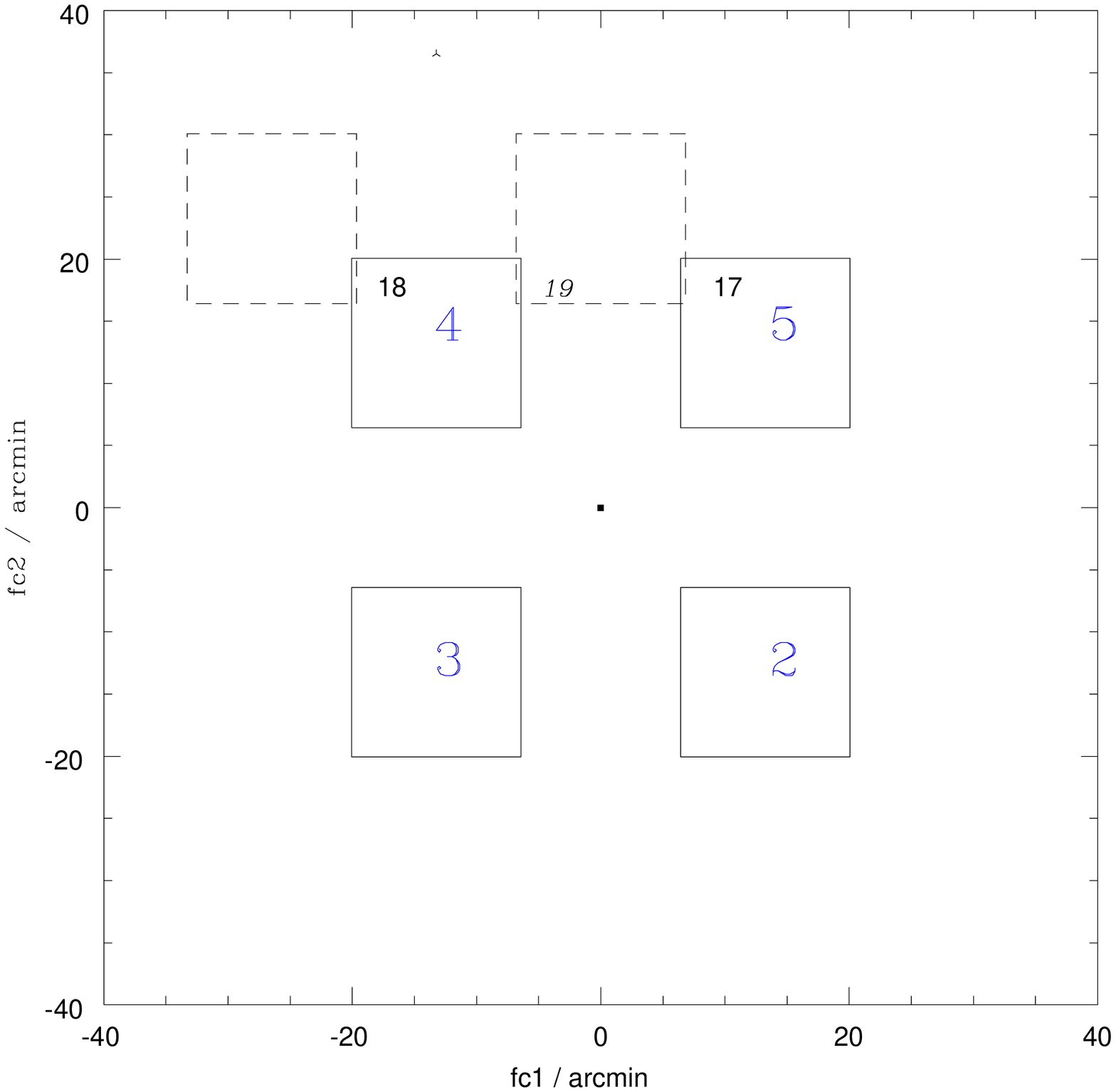}
\caption{Non-typical arrangements of overlaps in the DXS. These occurred when
good guide stars were not present at the original pointing positions and so the 
pointings were moved by a few arc-minutes. The solid line shows the
pointing and the dotted line shows the neighbouring pointing. 
${\it 19}$ is an overlap where there were no J or K sources (with at least 20
good observations).}
\label{fig:othoverlap}
\end{figure}

The overlaps between different pointings can be complicated, due to WFCAM having
4 non-buttable detectors. With the typical overlapping there are 16
possibilities (not counting ones across a diagonal from each other). We found
all these 16 possibilities (one along each side of each of the 4 detectors),
see Fig~\ref{fig:typoverlap}, with an additional 2 arrangements where the
pointings were not arranged as expected, see Fig~\ref{fig:othoverlap}.

To describe each of these overlaps, we need a consistent set of labels. We use
the following descriptor: X1\_X2\_N, where X1 is the {\bf extNum} of
the first frame, X2 is the {\bf extNum} of the second frame and N is
a number describing which borders of each detector are used. The
{\bf extNum}s do not uniquely describe an overlap by themselves as can be seen
in Fig~\ref{fig:typoverlap}. A further complication is that the x and y axes rotate
by 90 degrees from detector to detector \citep[see][Fig 1]{Dye06}. A typical
overlap lies along the full length of the x- or y- axis and the width of the
overlap is typically $3\%$ of the width of the detector, so the distribution
of the sources between any two frame sets can specify which overlap it is in.
We find frame set pairs by looking for sources which are in two different deep
stacks, i.e. {\bf priOrSec$>0$} and then using the \verb+dxsSourceNeighbours+
table and \verb+Source+ table to find the frame sets. We checked though all
overlaps (except at corners where the number of sources is minimal). In each
overlap, we calculate the mean and standard deviation of the x-coordinate 
and the mean and standard deviation of the y-coordinate for both frames. If
the overlap is long along the y-axis and narrow along the x-axis and close
to $x=0$ then $<x>\sim0.5w$, $\sigma(x)\sim0.25w$, $<y>\sim0.5l$, $<y>\sim0.25l$
($w$ is the width of the overlap, typically $0.03$ of the detector size $l$).
We assign one of three values to each of the four coordinates (x1, y1, x2, y2).
If the mean is close to the detector mid-point, then the value is 0, if it is
close to 0, then the value is 1, and if it is close to the maximum, then the
value is 2. We also make sure the standard deviation is high in the first case
and low in the later cases. The descriptor $N$ is therefore:

\begin{equation}
N=\sum_i a_i3^{4-i}
\end{equation}

where $a_i$ is the value of the $i^{th}$ coordinate. So the overlap of
extension 2, along the y-axis, at $x\sim0$ with extension 5, along 
the x-axis at $y\sim4000$ has $a_0=1$ ($<x1>$: minimum), $a_1=0$ ($<y1>$:
midpoint), $a_2=0$ ($<x2>$: midpoint), $a_3=2$ ($<y2>$: maximum). The
overlap is therefore 2\_5\_29. Table~\ref{tab:overlaps}
describes all the overlaps in Figs~\ref{fig:typoverlap} \& 
\ref{fig:othoverlap}. These overlaps all have mirror images (e.g.
2\_2\_11$=$2\_2\_19 and 2\_5\_55$=$5\_2\_15. We have always ordered them 
with the lowest extension number first.

\begin{table}
\label{tab:overlaps} 
\begin{tabular}{lccccc}
\hline
Index & ID & \multicolumn{2}{|c|}{No.} & \multicolumn{2}{|c|}{$\Delta\,$mag} \\
 & & J & K & J & K \\
\hline
1 & 2\_2\_11 & 13 & 13 & $-0.001\pm0.003$ & $0.000\pm0.005$ \\
2 & 2\_2\_33 & 25 & 32 &  ${\bf 0.011\pm0.002}$ & $0.009\pm0.003$  \\
3 & 2\_3\_15 & 11 & 16 & $-0.005\pm0.006$ & $0.006\pm0.003$ \\
4 & 2\_3\_21 & 24 & 62 & $0.001\pm0.007$ & $-0.003\pm0.002$ \\ 
5 & 2\_5\_29 & 26 & 29 & $0.001\pm0.003$ & $-0.007\pm0.002$ \\
6 & 2\_5\_55 & 58 & 45 & $0.004\pm0.001$ & ${\bf 0.010\pm0.002}$ \\
7 & 3\_3\_11 & 26 & 33 & $-0.009\pm0.002$ & ${\bf -0.017\pm0.002}$ \\
8 & 3\_3\_33 & 21 & 28 & ${\bf 0.017\pm0.003}$ & ${\bf 0.025\pm0.003}$ \\
9 & 3\_4\_15 & 44 & 62 & $0.007\pm0.002$ & $0.001\pm0.001$ \\
10 & 3\_4\_21 & 11 & 23 & ${\bf 0.017\pm0.003}$ & ${\bf -0.011\pm0.003}$ \\
11 & 4\_4\_11 & 17 & 32 & ${\bf -0.023\pm0.004}$ & ${\bf -0.034\pm0.003}$ \\
12 & 4\_4\_33 & 29 & 39 & $0.007\pm0.002$ & ${\bf 0.013\pm0.002}$ \\ 
13 & 4\_5\_15 & 10 & 16 & $-0.006\pm0.005$ & $-0.003\pm0.002$ \\
14 & 4\_5\_21 & 11 & 63 & $0.001\pm0.004$ & $0.003\pm0.003$ \\ 
15 & 5\_5\_11 & 35 & 32 & $-0.005\pm0.002$ & ${\bf -0.013\pm0.003}$ \\ 
16 & 5\_5\_33 & 30 & 21 & $-0.001\pm0.002$ & $0.007\pm0.004$ \\ 
\hline
17 & 2\_5\_24 & 4 & 2 & ${\bf -0.037\pm0.007}$ & $0.003\pm0.009$ \\
18 & 3\_4\_56 & 4 & 5 & ${\bf -0.029\pm0.005}$ & ${\bf -0.045\pm0.006}$ \\
19 & 2\_4\_10 & 0 & 0 & $---$ & $---$ \\
\hline
\hline
\end{tabular}
\caption{Table of the overlaps in the DXS. This gives the number of J and K
sources in the overlap regions, with at least 20 good observations per source.
$\Delta\,J$ and $\Delta\,K$ represent the typical difference in magnitude
across the offset. The bold numbers are offsets with $\Delta\,m>0.01$ mag}
\end{table} 
  
To get from $N$ to the borders, simply convert N into a 4-digit base 3 number (e.g. 11=0102). The first 2 digits tell you which is the border for the first frame (01: along the x-axis with $y\sim\,y_{min}$) and the last two digits tell you which is the border for the second frame: (02: along the x-axis with $y\sim\,y_{max}$).   

\subsection{Offsets}

To calculate the offsets between overlaps, we assume that each different type of
overlap is affected in the same way, wherever in the survey it is, i.e. a 
2\_2\_11 overlap between two observations in the ELIAS N1 region of the 
UKIDSS-DXS will have the same offset as another 2\_2\_11 overlap between two 
other observations in the ELIAS N1 region or a 2\_2\_11 overlap in the Lockman
Hole region. Finally we select objects that have a magnitude $(m_{limit}-7.)<m<(m_{limit}-3.)$ where $m_{limit}$ is the J or K-band limit in the \verb+VarFrameSetInfo+ table, see
\S\ref{sec:photometry}. The brighter limit should reduce effects of saturation
and the fainter limit should remove very noisy objects.

We calculate $\Delta\,mag$ very simply by finding the $3\sigma-$clipped
mean magnitude of each source from observations on one side of the overlap and
subtracting the equivalent value on the other side. Some of these sources 
vary in time and have different numbers of observations on each side,
corresponding to when the observations were taken. However, these variations
are different for each source and just add some addition random error if
there are enough sources in each overlap. 

The offsets for each overlap are shown in Table~\ref{tab:overlaps}.

\subsubsection{Variations with position and magnitude.}

\begin{figure} 
\includegraphics[width=84mm,angle=0]{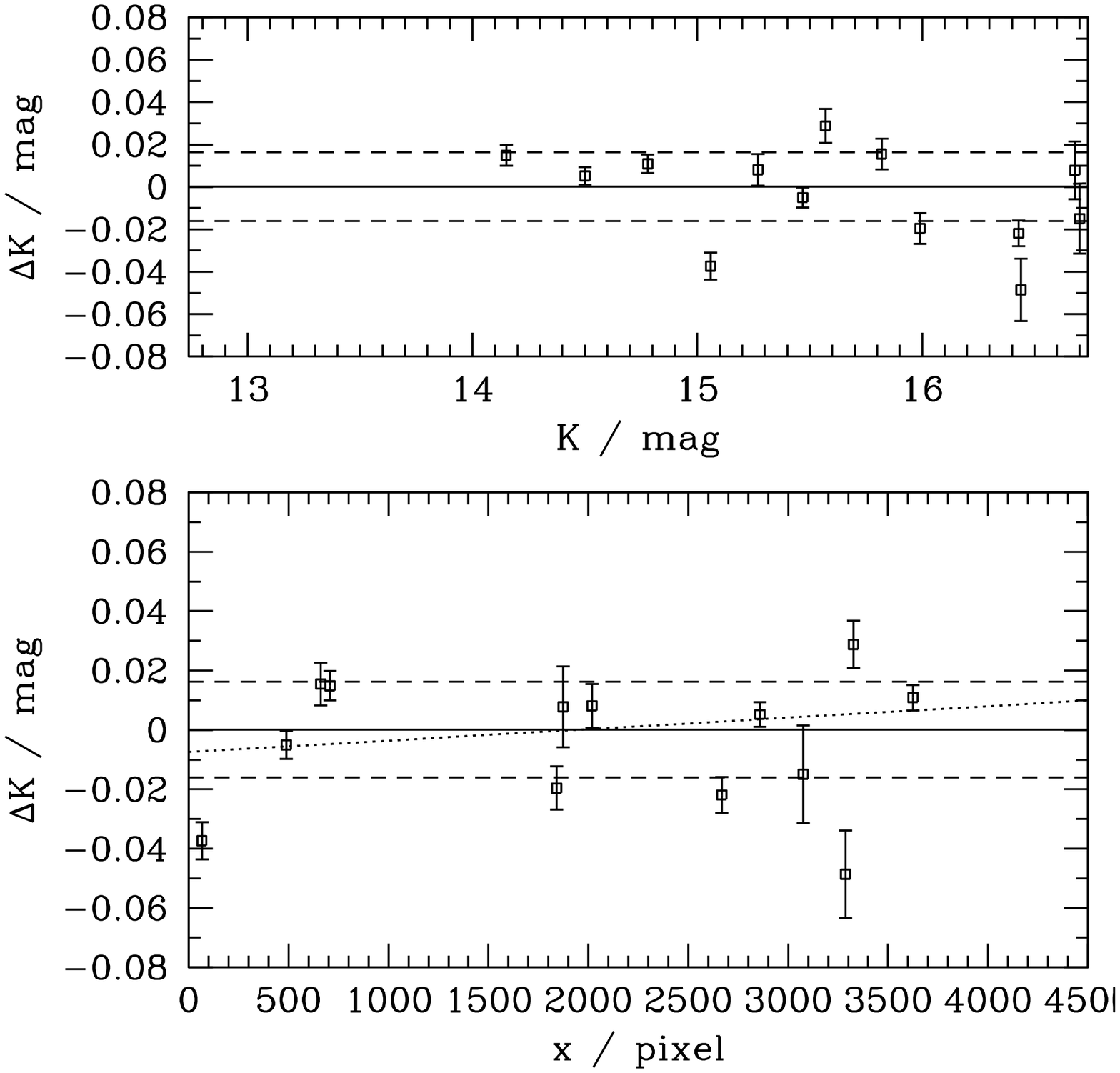}
\caption{Relative photometry of the sources on either side of the K-band
2\_2\_11 overlap. The lower plot shows the variation with the position, and the
upper plot shows variation with magnitude. The red solid line gives the mean
offset and the red dashed lines give the $3-sigma$ deviation from this offset.
The blue dotted line gives the best fit linear variation with position. This
overlap shows no significant offset and no significant trend
with magnitude or position.}
\label{fig:srcOver1}
\end{figure}

In each overlap region we have looked for variation with position or magnitude.
We look at the coordinate along the primary direction of the first extension
(i.e. along the side of the detector) We have plotted a subset of the overlaps
showing some with large offsets, some with small offsets in
Figs~\ref{fig:srcOver1} - Fig~\ref{fig:srcOver4}. The full set of plots can
be found at http://surveys.roe.ac.uk:8080/wsa/Overlaps/overlaps.html. While we
find some linear trends with position, there seems to be no noticeable variation with magnitude over the 4 magnitudes in question.
This suggests that the problem is not due to saturation or non-linearities in
the photometric solution, but is due to a spatial variations across the focal
plane which have not yet been eliminated.

\begin{figure}
\includegraphics[width=84mm,angle=0]{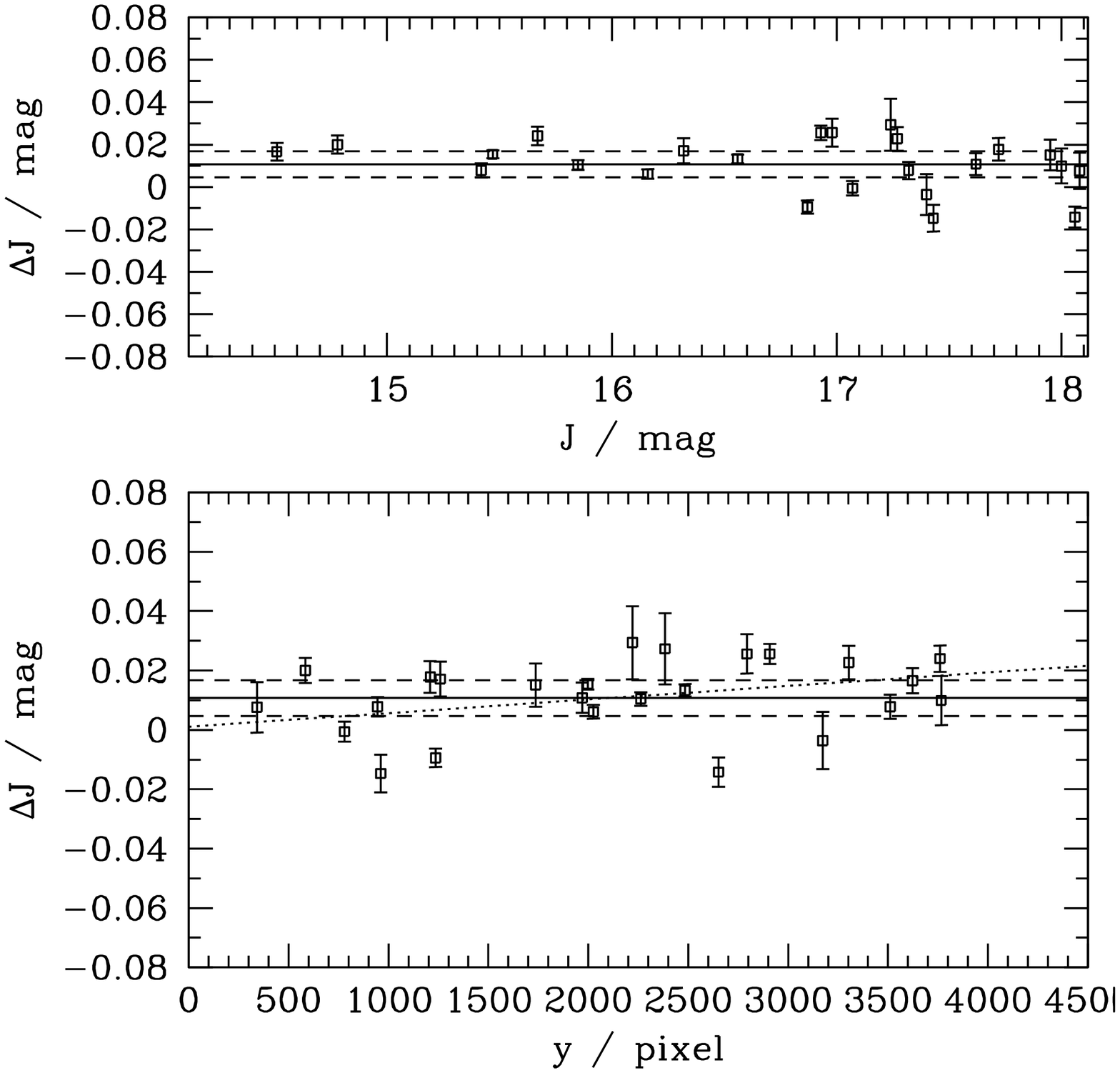}
\caption{Relative photometry of the sources on either side of the J-band
2\_2\_33 overlap. See Fig~\ref{fig:srcOver1} for details. This
overlap shows a significant offset $0.011\pm0.002$ mag. There is no significant
trend with magnitude or position.}
\label{fig:srcOver2}
\end{figure}

\begin{figure}
\includegraphics[width=84mm,angle=0]{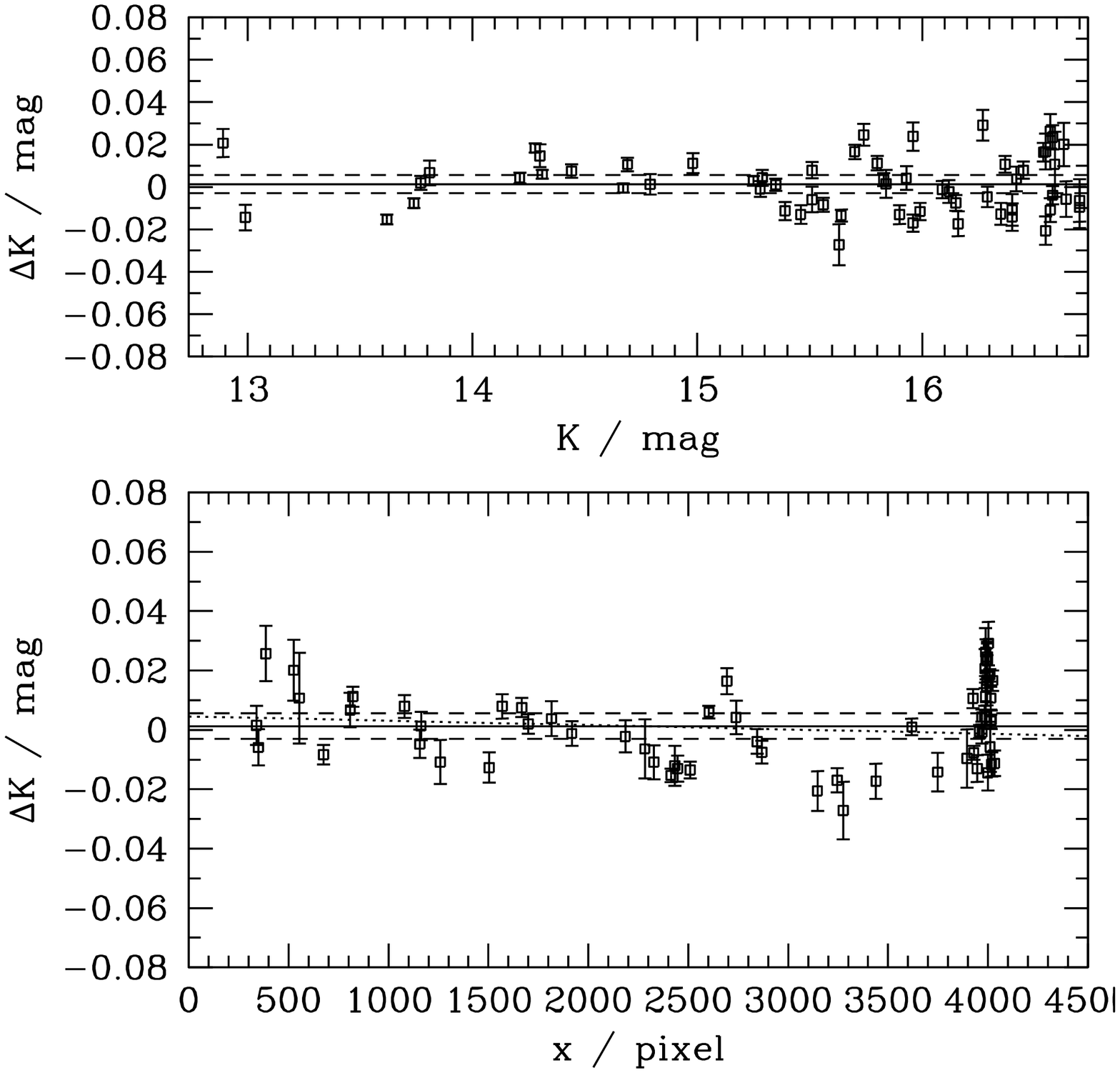}
\caption{Relative photometry of the sources on either side of the K-band
3\_4\_15 overlap. See Fig~\ref{fig:srcOver1} for details. This
overlap between two different detectors shows no significant offset or trends.
}
\label{fig:srcOver3}
\end{figure}  

\begin{figure}
\includegraphics[width=84mm,angle=0]{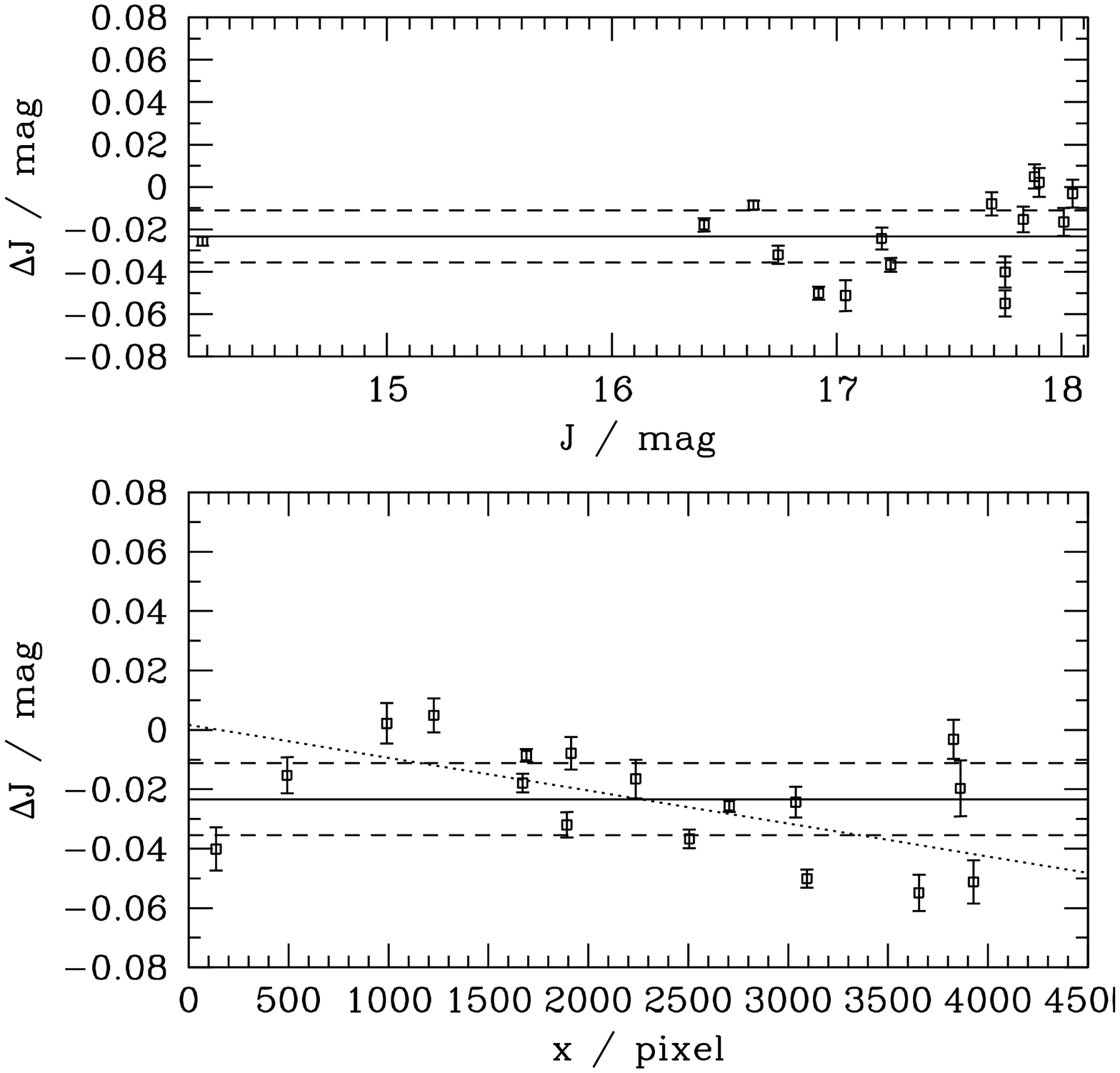}
\caption{Relative photometry of the sources on either side of the J-band
4\_4\_11 overlap. See Fig~\ref{fig:srcOver1} for details. There is a
significant offset $-0.023\pm0.004$ and a strong trend with position. Again
there is no trend with magnitude.}
\label{fig:srcOver4}
\end{figure}
 
\subsection{Correcting lightcurves}

\begin{figure} 
\includegraphics[width=84mm,angle=0]{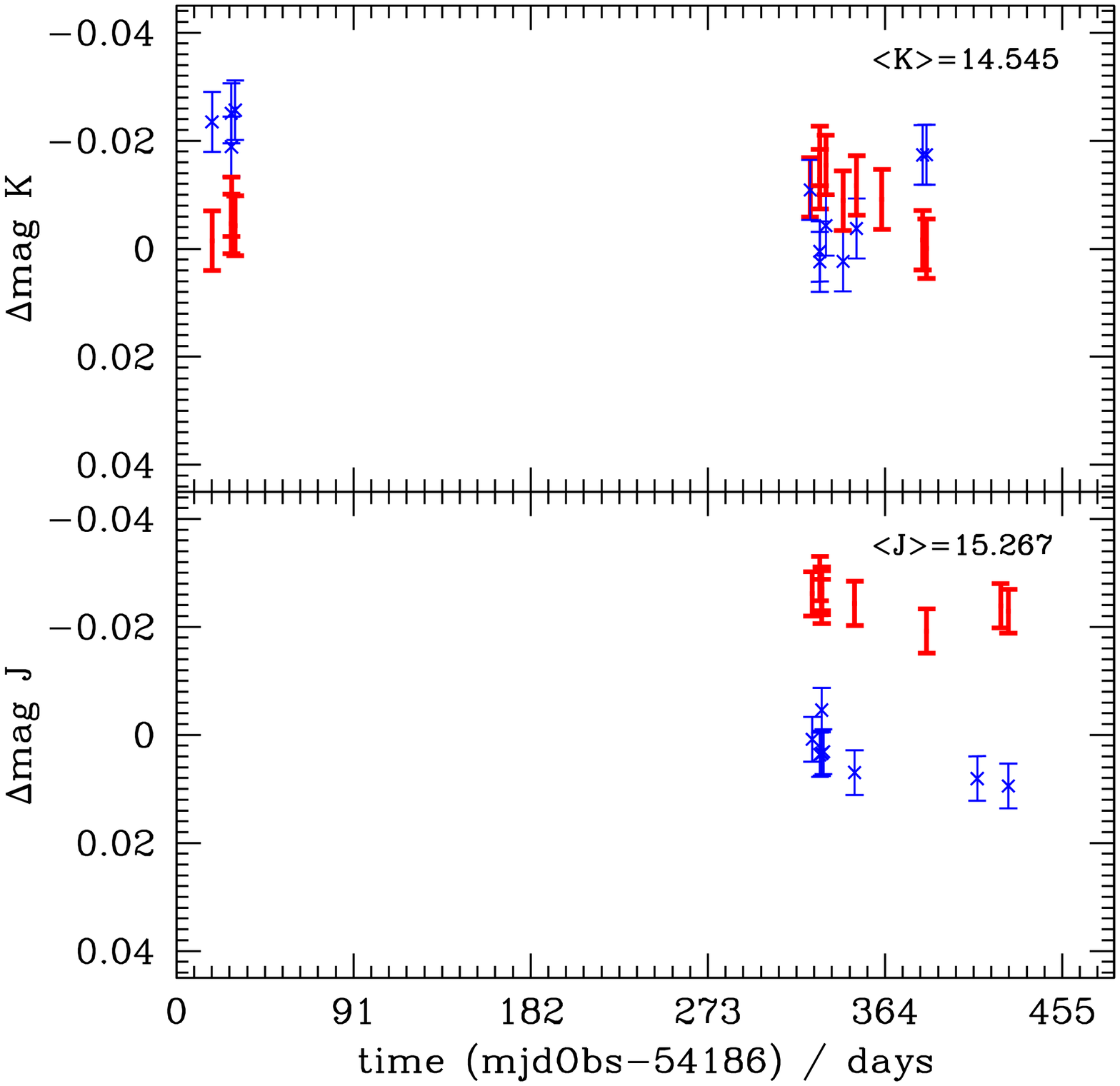}
\includegraphics[width=84mm,angle=0]{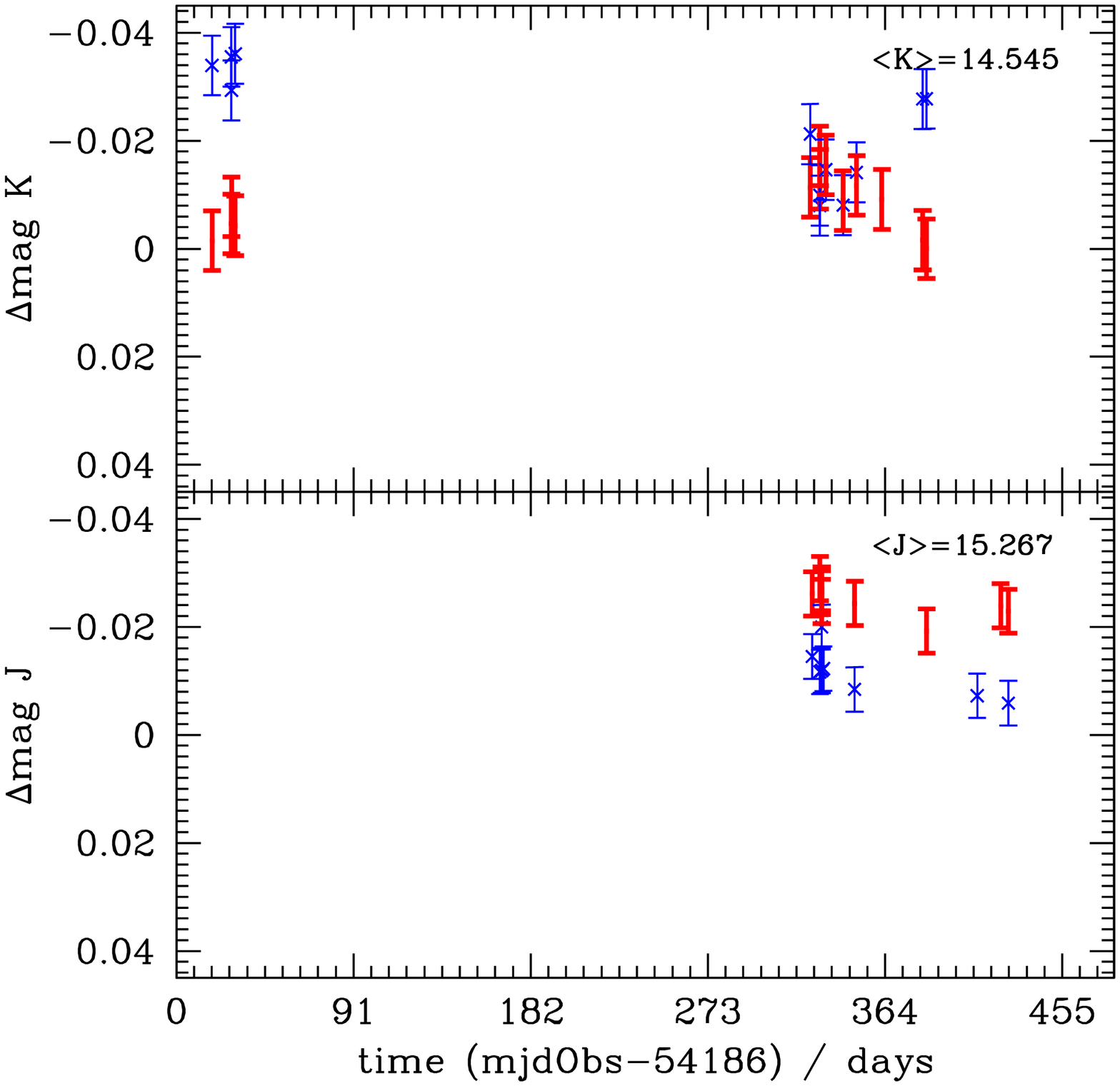}
\caption{Light curve for object, UDXS J105553.80+583930.7. Same labelling as for 
Fig~\ref{fig:lc4}, except that in the upper plot (a), the points having $y>3500.$ (crosses with thin error bars) have been corrected by a simple offset for the whole overlap, and in the lower plot (b), these points have been corrected using a linear fit to position. The points having $y<500.$ have not been altered in either case. With this second correction most of the difference is removed.}
\label{fig:lc5}
\end{figure}

Above, we have done a careful analysis of the overlaps and
conclude that the offsets across most overlap regions are low level $<0.01$ mag
\citep[as expected from][]{Hdg09}, but in a few cases, particularly between
extension number 4 on one frame and extension number 4 on the other, along the
x-axis, can be several hundredths of a magnitude --- almost 10 times the expected value (see Table~\ref{tab:overlaps}). There
is no magnitude dependence, but there is a positional dependence along some of
the overlaps, so this is not an effect of saturation. Fig~\ref{fig:lc5}a shows
the effect of correcting the light curve in Fig~\ref{fig:lc4} with the average
offset measured along the overlap. Some of the difference in the lightcurves is
removed. Fig~\ref{fig:lc5}b shows the effect of correcting with a positional
dependent offset, and in this case most of the difference is removed.
 
\label{lastpage}

\end{document}